\documentclass[10pt,a4paper,twoside]{report}
\usepackage[utf8]{inputenc}
\usepackage[T1]{fontenc}
\usepackage[english]{babel}
\usepackage{graphicx}
\usepackage{xcolor}
\usepackage{amsmath}  
\usepackage{amsthm}   
\usepackage{amsfonts}
\usepackage{amssymb}
\usepackage{url}
\usepackage{tabularx}
\usepackage{multirow}
\usepackage{lipsum}
\usepackage{bm}
\usepackage{bbm}
\usepackage{physics}
\usepackage{tikz}
\usepackage[compat=1.1.0]{tikz-feynman} 
\usepackage{ragged2e} 
\usepackage{lipsum}
\usepackage{enumerate}
\usepackage{hhline}
\usepackage{datatool}
\usepackage{booktabs}
\usepackage{verbatim}

\usepackage{geometry}	
\usepackage{setspace}

\usepackage{parskip}
\usepackage{hyperref} 
\usepackage[figure,table]{hypcap}
\usepackage{notoccite}
\hypersetup{colorlinks,
            linkcolor=black,  
            anchorcolor=black,
            citecolor=black,  
            filecolor=black,  
            menucolor=black,  
            pagecolor=black,  
            urlcolor=black,   
	        pdftitle={Thesis Lucas Sa},
            pdfauthor={Lucas Sá}}

\usepackage[numbers,sort&compress]{natbib} 

\newcommand{\acknowledgments}{@undefined}
\addto\captionsenglish{\renewcommand{\acknowledgments}{Acknowledgments}}
\addto\captionsenglish{\renewcommand{\bibname}{References}}

\newcommand{\coverThesis}{@undefined}
\newcommand{\coverSupervisors}{@undefined}
\newcommand{\coverExaminationCommittee}{@undefined}
\newcommand{\coverChairperson}{@undefined}
\newcommand{\coverSupervisor}{@undefined}
\newcommand{\coverMemberCommittee}{@undefined}

\addto\captionsenglish{\renewcommand{\coverThesis}{Thesis to obtain the Master of Science Degree in}}
\addto\captionsenglish{\renewcommand{\coverSupervisors}{Supervisors}}
\addto\captionsenglish{\renewcommand{\coverExaminationCommittee}{Examination Committee}}
\addto\captionsenglish{\renewcommand{\coverChairperson}{Chairperson}}
\addto\captionsenglish{\renewcommand{\coverSupervisor}{Supervisor}}
\addto\captionsenglish{\renewcommand{\coverMemberCommittee}{Member of the Committee}}

\def\FontLb{
  \usefont{T1}{phv}{b}{n}\fontsize{16pt}{16pt}\selectfont}
\def\FontMn{
  \usefont{T1}{phv}{m}{n}\fontsize{14pt}{14pt}\selectfont}
\def\FontMb{
  \usefont{T1}{phv}{b}{n}\fontsize{14pt}{14pt}\selectfont}
\def\FontSn{
  \usefont{T1}{phv}{m}{n}\fontsize{12pt}{12pt}\selectfont}

\newcommand{\kket}[1]{\ket{\ket{#1}}}
\newcommand{\bbra}[1]{\bra{\bra{#1}}}
\renewcommand*\d{\mathop{}\!\mathrm{d}}
\newcommand{\dirac}[1]{\,\delta\!\left(#1\right)}
\newcommand{\heav}[1]{\,\Theta\!\left(#1\right)}
\renewcommand{\vec}[1]{\underline{#1}}
\newcommand{\s}{\mathfrak{s}}
\global\long\def\T{\sf{T}}

\global\long\def\Tr{\mathrm{Tr}}

\global\long\def\im{\mathrm{Im}}
\global\long\def\re{\mathrm{Re}}

\global\long\def\det{\mathrm{det}}
\global\long\def\av#1{\left\langle #1 \right\rangle }

\begin{document}

\pagestyle{plain}

\pagenumbering{roman}


\thispagestyle{empty}

\flushleft\includegraphics[width=0.5\textwidth]{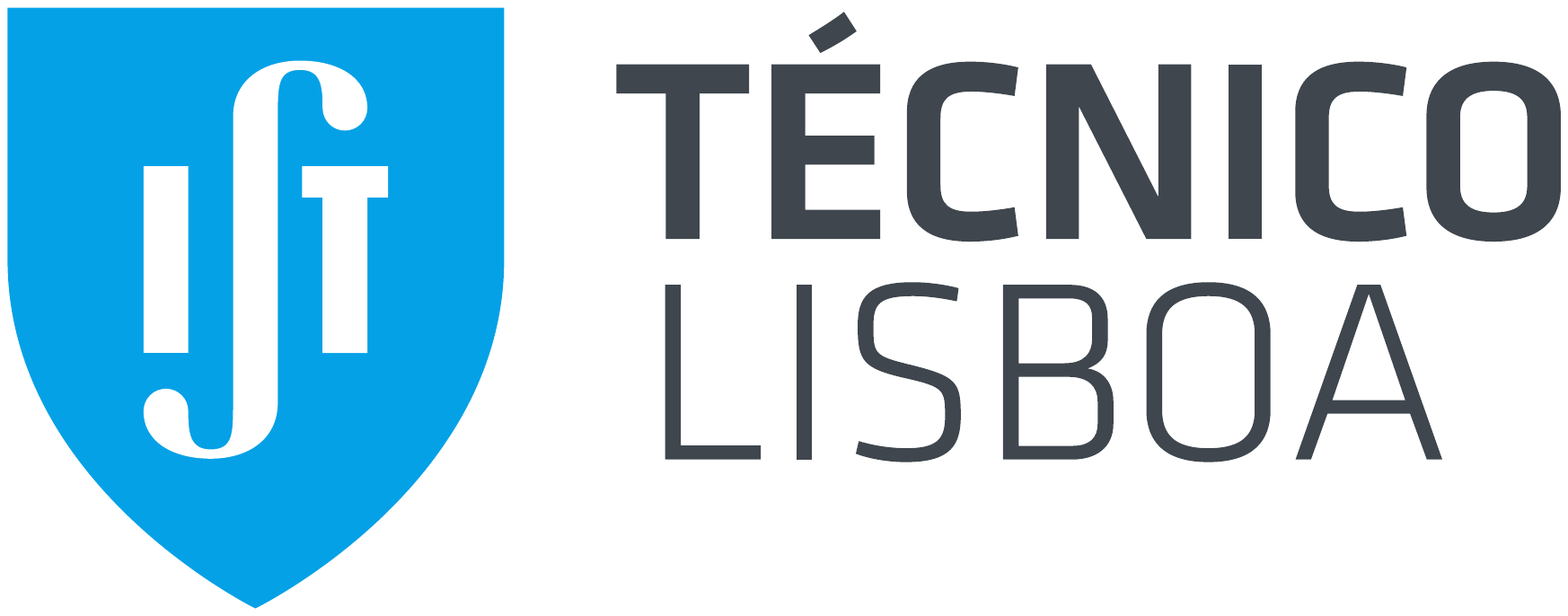}

\begin{center}

\vspace{1.5cm}
{\FontLb Dissipation and decoherence for generic\\ open quantum systems:}\\
\vspace{+0.25cm}
{\FontMn Spectral and steady-state properties of random Liouvillian operators}\\
\vspace{2.6cm}
{\FontMb Lucas de Barros Pacheco Seara de Sá} \\
\vspace{2.0cm}
{\FontSn \coverThesis}\\
\vspace{0.3cm}
{\FontLb Engineering Physics}\\
\vspace{1.0cm}
{\FontSn %
\begin{tabular}{ll}
 \coverSupervisors: & Prof.\ Pedro José Gonçalves Ribeiro \\
                    & Prof.\ Toma\v{z} Prosen
\end{tabular} } \\
\vspace{1.0cm}
{\FontMb \coverExaminationCommittee} \\
\vspace{0.3cm}
{\FontSn %
\begin{tabular}{c}
\coverChairperson:     Prof.\ Pedro Miguel Félix Brogueira \\ 
\coverSupervisor:      Prof.\ Pedro José Gonçalves Ribeiro \\
\coverMemberCommittee: Prof.\ Vítor João Rocha Vieira
\end{tabular} } \\
\vfill
{\FontMb September 2019} \\ 
\end{center}

\cleardoublepage



\section*{Abstract}

\justifying{
We study generic open quantum systems with Markovian dissipation, focusing on a class of stochastic Liouvillian operators of Lindblad form with independent random dissipation channels (jump operators) and a random Hamiltonian.
We perform a thorough numerical study, focusing on global spectral features, the spectral gap, and the steady-state purity and statistics. We establish that all properties follow three different regimes as a function of the dissipation strength, whose boundaries depend on the particular observable. Within each regime, we determine the scaling exponents with the dissipation strength and system size. On the analytical side, we compute the average spectral gap at arbitrary dissipation and provide simple closed-form expressions for the asymptotic values at strong and weak dissipation strength. 
We also consider spectral correlations in generic complex spectra, such as that of the Liouvillian. We introduce and study, both analytically and numerically, the complex ratio of the nearest-neighbor spacing by next-to-nearest-neighbor spacing. 
Besides the usual level repulsion, we find that complex spacing ratios have a nontrivial angular dependence which offers a distinct signature of quantum chaos. It distinguishes integrable systems, which support the flat ratio distribution characteristic of Poisson statistics, and chaotic dissipative quantum systems, which conform to Random Matrix Theory statistics. We apply this new signature of quantum chaos to random Liouvillians and boundary-driven dissipative spin-chains.
The results of this thesis can help understand the long-time dynamics, steady-state properties, and spectral correlations of generic dissipative systems.
}

\vfill

\noindent\textbf{\Large Keywords:} open quantum system, Liouvillian operator, Lindblad equation, random matrix theory, quantum chaos.

\cleardoublepage


\section*{Resumo}


\justifying{
Estudamos sistemas quânticos abertos genéricos com dissipação Markoviana, focando a nossa atenção numa classe de operadores Liouvillianos estocásticos do tipo Lindbladiano, com canais de dissipação aleatórios independentes e um Hamiltoniano aleatório.
Realizamos um estudo numérico extensivo, considerando propriedades espectrais globais, o hiato espectral e a pureza e estatística do estado estacionário. Estabelecemos que todas estas propriedades seguem três regimes diferentes, em função da intensidade de dissipação, cujas fronteiras dependem da observável em questão. Dentro de cada regime, determinamos os expoentes de \textit{scaling} com intensidade de dissipação e tamanho do sistema. Analiticamente, calculamos o hiato espectral médio para dissipação arbitrária, dando igualmente expressões simples em forma fechada para os valores assintóticos com dissipação forte ou fraca.
Consideramos igualmente correlações espectrais em espectros complexos genéricos, tais como o do Liouvilliano. Introduzimos e estudamos, tanto numerica como analiticamente, o rácio complexo do espaçamento a primeiros vizinhos pelo espaçamento a segundos vizinhos. Além da usual repulsão de níveis, determinamos que rácios de espaçamentos complexos têm uma dependência angular não trivial que oferece uma assinatura de caos quântico bastante clara. Permite distinguir sistemas integráveis, que suportam a estatística de Poisson, de sistemas quânticos dissipativos caóticos, que seguem as previsões da teoria das matrizes aleatórias. Aplicamos esta nova assinatura de caos quântico a Liouvillianos aleatórios e a cadeias de spin dissipativas forçadas nos extremos.
Os resultados desta tese podem ajudar a compreender melhor a dinâmica a tempos longos, propriedades estacionárias e correlações espectrais de sistemas dissipativos genéricos.
}

\vfill

\noindent\textbf{\Large Palavras-chave:} sistemas quânticos abertos, operador Liouvilliano, equação de Lindblad, teoria de matrizes aleatórias, caos quântico.

\cleardoublepage


\section*{\acknowledgments}


I would like to start by thanking my thesis supervisors, Professor Pedro Ribeiro and Professor Toma\v{z} Prosen, for guiding me throughout this past year, for their enthusiastic encouragement on my progress, sharp critique, general advice, and for always finding some time for me in their (quite) busy schedules. I have thoroughly enjoyed the many discussions we had. This work could not exist without them.

\noindent I also thank Dr.\ Tankut Can for enlightening discussions on the results of this thesis, and for sharing his notes, which allowed for a great improvement on the analytical computations of the spectral gap.

\noindent I extend my gratitude to Francisco Faro for proofreading parts of this thesis and to Henrique Silvério for helping me with tricky normalization factors and for general office comradeship.

\noindent I acknowledge the computer resources of CeFEMA and, particularly, of FMF in Ljubljana. The numerical computations of this work were mainly performed on FMF's cluster \textit{Olimp} and supported by the H2020 ERC Advanced Grant 694544-OMNES. I also acknowledge partial support by FCT through Pedro Ribeiro's Investigator contract IF/00347/2014 and Grant No.\ UID/CTM/04540/2019.
\cleardoublepage


\chapter*{List of Publications}

{\justifying
\setlength\parskip{6pt}
\setlength\parindent{18pt}
The work of this thesis lead to the following two publications:

\begin{itemize}
    \item[] \hspace{-18pt}\cite{sa2019}\hspace{+5pt} Lucas Sá, Pedro Ribeiro, and Toma\v{z} Prosen\\
    \textit{Spectral and Steady-State Properties of Random Liouvillians}\\
    \href{https://arxiv.org/abs/1905.02155}{arXiv:1905.02155} (2019)
\end{itemize}

\textit{Abstract:} We study generic open quantum systems with Markovian dissipation, focusing on a class of stochastic Liouvillian operators of Lindblad form with independent random dissipation channels (jump operators) and a random Hamiltonian. We establish that the global spectral features, the spectral gap, and the steady-state properties follow three different regimes as a function of the dissipation strength, whose boundaries depend on the particular observable. Within each regime, we determine the scaling exponents with the dissipation strength and system size. We find that, for two or more dissipation channels, the spectral gap increases with the system size. The spectral distribution of the steady-state is Poissonian at low dissipation strength and conforms to that of a random matrix once the dissipation is sufficiently strong. Our results can help to understand the long-time dynamics and steady-state properties of generic dissipative systems.

\vspace{+1em}
\begin{itemize}
     \item[] \hspace{-18pt}\cite{sa2019CSR}\hspace{+3.5pt} Lucas Sá, Pedro Ribeiro, and Toma\v{z} Prosen\\
     \textit{Complex spacing ratios: a signature of dissipative quantum chaos}\\
     \href{https://arxiv.org/abs/1910.12784}{arXiv:1910.12784} (2019)
\end{itemize}

\textit{Abstract:} We introduce a complex-plane generalization of the consecutive level-spacing ratio distribution, used to distinguish regular from chaotic quantum spectra. 
Our approach features the distribution of complex-valued ratios between nearest- and next-to-nearest neighbor spacings. We show that this quantity can successfully detect the chaotic or regular nature of complex-valued spectra. This is done in two steps. 
First, we show that, if eigenvalues are uncorrelated, the distribution of complex spacing ratios is flat within the unit circle, whereas random matrices show a strong angular dependence in addition to the usual level repulsion. The universal fluctuations of Gaussian Unitary and Ginibre Unitary universality classes in the large-matrix-size limit are shown to be well described by Wigner-like surmises for small-size matrices with eigenvalues on the circle and on the two-torus, respectively. To study the latter case, we introduce the Toric Unitary Ensemble, characterized by a flat joint eigenvalue distribution on the two-torus. 
Second, we study different physical situations where nonhermitian matrices arise: dissipative quantum systems described by a Lindbladian, non-unitary quantum dynamics described by nonhermitian Hamiltonians, and classical stochastic processes. We show that known integrable models have a flat distribution of complex spacing ratios whereas generic cases, expected to be chaotic, conform to Random Matrix Theory predictions. Specifically, we were able to clearly distinguish chaotic from integrable dynamics in boundary-driven dissipative spin-chain Liouvillians and in the classical asymmetric simple exclusion process and to differentiate localized from delocalized regimes in a nonhermitian disordered many-body system.
}
\phantomsection
\addcontentsline{toc}{chapter}{List of Publications}
\cleardoublepage

\tableofcontents
\cleardoublepage 

\listoftables
\cleardoublepage 

\listoffigures
\cleardoublepage 


\chapter*{List of Abbreviations}
 
\newcommand*{\addacronym}[2]{%
    \DTLnewrow{acronyms}%
    \DTLnewdbentry{acronyms}{Acronym}{#1}%
    \DTLnewdbentry{acronyms}{Description}{#2}%
}
 
\DTLnewdb{acronyms}
\addacronym{RMT}{Random Matrix Theory}
\addacronym{OQS}{Open quantum system}
\addacronym{GUE}{Gaussian Unitary Ensemble}
\addacronym{GOE}{Gaussian Orthogonal Ensemble}
\addacronym{GSE}{Gaussian Symplectic Ensemble}
\addacronym{CUE}{Circular Unitary Ensemble}
\addacronym{COE}{Circular Orthogonal Ensemble}
\addacronym{CSE}{Circular Symplectic Ensemble}
\addacronym{GinUE}{Ginibre Unitary Ensemble}
\addacronym{GinOE}{Ginibre Orthogonal Ensemble}
\addacronym{GinSE}{Ginibre Symplectic Ensemble}
\addacronym{TUE}{Toric Unitary Ensemble}
\addacronym{SUE}{Spherical Unitary Ensemble}
\addacronym{ED}{Exact diagonalization}
\addacronym{NN}{Nearest-neighbour}
\addacronym{NNN}{Next-to-nearest-neighbour}
\addacronym{kNN}{$k^\mathrm{th}$-nearest-neighbor}

\DTLsort{Acronym}{acronyms}

\begin{itemize}
\DTLforeach*{acronyms}{\thisAcronym=Acronym,\thisDesc=Description}%
   {\item[] \textbf{\thisAcronym}\quad \thisDesc}%
\end{itemize}

\cleardoublepage 

\setcounter{page}{1}
\pagenumbering{arabic}

\justifying
\setlength\parskip{6pt}
\setlength\parindent{18pt}


\chapter{Introduction}
\label{chapter:introduction}

\section{Background and motivation}
\label{section:background_motivation}
No physical system is truly isolated. Any realistic quantum system will be influenced by its environment, which typically has a much larger number of degrees of freedom. The physical properties of a set of interacting quantum mechanical degrees of freedom can change fundamentally in the presence of this macroscopic environment, which induces dissipation and decoherence. It is the theory of open quantum systems (OQS) which addresses these important phenomena~\cite{breuerpetruccione,weiss2012,braun2001,rivas2012,alicki2007}.

The idealization of a closed system is an extremely useful one but it fails in several situations. Experimental techniques and technologies such as inelastic neutron scattering, nuclear magnetic resonance or optical and electronic spectroscopy are applied to nonequilibrium OQS, therefore a thorough understanding of its dynamics is needed. Information theory and information processing are heavily influenced by it since coherent manipulation of quantum states is a key ingredient in them. It is also crucial on a more fundamental level, namely for the conceptual understanding of the act of quantum measurement. The very act of measuring requires the coupling of the system to an external measuring device, rendering the system open and, as is known, this coupling to the measuring apparatus influences the results of the measurement itself.

In general, a microscopic description in terms of the whole set of degrees of freedom of both the system and its environment is impossible. Even if we could proceed with such a complete description, it would contain much more information than we are interested in. One thus seeks an effective description in terms of the open system degrees of freedom, by tracing out the bath degrees of freedom~\cite{breuerpetruccione}. One possible way to accomplish this is by modeling the evolution of an OQS by a master equation for its density matrix, generalizing the Schr\"odinger equation, and which in our work we shall take to be of the Lindblad form~\cite{lindblad1976,gorini1976,alicki2007}. In this description, the dynamics of the system are described by some effective Hamiltonian and a set of jump operators connecting the system to the bath. 

For closed systems, the Schr\"odinger equation can be solved in essentially two cases: integrable (exactly solvable) systems and chaotic (fully random) ones, where we have adopted the nomenclature from their limiting classical behavior. The former have a high enough degree of symmetry and infinite sets of conservation laws such that explicit solutions for particular systems can be found; the latter are complex enough such that microscopic details of particular systems get averaged out and only universal behavior survives. 

For open systems, many explicit solutions of the first type have been constructed for the Lindblad equation~\cite{prosen2008,prosen2010,ribeiro2019,rowlands2018,medvedyeva2016,eisler2011,banchi2017,prosen2011,prosen2014,prosen2015,karevski2013,ilievski2017} over the years. However, only very recently~\cite{sa2019,denisov2018,can2019a,can2019} has the second, i.e.\ the generic, type started being addressed. As enlightening as exact solutions to simple models are, they do not tell us much about the generic case. They are undoubtedly important in partially characterizing the behavior of limited classes of systems, but they do not provide the full description of the whole spectrum of OQS nor of universal behavior common to several classes. The solution for chaotic OQS must, then, resort to a statistical theory in the thermodynamic (infinitely-large system) limit. Since our quantum system is described by some set of operators (the Hamiltonian and the jump operators), we must turn to Random Matrix Theory (RMT)~\cite{haake2013,mehta2004,akemann2011,forrester2010,livan2018,guhr1998,schomerus2017}. The thermodynamic limit is, then, implemented by considering the large-$N$ limit, i.e.\ $N\times N$ matrices with $N\to\infty$.

To understand the objectives of this thesis it is important to explain Lindbladian dynamics and RMT in some more detail first. With these two key ingredients, we can formulate a theory of generic open quantum systems.

\subsection{Lindbladian dynamics}
\label{subsection:lindbladian_dynamics}
The Lindblad equation can be obtained as an approximation to the Schr\"odinger equation for the joint evolution of the system plus the environment. From another perspective, it can also be seen as a generalization of the Schr\"odinger equation for the system alone. 

The Schr\"odinger equation for the time evolution of the state of the system, under the action of an Hamiltonian $H$, can be rewritten as the von Neumann equation for the system's density matrix $\rho$ (which contains the probabilities of each state as its diagonal entries):
\begin{equation}
    \frac{\partial}{\partial t}\rho(t)=-i\comm{H}{\rho}(t)\,.
\end{equation}
One then assumes that the bath is much larger than the system and that the coupling between them is weak (Born approximation), leading to a factorization of the total density matrix into system and bath parts. One further assumes that there is a clear separation of the system and environment timescales (Markov approximation), leading the future state to depend only on the present and not the past states, effectively rendering the bath memoryless. A standard procedure~\cite{breuerpetruccione} is then to apply second-order perturbation theory, take the trace over the bath degrees of freedom and obtain the Lindblad equation for the system's density matrix. It is a master equation $\partial_t \rho(t)=\mathcal{L}[\rho]$, where the Liouvillian superoperator\footnote{Here we are describing the state of the system by a linear operator~$\rho$ (its density matrix) and not by a vector~$\ket\psi$ in the Hilbert space. Hence, the evolution is determined by a \emph{superoperator} $\mathcal{L}$ acting on operators and not by an operator acting on vectors as usual. This is discussed in Section \ref{section:superoperators} below.} $\mathcal{L}$ acts on $\rho$ as
\begin{equation}\label{eq:Liouvillian_diagonal}
    \mathcal{L}[\rho]=-i\comm{H}{\rho}+\sum_\ell\left(W_\ell\,\rho W_\ell^\dagger-\frac{1}{2}W_\ell^\dagger W_\ell\rho-\frac{1}{2}\rho W_\ell^\dagger W_\ell\right)\,,
\end{equation}
with $H$ the Hamiltonian and $W_\ell$ traceless jump operators. The number of linearly independent jump operators (i.e.\ the number of different decay channels) is $r\in[1,N^2-1]$ ($N$ the dimension of the Hilbert space), i.e.\ $\ell=1,\dots,r$. Note that $H$ need not be the bare Hamiltonian of the system alone, but may be ``renormalized'' by the trace operation. 

It can be shown~\cite{alicki2007} that the evolution of the density matrix under the master equation with Liouvillian operator of the Lindblad type is hermitian, trace-preserving and completely positive\footnote{A positive map sends positive density operators to positive density operators. Complete positivity is a stronger condition meaning that the map is positive even when acting only on part of a larger system. That is, the map $\mathcal{L}$ is completely positive if $\mathcal{L}\otimes \mathbbm{1}_M$ is positive for all $M$, where $\mathbbm{1}_M$ is the $M$-dimensional identity.}. This means that, if we start with a density matrix $\rho(0)$ which is physical, i.e.\ is hermitian and has probabilities as diagonal entries (positive numbers which add up to one), then those three properties ensure it will remain physical at all times. 
Alternatively, the Lindblad equation can also be derived as the most general linear equation which has those important three properties~\cite{hall2014}. 

In a closed quantum system, the spectrum of the Liouvillian is purely imaginary, with its eigenvalues given by the differences between the eigenenergies of the system. The states only oscillate, with frequency dictated by these energy differences. In an open system, the decay channels induce a renormalization of the energy levels and, more importantly, introduce a real part of the eigenvalues, the latter being related to dissipation and decoherence. We can decompose the density matrix of the system into a superposition of its eigenmodes, with each eigenmode decaying at a rate set by the inverse of the real part of the corresponding eigenvalue. From the form of the Lindblad equation, it results that there is always a zero eigenvalue (in generic systems there is only one, although more could exist). The corresponding eigenmode is the only one that does not decay in time and hence it is the steady-state density matrix. This special state completely determines the long-time asymptotics and it can be characterized, for instance, by its level statistics or its entropy. Furthermore, the introduction of dissipation opens a spectral gap between the steady-state and the bulk of the spectrum. The spectral gap sets the (inverse) timescale it takes the system to relax to the steady-state configuration.

In this thesis, we consider a generic Liouvillian superoperator, with random Hamiltonian and jump operators, as a means to study the properties of generic Lindblad-type Liouvillians. Since $H$ and the $W_\ell$ are represented by matrices in some basis, we next introduce Random Matrix Theory.

\subsection{Random Matrix Theory}
\label{subsection:RMT}
Random matrices were first considered by Wishart~\cite{wishart1928} in a purely mathematical context. In the 1950s, Wigner showed~\cite{wigner1955,wigner1958} that, for complex enough systems, such as heavy nuclei, the many-particle interactions lead the Hamiltonian of the system to statistically behave like a large random matrix. The original work of Wigner thus linked the areas of complex quantum many-body physics and RMT in a deep and intricate way.

The appeal of the RMT approach relied on the fact that it allowed a minimal description of the problem, solely in terms of the symmetries of the Hamiltonian (the relevant operator in that case) and not of the microscopic details of any particular model. Originally, the behavior of the system under time-reversal transformations, described in terms of the (anti-unitary) time-reversal operator $\mathcal{T}$, was the defining symmetry property. If the system has no time-reversal symmetry ($\comm{H}{\mathcal{T}}\neq0$), then the Hamiltonian has complex entries, is hermitian and is diagonalized by unitary transformations. If there exists time-reversal symmetry ($\comm{H}{\mathcal{T}}=0$) and $\mathcal{T}^2=+1$ then the Hamiltonian is real-symmetric with real entries and is diagonalized by orthogonal transformations. Finally, for a $\mathcal{T}$-symmetric system with $\mathcal{T}^2=-1$ the Hamiltonian has quaternionic entries, is quaternionic self-dual and is diagonalized by unitary symplectic transformations~\cite{haake2013}. Note that, in all these cases, the eigenvalues of the Hamiltonian are real, as usual. If we further require that the entries of the matrices are all independent random variables, then the ensembles are restricted to have Gaussian distribution functions, the Gaussian Unitary Ensemble (GUE), Gaussian Orthogonal Ensemble (GOE) and Gaussian Symplectic Ensemble (GSE), respectively, with the names deriving from the transformations which diagonalize the matrices of the ensembles. This is the classical Wigner-Dyson classification or Dyson's threefold way~\cite{dyson1962i,dyson1962ii,dyson1962iii,dyson1962a}. Further symmetries of the Hamiltonian enlarge the set of possible ensembles to the ten Altland-Zirnbauer classes~\cite{zirnbauer1996,altland1997,beenakker2015,haake2013,schomerus2017}, the tenfold way. The new classes are three chiral ensembles, resulting from combinations of time-reversal and chiral symmetry and four Bogoliubov-de Gennes ensembles, arising from combinations of time-reversal and particle-hole (or charge conjugation) symmetries.

The Wigner-Dyson ensembles are a good point for a small detour to elaborate on typical questions one usually asks in an RMT approach and the kind of answers one gets. They also illustrate the difference between integrable and chaotic systems discussed above. An $N\times N$ matrix $H$ of one of the Gaussian ensembles has Gaussian probability distribution function, $P(H)\propto\exp{-(N\beta/2)\Tr[H^2]}$ (which is to be understood component-wise), with $\beta=1$, $2$, $4$ for the GOE, GUE, GSE respectively ($\beta$ counts the number of real degrees of freedom each entry of $H$ has). This distribution function is invariant under rotations of the matrix $H$, i.e.\ $P(H)=P(U^{-1}HU)$ with $U$ orthogonal, unitary, or symplectic for the GOE, GUE, or GSE, respectively. In fact, it is a general result that the only ensembles with this invariance property as well as independent entries are the Gaussian ensembles~\cite{mehta2004,fyodorov2004}. From this rotational invariance, it follows that the eigenvectors are irrelevant for the spectral properties and can be integrated out.

Several quantities can be used to characterized the spectral properties of the Hamiltonian. One of the most important is the spectral density (the density of states) $\varrho(x)=(1/N)\sum_i\langle\delta(x-E_i)\rangle$, such that $\int_\mathcal{R}\dd x\ \varrho(x)$ counts the fraction of eigenvalues $E_i$ of $H$ inside the interval $\mathcal{R}$. Methods to obtain the spectral density are discussed in Chapter~\ref{chapter:state_of_the_art}. At any rate, for all three Gaussian ensembles one obtains, in the limit $N\to\infty$ and after rescaling the eigenvalues to $x'=x/\sqrt{\beta N}$, the famous Wigner semicircle law~\cite{wigner1955}, $\varrho_\mathrm{W}(x')=(1/\pi)\sqrt{2-{(x')}^2}$, $-\sqrt{2}<x'<\sqrt{2}$.

The spectral density contains information on the (global) distribution of the energy levels but not on their (local) correlations, which are important, for instance, for the distinction between integrable and chaotic systems. The joint eigenvalue distribution, given by
\begin{equation}\label{eq:joint_prob}
    P(E_1,\dots,E_N)\propto \prod_{i<j}\abs{E_i-E_j}^\beta\prod_k\exp{-\frac{\beta N}{2}E_k^2}\,,
\end{equation}
obviously contains all the information about the correlations\footnote{The correlations arise because of the term $\abs{E_i-E_j}^\beta$ (the Vandermonde interaction), which renders the probability nonfactorizable, $P(E_1,\dots,E_N)\neq P_1(E_1)\cdots P_N(E_N)$. While the exponential in Eq.~(\ref{eq:joint_prob}) follows directly from the entry distribution, the correlation term arises from the Jacobian of the change of variables from entries to eigenvalues and eigenvectors. It is related to a rotation in eigenvector space, see Ref.~\cite{schomerus2017}.}. By integrating out $N-k$ eigenvalues we obtain the $k$-point distribution function (in particular, the spectral density is the 1-point distribution function). However, the joint eigenvalue distribution is often too cumbersome to work with. 

Alternatively, we can consider the level spacing statistics, $P(s)$, (statistics of the distance~$s$ between two consecutive eigenvalues), which also contains correlation information. One finds that~\cite{guhr1998}
\begin{equation}\label{eq:Wigner_surmise}
    P_\beta(s)=a_\beta s^\beta e^{-b_\beta s^2}\,,\quad\quad
    a_\beta=2\,\frac{\Gamma^{\beta+1}\left(\frac{\beta+2}{2}\right)}{\Gamma^{\beta+2}\left(\frac{\beta+1}{2}\right)}\,,\quad\quad
    b_\beta=\frac{\Gamma^{2}\left(\frac{\beta+2}{2}\right)}{\Gamma^{2}\left(\frac{\beta+1}{2}\right)}\,,
\end{equation}
a result known as Wigner's surmise.\footnote{To be precise, this result is derived for $2\times2$ matrices; the formulas for arbitrary-$N$ are only know in terms of infinite products~\cite{mehta2004}. Numerical results show, however, that the $N\to\infty$ level statistics differ little from Wigner's surmise, at most $1-2\%$.} If one instead computes the level spacing statistics for uncorrelated energy levels, one obtains Poisson statistics $P(s)= \exp{-s}$.\footnote{The set of uncorrelated variables is called a Poisson (point) process because the probability of finding $n$ levels in a given interval of length $s$ is given by a Poisson distribution $P_n(s)=(s^n/n!)e^{-s}$. The special case of $n=0$ (i.e.\ there being an empty spacing) gives the exponential distribution.} From the behavior of $P(s)$ as $s\to0$ we see that the Wigner-Dyson statistics lead to level repulsion, since $P(s)$ vanishes at small $s$, while Poisson statistics lead to level attraction (or clustering) since $P(s)$ is at its maximum at at small $s$.

\begin{figure}[htbp]
    \centering
    \includegraphics[width=0.4\textwidth]{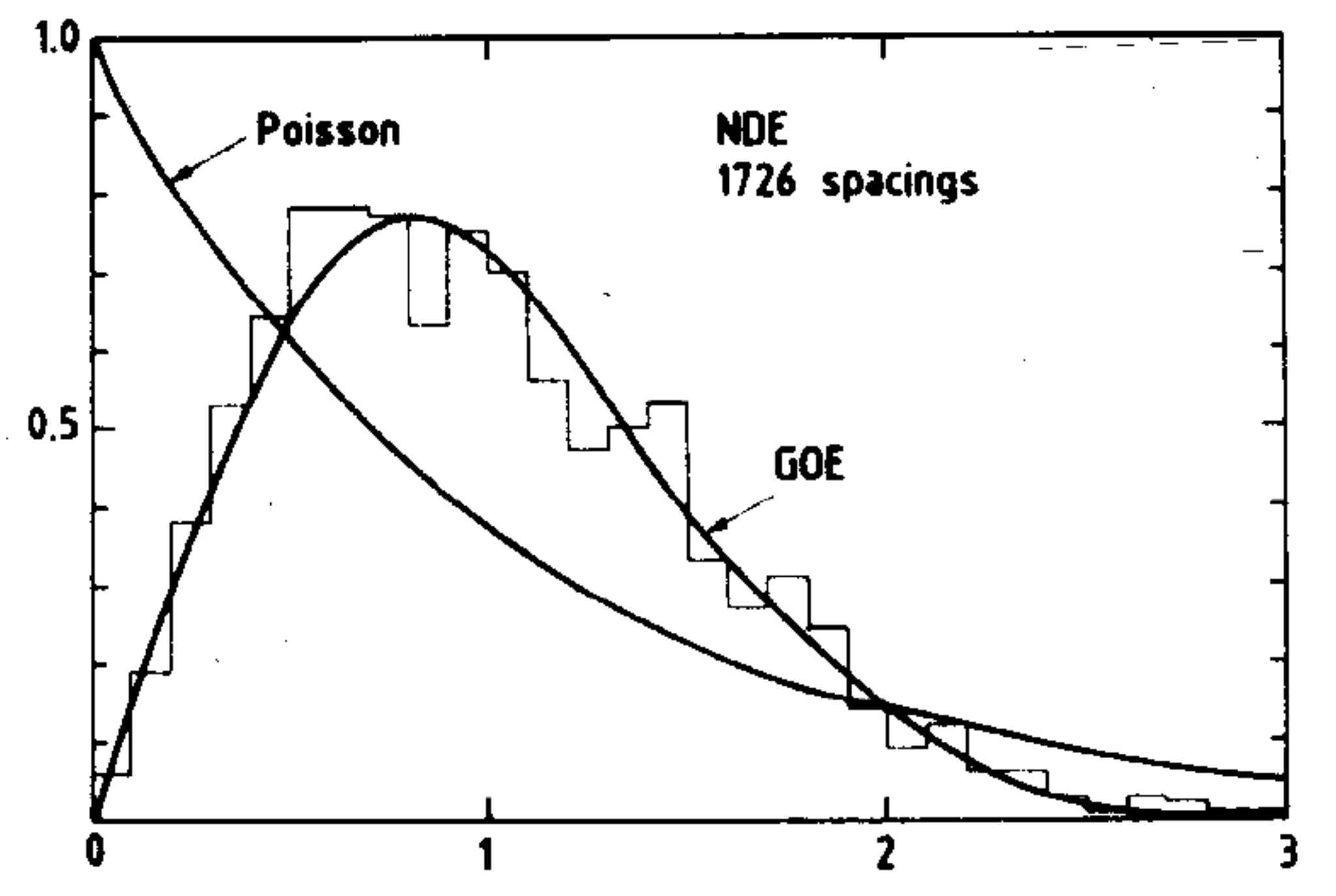}
    \caption{Comparison of GOE and Poisson level spacings with real data from 1762 resonance energies corresponding to 36 sequences of 32 different nuclei. Image taken from O.\ Bohigas, R.\ U.\ Haq, and A.\ Pandey, ``Fluctuation properties of nuclear energy levels and widths comparison of theory with experiment'', in \textit{Nuclear Data for Science and Technology}, K. H. B\"ochhoff (Reidel, 1983), p. 809.}
    \label{fig:comparison}
\end{figure}

Two famous conjectures by Berry and Tabor~\cite{berry1977} and by Bohigas, Giannoni, and Schmit~\cite{bohigas1984} assert that the quantum spectra of systems which are integrable in the classical limit follow Poisson statistics while quantum systems with a chaotic semiclassical limit follow RMT statistics. Although not proven until today, the conjectures are supported by a vast set of experimental and numerical data~\cite{guhr1998,haake2013,stockmann2000}, starting from the comparison of nuclear spectra with the GOE level statistics, see Fig.~\ref{fig:comparison}.

These observations have evolved into the field known as \emph{quantum chaos}, of great current interest~\cite{guhr1998,haake2013,braun2001,stockmann2000}. An apparent paradox seems, nevertheless, to riddle the field. On the one hand, quantum mechanics is a linear theory and there seems to be no place for chaotic dynamics. On the other hand, by the correspondence principle, one expects that, as $\hbar\to0$, quantum mechanics goes into classical mechanics, which exhibits chaos. Hence, there must be some quantum mechanisms underlying classical chaos, which manifest themselves through \emph{quantum signatures of chaos} (one primary example being the level repulsion/attraction discussed above). Furthermore, given the close relationship of quantum mechanics with wave mechanics (compare the Schr\"odinger equation with the Helmholtz equation), many of the results from quantum chaos are also verified at the level of classical wave phenomena, particularly in microwave cavities~\cite{stockmann2000}.

Returning to the possible classes of random matrices, hermitian matrices are not the only relevant matrices with real eigenvalues.\footnote{The eigenvalues of a unitary matrices are on the complex unit circle, $e^{i\theta_j}$, but are fully determined by the real angles $\theta_j\in(-\pi,\pi]$.} In fact, motivated by the study of the $S$-matrix in scattering problems, Dyson considered random unitary matrices~\cite{dyson1962i,dyson1962ii,dyson1962iii,dyson1962iv}. Depending on whether the entries of the unitary matrix are real, complex or quaternionic, one draws the matrices from the Circular Orthogonal, Unitary, or Symplectic Ensemble (COE/CUE/CSE), respectively, in analogy with the Wigner-Dyson classes. Since the support of the spectrum (the unit circle in the complex plane) is compact no convergence issues for the integrals arise and one may take a flat distribution for the angles, instead of a Gaussian one. Of course, one still has to include the Vandermonde interaction, which now reads $\prod_{j<k}\abs{e^{i\theta_j}-e^{i\theta_k}}^\beta$. The resulting spectral density is, then, flat on the unit circle, $\varrho(\theta)=1/(2\pi)$. Although this 1-point function is very different from the Wigner semicircle, correlations of eigenvalues (fluctuations around the mean) are the same for the Gaussian and circular ensembles. More generally, one verifies that spectral correlations are highly universal~\cite{ambjorn1990,guhr1998}, i.e.\ independent of the particular choice of weight for the ensemble, within a given symmetry class. Indeed, while the (global) density of states of, say, a heavy nucleus is definitely \emph{not} described by a semicircle, its local fluctuations (say, the spacings) are extremely well described by RMT. This justifies why we may choose the mathematically most convenient weight (Gaussian or flat) without sacrificing broad applicability of the predictions.

We may further relax the constraints on our random matrices and consider matrices with general complex eigenvalues (i.e.\ which are neither hermitian nor unitary). Imposing a Gaussian weight, $P(H)\propto\exp{-N\Tr[H^\dagger H]}$, but with a matrix $H$ where all entries are independent (not only the diagonal plus upper triangle as in the Wigner-Dyson classes), we are lead to Ginibre's ensembles~\cite{ginibre1965}. Again, depending on whether the entries are real, complex or quaternionic, one gets the Ginibre Orthogonal, Unitary or Symplectic Ensemble (GinOE/GinUE/GinSE), respectively. Since the Ginibre matrices are no longer hermitian, their spectrum is complex. All three ensembles have a uniform distribution on the unit disk in the complex plane, with the spectral distribution, $\varrho(z)=(1/\pi)\Theta(1-\abs{z})$, with $\Theta$ the Heaviside step-function, which is known as the circular law~\cite{ginibre1965,haake2013}. Further imposing symmetries on the nonhermitian matrices leads to a broad set of classes of random matrices~\cite{schomerus2017}, and nonhermitian extensions of the Altland-Zirnbauer classification have been introduced~\cite{bernard2002,magnea2008,kawabata2018ST,kawabata2019,hamazaki2019}.

The above ensembles of random matrices have been constructed on the assumption of independence of all its entries (excluding the ones imposed by symmetry) and we were lead to rather strong restrictions on the possible classes. If one lifts the restriction of independence of entries, even more ensembles arise. Of prominent importance for the Liouvillian case is the Wishart ensemble~\cite{wishart1928} of positive semidefinite matrices of the form $X=W^\dagger W$. Here, $W$ is a $M\times N$ matrix with $M\geq N$ (otherwise there are $\nu=M-N$ eigenvalues identically zero). Because of the correlation of entries of $X$, there arises a term $(\det X)^{\beta/2(N-M+1)-1}$ in the joint probability distribution. The spectral density of the Wishart ensemble, in the limits $N,M\to\infty$ with $m=M/N\geq1$ fixed, is not the Wigner semicircle but the Marchenko-Pastur law~\cite{marchenko1967}, which is discussed in Sections~\ref{section:random_ensembles} and \ref{subsection:diagrammatics_Wishart}.

Finally, also the Gaussian character of the statistics can be lifted and, for the case of rationally invariant matrices, one obtains the invariant matrix ensembles, with weight $P(H)\propto\exp{-N\Tr[V(H)]}$, where the potential~$V$ is not necessarily quadratic. Although universal fluctuations are not affected, global spectral properties are.

To finalize this brief introduction to RMT, we remark that its applications are not restricted to quantum chaos. In fact, RMT has been successfully applied to a great variety of other physical systems and problems (for reviews see Refs.~\cite{guhr1998,stephanov1999,akemann2011}), e.g., mesoscopic physics \cite{beenaker1997}, localization and disordered systems~\cite{efetov1999}, random growth models~\cite{ferrari2010}, quantum chromodynamics \cite{verbaarschot2010} and quantum gravity in two dimensions \cite{gross1990}.

Having introduced the random Liouvillian superoperator, Eq.~(\ref{eq:Liouvillian_diagonal}), and the RMT formalism to study it, we have all the ingredients to state the objectives of this thesis.

\section{Objectives}
\label{section:objectives}

The central question encompassing all of our work is: 

\textit{``What properties of an open quantum system are independent of the microscopic details of particular systems?''}

\noindent As in the case of closed quantum systems, the dynamics of complex OQS are expected to exhibit some universal properties, solely determined by symmetry, described by random matrix models. Determining the properties of these stochastic models is an essential first step to identify such universal features. It opens a new avenue for understanding the long-time dynamics and steady-states of generic complex quantum dissipative systems. The ubiquity of quantum dissipation renders the developments in this field of great interest to areas ranging from condensed matter to quantum optics and of potential technological impact in the fabrication of complex quantum structures and, ultimately, quantum computers.

The main goal of the thesis was therefore to establish the spectral and steady-state properties of generic open quantum systems. Our approach considered the construction of Liouvillian superoperators and steady-state density matrices out of random matrices of known statistics and aimed at studying their statistical behavior. Specific goals were as follows: 
\begin{itemize}
    \item Specify a stochastic Lindblad operator consisting of a random Hamiltonian and random jump operators drawn from the appropriate ensembles.
    \item Determine numerically the spectral properties of the stochastic Liouvillian--including the average spectral gap and the low-lying eigenvalues that determine the long-time asymptotics.
    \item Study the steady-state properties--in particular, the spectrum of the steady-state density matrix and the correlations of its entries.
    \item Explore the use of diagrammatic expansions, to obtain spectral and steady-state properties analytically.
    \item Construct and study quantities that readily quantify the (universal) degree of chaoticity or integrability of a quantum system; apply this procedure to determine in which class of models realistic physical systems are found.
\end{itemize}

\section{Thesis Outline}
\label{section:outline}
To achieve the goals set forward in the previous section, we will build upon the vast body of literature on RMT and quantum chaos, reviewed in Chapter~\ref{chapter:state_of_the_art}.
In Chapter~\ref{chapter:model}, we set up the model to be used in the remaining chapters. Specifically, we discuss in more detail the concept of a quantum superoperator, we look at the importance of its spectrum and eigenstates, we describe two matrix representations of Liouvillian superoperators, and we discuss some relevant RMT ensembles.
Then, in Chapter~\ref{chapter:numerics}, we numerically study the properties of the spectrum and the steady-state of a general Liouvillian, looking at the shape of its spectrum, the spreading of its decay rates, the spectral gap, and the steady-state purity and spectral statistics.
Chapter~\ref{chapter:analytics} presents a detailed analytical study of the spectral gap. Employing diagrammatic expansions, we compute the gap exactly for all values of dissipation strength. Simple expressions for the asymptotic values at strong and weak dissipation are also derived.
Next, in Chapter~\ref{chapter:spacings}, we introduce complex spacing ratios as a new signature of dissipative quantum chaos. We use it extensively to distinguish integrable and chaotic Liouvillians. Besides the numerical work, we provide a thorough theoretical analysis: we compute the ratio distribution exactly for Poisson processes in $d$ dimensions, write down the exact distribution for hermitian and nonhermitian ensembles, and provide Wigner-like surmises for both cases. 
Chapters~\ref{chapter:numerics}--\ref{chapter:spacings} comprise the original results of this thesis and can be seen as two different ways to address our central question: while in Chapters~\ref{chapter:numerics} and \ref{chapter:analytics} we look at spectral densities and extremal eigenvalues and eigenstates, which are expected to be modeled by random Liouvillians, in Chapter~\ref{chapter:spacings} we examine the truly universal spectral correlations in complex spectra, of which Liouvillians are a prototypical example.
We draw our conclusions and point towards possible generalizations and extensions of this work in Chapter~\ref{chapter:conclusions}.
Appendix~\ref{appendix:polynomials} gives the explicit expressions for some lengthy polynomials encountered in Chapter~\ref{chapter:spacings}.


\cleardoublepage


\chapter{State of the art and methods}
\label{chapter:state_of_the_art}

The aim of this chapter is to provide a literature review of relevant topics of RMT and quantum chaos. More specifically we will start with the most simple available objects in RMT, the Gaussian ensembles, and detail some of the techniques employed in studying their spectral properties (Section~\ref{section:hermitian_RMT}). Two extensions of those methods, for nonhermitian RMT (Section~\ref{section:nonhermitian_RMT}) and free probability (Section~\ref{section:free_probability}), are needed to construct random Liouvillians (Section~\ref{section:review_random_Liouvillian}). In these sections, we will focus on diagrammatic expansions, but some other methods are also briefly mentioned. Finally, we address spectral observables (Section~\ref{section:review_spacing_distributions}), which are quantities allowing comparison of RMT predictions with actual numerical or experimental data. Of special importance for this work are spacing and spacing ratio distributions and the unfolding procedure, all of which are explained in detail.

\section{Hermitian RMT}
\label{section:hermitian_RMT}
An RMT study usually starts by identifying the $N\times N$ random matrices $\phi$ of the model and specifying their probability distribution function $P(\phi)$. By a orthogonal/unitary/symplectic transformation one can rotate $\phi$ into its diagonal basis, with eigenvalues $E_1,\dots,E_N$ and eigenvectors organized into an $N\times N$ matrix $U$. Assuming a rotationally invariant matrix ensemble, the eigenvectors can be trivially integrated out and one works with the joint eigenvalue distribution $P(E_1,\dots,E_N)$ (Eq.~(\ref{eq:joint_prob}) is the special case for Gaussian ensembles). The change of basis introduces the ubiquitous Vandermond interaction $\abs{\Delta}^\beta\equiv\prod_{j<k}\abs{E_j-E_k}^\beta$, independently of the specific choice of weight $P(\phi)$ (modulo a possible $\beta$-dependence of $P(\phi)$). In particular, if the spectral support of $\phi$ is compact, the weight may be chosen flat (for instance, for the circular ensembles). As mentioned in Chapter~\ref{chapter:introduction}, one is interested in $k$-point correlators,
\begin{equation}
    R_k(E_1,\dots,E_k)=\frac{N!}{(N-k)!}\int_{-\infty}^{+\infty}\d E_{k+1}\cdots\d E_NP(E_1,\dots,E_N)\,,
\end{equation}
which gives the probability of finding any $k$ eigenvalues with values $E_1,\dots,E_k$ independently of the values of the remaining $N-k$ eigenvalues. As discussed before, the 1-point correlation function (the density of states) is highly dependent on the choice of weight of the ensemble, but higher correlation functions should display universality, after eliminating the dependence on the local density (what is called \emph{unfolding} the spectrum, see Section~\ref{section:review_spacing_distributions} below). A very interesting discussion of universality in RMT can be found in Ref.~\cite{guhr1998}.

Early methods to compute correlation functions (and which are still widely used today) were the Orthogonal Polynomials Method~\cite{mehta2004} and the Coulomb gas (or Dyson gas) analogy~\cite{dyson1962ii,forrester2010}. In the former, one can express $R_k$ as a determinant of a $k\times k$ matrix (the kernel) which allows the multiple integrals in the definition of the correlation function to be performed iteratively, using orthogonal polynomials. In the latter method, one rewrites the joint eigenvalue distribution as 
\begin{equation}\label{eq:def_Rk_1}
    P(E_1,\dots, E_N)\propto \exp{-\beta N\left(\frac{1}{2}\sum_{j=1}^NE_j^2-\frac{1}{2N}\sum_{j\neq k}\log\abs{E_j-E_k}\right)}\equiv e^{-\beta N V(E_1,\dots,E_N)}\,,
\end{equation}
which we can identify as the Boltzmann factor for a two-dimensional classical gas, with a potential $V$ containing two terms: the single-particle confining potential (quadratic in this case because of the Gaussian weight of the ensemble) and a logarithmic Coulomb interaction between pairs of eigenvalues (which are confined to move on the line because of the hermiticity condition of the ensemble). The methods of electrostatics and statistical methods can then be used to study the gas of eigenvalues.

The $k$-point correlation functions can also be written as~\cite{guhr1998,guhr2010}
\begin{equation}\label{eq:def_Rk_2}
    R_k(E_1,\dots,E_k)=\frac{1}{\pi^k}\int \d \phi\, P(\phi)\prod_{j=1}^k\Im\Tr\frac{1}{E_j^--\phi}\equiv\frac{1}{\pi^k}\left\langle \Im G(E_1^-,\dots,E_k^-)\right\rangle\,,
\end{equation}
where $\d \phi$ is the (flat) measure in matrix space, $E_j^-= E_j-i\varepsilon$, with $\varepsilon\to0^+$, and $\langle\cdot\rangle$ denotes the ensemble average over $\phi.$\footnote{The two definitions of correlation functions, Eqs.~(\ref{eq:def_Rk_1}) and (\ref{eq:def_Rk_2}), differ slightly due to the presence of some Dirac-$\delta$ terms, see Ref.~\cite{guhr1998}. This technical subtlety is irrelevant here and will not be considered further.} This definition introduces an auxiliary quantity, the $k$-point Green's function, $G(z_1,\dots,z_k)=\prod_{j=1}^k\Tr(z_j-\phi)^{-1}$.
In particular the 1-point Green's function (often called simply \emph{the} Green's function or the resolvent) $G(z)=(1/N)\Tr(z-\phi)^{-1}$ is related to the spectral density of $\phi$ via $\varrho(x)=(1/\pi) \langle\Im[G(x-i0^+)]\rangle$, and will be heavily used in Chapter~\ref{chapter:analytics}.

Over the years, the proliferation of physical applications of RMT and the development of a vast set of analytical tools have been driving each other on. Ideas originally developed by t'Hooft~\cite{thooft1974}, in the context of the strong interaction, lead Brézin \textit{et al.}~\cite{brezin1978} to propose an expansion of the Green's function in terms of planar Feynman diagrams. In fact, $G$ can be written as an expansion in moments of $\phi$, $G(z)=(1/N)\sum_{n}\langle\Tr\phi^n\rangle/z^{n+1}$. The moments can, in turn, be obtained from a generating function, which is formally the path integral of a quantum field theory in 0+0 dimensions for a matrix field and standard field-theoretic methods apply, in particular, diagrammatic expansions (see Chapter~\ref{chapter:analytics}). The diagrams can be classified by the topology (namely by the genus) of the surfaces where they can be drawn on, with diagrams on surfaces of higher genus suppressed by powers of $1/N$. In the large-$N$ limit, only planar diagrams (with as many loops as propagators, which we can draw on a sphere or a plane) survive. The problem is then reduced to counting the number of planar diagrams at each order $n$ of the expansion and resum the infinite series. The perturbative expansion of the resolvent proved to be extremely useful in establishing the universality of fluctuations~\cite{ambjorn1990,brezin1993,brezin1995,akemann1996}. It was shown that, although the one-point Green's function (and hence the spectral density) depends strongly on the details of the distribution function of the ensemble, higher-order Green's functions (and hence correlations of eigenvalues) depend on the particular potential of a rotationally-invariant ensemble only through its endpoints and are, therefore, highly universal.

Another major breakthrough came when Efetov's supersymmetric method for the theory of disordered systems and localization~\cite{efetov1983,efetov1999} was shown to also apply to RMT~\cite{verbaarschot1985,guhr2010,haake2013}. Indeed, one is allowed to average a generating functional for the Green's function by writing it as a ratio of determinants and employing Gaussian integration identities for commuting and anticommuting variables. More concretely, the determinant in the denominator becomes a standard bosonic Gaussian integral, while the one in the numerator becomes a Gaussian integral over Grassmann variables. Performing the average over the matrix ensemble, the problem is then formally a saddle-point evaluation of a zero-dimensional nonlinear sigma-model. Initially, it was only considered for Gaussian ensembles, because the averaging introduced a quartic interaction followed by a Hubbard-Stratonovich decoupling to turn the integrals into quadratic ones again, which can then be solved by saddle point approximations. The method has by now been extended to arbitrary rotation invariant ensembles~\cite{hackenbroich1995,guhr2006a,guhr2006b,littelmann2008,sommers2008,kieburg2009a,kieburg2009b}. In particular, Hackenbroich and Weidenm\"uller~\cite{hackenbroich1995} showed that all local correlation functions are independent of the precise potential $V(\phi)$. The supersymmetric method provides a nonpertubative approach to the computation of the Green's function, in contrast to the diagrammatic method. Other approaches based on a matrix nonlinear sigma-model are the replica trick, initially arising in the theory of spin glasses~\cite{edwards1975,edwards1976}, and the Keldish method~\cite{altland2000,kurchan2002}.

The diagrammatic, supersymmetric and replica methods allowed tackling some problems not accessible via classical methods, notably scattering problems, see Ref.~\cite{schomerus2017} for a recent review. One particular way to model scattering systems is through effective nonhermitian Hamiltonians~\cite{haake1992,lehmann1995,janik1997a,janik1997b}. The results obtained in these references will be heavily built on in subsequent chapters, so we next consider RMT for operators with general complex eigenvalues.\footnote{In the literature all ensembles of matrices with one-dimensional eigenvalues are called, somewhat inaccurately, hermitian ensembles (even if the matrices not hermitian but, say, unitary), while the theory of matrices with general complex eigenvalues is called nonhermitian RMT.}

\section{Nonhermitian RMT}
\label{section:nonhermitian_RMT}

The Green's function methods above were initially developed for random matrices with a one-{dimension-al} spectrum. That is the case of hermitian matrices (real spectrum) and unitary matrices (unit circle as the spectrum). For matrices whose spectrum is a finite region in the complex plane, direct application of the above methods usually fails~\cite{feinberg1997a,feinberg1997b,janik1997a,janik1997b,feinberg2006,jarosz2004,jarosz2006}. 

The reason is that the Green's function is a holomorphic function of $z$ outside the spectral support of the operator only.\footnote{We, therefore, refer to the Green's function of the previous section as the holomorphic Green's function.} When the spectrum is one-dimensional, the domain of analyticity is thus the whole complex plane modulo some cuts and the spectral density is obtained exactly from the discontinuities of the Green's function along those cuts. When the spectral support becomes two-dimensional, the information it contains is not accessible by coming from infinity where $G$ is again an analytic function of $z$. In particular, the spectral density cannot be obtained from a discontinuity along the branch cuts nor is a diagrammatic expansion in powers of $1/z$ (i.e.\ for large $z$) applicable. 

The first issue (relation between the Green's function and the spectral density) can be easily solved by promoting $G$ to a function of both $z$ and $z^*$ noting that the identity $\varrho(z)=(1/\pi)\partial_{z^*}G(z,z^*)$ holds, which can be seen as follows. Using $\partial_z\partial_{z^*}\log(zz^*)=\pi\dirac{x}\dirac{y}$~\cite{feinberg1997a}, where $z=x+iy$, by definition we have $\varrho(z)\equiv\varrho(x,y)=(1/N)\av{\sum_i\dirac{x-\re E_i}\dirac{y-\im E_i}}=1/(N\pi)\av{\sum_i\partial_z\partial_{z^*}\log[(z-E_i)(z^*-E_i^*)]}=1/(N\pi)\partial_z\partial_{z^*}\av{\Tr\log(z-\phi)(z^*-\phi^\dagger)}=1/(N\pi)\partial_{z^*}\av{\Tr(z-\phi)^{-1}}=(1/\pi)\partial_{z^*}G(z,z^*)$, as stated. 

To circumvent the second problem (perturbative computation of $G$), two distinct approaches have been suggested such that the hermitian diagrammatic methods above can be used, both by adding a block-matrix structure to the Green's function.

First, if one uses the method of hermitization~\cite{feinberg1997a,feinberg2006}, one constructs a block-matrix $V=\left(\begin{smallmatrix}0&\phi\\\phi^\dagger&0\end{smallmatrix}\right)$ which is hermitian and hence amenable to hermitian perturbative methods, namely an expansion in a new variable $\mathcal{G}_0=\left(\begin{smallmatrix}w&z\\z^*&w\end{smallmatrix}\right)^{-1}$. The Green's function also acquires a block structure $\mathcal{G}=\left(\begin{smallmatrix}\mathcal{G}_{11}&\mathcal{G}_{12}\\\mathcal{G}_{21}&\mathcal{G}_{22}\end{smallmatrix}\right)$. As usual, the Green's function $\mathcal{G}$, the free part of the Green's function $\mathcal{G}_0$, and the perturbation (potential)~$V$ are related by $\mathcal{G}^{-1}=\mathcal{G}_0^{-1}+V$. The Green's function $G$ is given by the block $\mathcal{G}_{21}$ upon taking $w\to0$, $G(z,z^*)=\lim_{w\to0}\mathcal{G}_{21}(z,z^*,w)$. Since $G$ is nonanalytic in the thermodynamic limit  only, it is crucial that the limit $w\to0$ is taken only after the limit $N\to\infty$.

The second method is a quaternionic extension~\cite{janik1997b,jarosz2004,jarosz2006} of the Green's function. Just as the real spectrum needs the Green's function to be defined on the complex numbers, a complex spectrum can be handled by defining the Green's function on the field of the quaternions. In fact, one again defines $\mathcal{G}=\langle(Q-X)^{-1}\rangle$ as before, but where now the expansion variable is a quaternion $Q=\left(\begin{smallmatrix}z&iw^*\\iw&z^*\end{smallmatrix}\right)$ written in block-matrix form and $X=\mathrm{diag}(\phi,\phi^\dagger)$. The Green's function $G$ is obtained from the upper-left block of $\mathcal{G}$ when $w\to0$, just as before, and again perturbation theory in $Q^{-1}$ can be applied. 

Yet another different approach to tackle the nonhermitian problem~\cite{sommers1988} is to exploit the electrostatic analogy alluded before by defining a potential $\Phi(z)=-(1/N)\log\det[(z^*-\phi^\dagger)(z-\phi)]$ (which we had already encountered in an intermediate step above) and an ``electric field'' $\bm E=(2\Re G,-2\Im G)$. The relation between the potential, the Green's function, and the spectral density is then given by the Poisson equation, $\nabla^2\Phi=-\bm{\nabla}\cdot\bm{E}=-4\pi\varrho$. Here, one computes the potential instead of the Green's function. This computation can be done perturbatively since the matrix $(z^*-\phi^\dagger)(z-\phi)$ in the potential is again hermitian. Alternatively, $\Phi$ can also be evaluated by a replica trick, as is done in Ref.~\cite{sommers1988}.

\section{Addition and multiplication of random matrices}
\label{section:free_probability}
The methods discussed up until now have been concerned with matrices drawn from a single ensemble, but nothing has been said about sums or products of matrices drawn from different ensembles (for instance, a random Liouvillian). Just as for ordinary random variables the distribution of the sum or product can be computed from the individual distribution if the random variables are independent, the same applies to random matrices if they obey stronger requirements of free independence~\cite{speicher2009,mingo2017}. Those conditions are met if the matrices are drawn from independent ensembles. The issue is then addressed by the theory of free probability, which was first introduced in physics by Zee~\cite{zee1996}. 

Here, one wants the spectral density (or the corresponding Green's function) of $M=M_1+M_2$, knowing the spectral densities (or the Green's functions) of $M_1$ and $M_2$. The spectral density of the sum is then computed with the following algorithm~\cite{zee1996,janik1997b,nowak2017}: one obtains the functional inverses of the individual Green's functions, dubbed Blue's functions, $G_i(B_i(z))=B_i(G_i(z))=z$ (with $i=1,2$), and from them constructs the $\mathcal{R}$-transforms $R_i(z)=B_i(z)-1/z$. It can be shown~\cite{mingo2017} that it is the $\mathcal{R}$-transform (and not the Green's function) which has the desired additive property, i.e.\ $R(z)=R_1(z)+R_2(z)$ but $G(z)\neq G_1(z)+G_2(z)$. One then runs the procedure in reverse, obtaining $B(z)$ from $R(z)$ and afterwards $G(z)$ from $B(z)$. For the product of matrices from different ensembles, the $\mathcal{S}$-transform, defined by $S(z)=1/R(zS(z))$ is to be used, since it has the multiplicative property $S(z)=S_1(z)S_2(z)$~\cite{mingo2017}. 

The $\mathcal{R}$-transform is also known in the physics literature as the self-energy $\Sigma(z)$, which is defined via its relation to the Green's function, $G=1/(z-\Sigma)$ (see section~\ref{subsection:diagrammatics_Wishart} for a diagrammatic interpretation of this relation). Indeed, by evaluating $R(G(z))=B(G(z))-1/G(z)=z-1/G(z)$, it follows that the Green's function and the $\mathcal{R}$-transform are related by $G(z)=1/[z-R(G(z))]$, which is precisely the definition of the self-energy. We also note that knowledge of the self-energy allows one to obtain the Green's function through the solution to the algebraic equation $G(z)\Sigma[G(z)]-zG(z)+1=0$.

The free probability approach applies to both hermitian and nonhermitian matrices. In the latter case, a quaternionic extension, in which the Blue's function also acquires a block structure~\cite{janik1997b}, has to be considered, but the bulk of the algorithm is as above.

\section{Random Liouvillians}
\label{section:review_random_Liouvillian}
Given the tremendous success of Random Matrix Theory in the description of complex closed quantum systems, starting with Wigner's seminal work six decades ago, it is perhaps surprising that the same ideas in the context of dissipative systems have remained mostly unexplored. Although RMT has been successfully applied in the past to the scattering description of OQS~\cite{haake1992,lehmann1995,janik1997a,janik1997b,schomerus2017} and to discrete OQS~\cite{bruzda2009,bruzda2010}, only very recently has a RMT approach started being applied to the Lindblad description of OQS~\cite{sa2019,denisov2018,can2019,can2019a}.

Following the general ideas of closed-system RMT, Ref.~\citep{denisov2018} studies an ensemble of random Lindblad operators without Hamiltonian and with a maximal number of independent decay channels by using the quaternionic extension of free probability and finds that, in this case, the spectrum acquires a universal lemon-shaped form. The spectral boundary is explicitly computed using the quaternionic extension of free probability. This work considered, however, the number of decay channels $r\propto N^2$, that is, that as more states are added to the system, the number of decay modes of each state also grows. Although mathematically appealing, this behavior seems somewhat unphysical. A system with the number of jump operators of $\mathcal{O}(N^0)$ (only some states of the system are coupled to the environment by a finite number of channels each) or $\mathcal{O}(N)$ (all states are coupled to the environment but with fixed and finite number of channels) seems more physically plausible. In this thesis, we focus on the case of a small number of channels, although we also perform some analytic computations for all $r$. 

In a similar spirit, Ref.~\cite{can2019a} considers an ensemble of random Lindbladians consisting of an Hamiltonian part and a finite number of hermitian jump operators and finds a sharp spectral transition as a function of the dissipation strength. Finally, Ref.~\cite{can2019} studies in detail the spectral gap for several different Liouvillians. Specifically, it addresses Liouvillians without Hamiltonian and a single jump operator, either hermitian, normal or arbitrary (Ginibre); without Hamiltonian and several arbitrary jump operators; and with Hamiltonian and one jump operator. The findings of Ref.~\cite{can2019} agree with our analytic findings as well as with our numerical investigations~\cite{sa2019}.

Yet, there a several open questions related to the nature of the spectrum, specifically when the dissipative and the Hamiltonian components are comparable. Additionally, the properties of the non-trivial steady-state, ensuing in the presence of nonhermitian jump operators, are completely unexplored. We address these questions in Ref.~\cite{sa2019} and in Chapters~\ref{chapter:numerics} and \ref{chapter:analytics} of this thesis.

\section{Spectral observables and spacing distributions}
\label{section:review_spacing_distributions}
In Chapter~\ref{chapter:introduction}, we noted that spectral fluctuations allow us to distinguish integrable form chaotic dynamics. This is possible because of the universality of the correlation functions $R_k$ with $k\geq 2$. Despite the correlation functions being the natural object to work with theoretically, when performing experiments or numerical simulations, other spectral observables (which are built out of one or more correlation functions) are easier to access. It is therefore usual to consider these spectral observables instead of the correlation functions. We again consider first real spectra and, afterwards, complex spectra (our real case of interest here).

The most popular spectral observable is, arguably, the spacing distribution $P(s)$. Recall that it is defined as the distribution of the distances between two neighboring eigenvalues. Although it may seem at first that the spacing distribution is related to the 2-point function $R_2$ only, it is, in fact, a function of all the $R_k$ with $k\geq2$. We have already seen in Chapter~\ref{chapter:introduction} that, for the Gaussian ensembles, $P(s)$ is very well described (both qualitatively and quantitatively) by the distribution for $2\times2$ matrices, the Wigner surmise, Eq.~(\ref{eq:Wigner_surmise}).
Because of universality, this result holds for any ensemble with the same symmetries of the GOE/GUE/GSE (note that the Gaussian decay at large $s$ is \emph{not} due to the Gaussian distribution of the entries). In obtaining the Wigner surmise, not only the distribution itself, but also its first moment were normalized to one, $\int_0^\infty\d s P(s)=\int_0^\infty \d s\, sP(s)=1$. By setting $\langle s\rangle=1$, we are changing the scale to the so-called unfolded scale, on which different RMT predictions can be compared with each other and with numerical and experimental data (the procedure for unfolding an actual dataset is given below). Given the practical interest in level spacings, they have been studied extensively over the years and are very well understood~\cite{guhr1998,mehta2004,haake2013,forrester2010,forrester2004}. Spacing distributions further allow one to study intermediate statistics, either with crossovers between Poisson and RMT statistics~\cite{brody1973,berry1984,lenz1991b,prosen1993,prosen1994,bogomolny1999,bogomolny2001} or transitions between different RMT universality classes~\cite{mehta1983,pandey1983,lenz1991a,lenz1991b,schierenberg2012}. Higher-order spacings (i.e. distance between $k$th-nearest-neighbors) have also been considered over the years~\cite{dyson1962iii,gunson1962,sakhr2006,srivastava2018}.

While the spacing statistics measure short-range correlations, long-range fluctuations are captured, for instance, by the number variance $\Sigma^2(L)$~\cite{dyson1962iv,guhr1998}, which is defined as the variance in the number of levels in an interval of length $L$. For random matrices one finds $\Sigma^2(L)\propto \log L$ at large $L$, while for Poisson statistics $\Sigma^2(L)\propto L$. One can, therefore, also distinguish a chaotic system from an integrable one by the large-$L$ behavior of the number variance. $\Sigma^2(L)$ can be reduced to an integral of the 2-point function and hence measures only 2-level correlations. 

We now show how to unfold an actual spectrum~\cite{haake2013,guhr1998}, i.e.\ how to eliminate its dependence on the local mean spectral density, which is nonuniversal and system-dependent. Denote the ordered sequence of measured levels by $E_1,\dots,E_N$. Its spectral function is\footnote{We are using a definition differing by a factor of $1/N$ relative to before, to avoid cluttering the argument with factors of $N$.} $\sum_{j=1}^N\dirac{E-E_j}\equiv S(E)=\varrho(E)+\delta S(E)$, where we decomposed $S$ into the \emph{mean} spectral density, $\varrho(E)=\langle S(E)\rangle$, and the fluctuations around the mean, $\delta S(E)$.\footnote{Here, $\langle\cdot\rangle$ is an energy average over the spectrum (also called a running average). That is, we are considering an interval $\Delta E$ with $\Delta N$ levels ($1\ll\Delta N\ll N$) inside which the mean level density is approximately constant (and equal to $\overline{\varrho})$ and defining the average of some function $g(E)$ to be $\langle g(E)\rangle=(1/\Delta E)\int_E^{E+\Delta E}\d E' g(E')=(1/\Delta N)\int_0^{\Delta N}\d n\, g(E+\overline{\varrho} n)$. This is \emph{not}, \textit{a priori}, the same average which we defined before, the ensemble average $\int\d\phi P(\phi)g(\phi)$. However, for us to be able to compare the predictions of RMT with actual data, we must assume that both averages are actually equal. This assumption goes by the name of ergodic hypothesis~\cite{guhr1998}.}
We also consider the cumulative spectral function (also known as the level staircase) $\int_{-\infty}^E \d E'S(E')\equiv F(E)=\mathcal{N}(E)+\delta F(E)$, which we also decomposed into the mean level number $\mathcal{N}(E)=\int_{-\infty}^E \d E'\varrho(E')$ and the fluctuating level number $\delta F$. It is relevant to note that $S(E)=\d F(E)/\d E$ and $\varrho(E)=\d \mathcal{N}(E)/\d E$.
To eliminate the local mean level density we want to change variables to a new level sequence, $e_1,\dots,e_N$, with $e_j=f(E_j)$ for some function $f$, for which the mean level density is flat and thus fluctuations can be uniformly compared. Hence, the appropriate choice of $f$ is such that the mean level density in the new variables is unity, i.e.\ $\Tilde{\varrho}(e)=1$, where $\Tilde{\varrho}=\varrho\circ f^{-1}$. This condition can also be written as $\Tilde{F}(e)=e+\delta\Tilde{F}(e)$, or $\mathcal{N}(f^{-1}(e))=e$. It then immediately follows that the correct choice is $f=\mathcal{N}$, i.e.\ $e_j=\mathcal{N}(E_j)$.

Unfolding is a nontrivial procedure. It requires an analytic expression (or accurate estimate) of the level density $\varrho$, which one does not have in general. Unfolding can be easily achieved in theoretical calculations for the classical random matrix ensembles and for simple systems like quantum billiards~\cite{bohigas1984,bohigas1984b,stockmann2000,prosen1993,prosen1994}, but it is not possible for more complex systems, like many-body systems or random Liouvillians. Furthermore, the numerical procedure to actually implement the unfolding sometimes proves ambiguous and numerically unreliable.

An alternative way to overcome the local dependence on the level density is to consider ratios of spectral observables. It is expected that the ratios are automatically independent of the local density of states (the dependence of numerator and denominator should ``cancel out''), rendering the unfolding unnecessary. In this spirit, ratios of consecutive spacings were first considered in Ref.~\cite{oganesyan2007} instead of the spacings themselves. They were extensively applied in comparing real data to the theoretical predictions, in cases previously impossible~\cite{oganesyan2007,pal2010,cuevas2012,iyer2013,laumann2014,luitz2015,johri2015,argawal2015,chen2018,buijsman2019,d'alessio2014,kollath2010,collura2012}.

In Refs.~\cite{atas2013,atas2013long}, the concept was further refined and analytic expressions for the ratio distributions were obtained for the first time. If we again denote the ordered level sequence by $E_1,\dots, E_N$ and the spacings between consecutive spacings by $s_n=E_{n+1}-E_n$, these authors studied the statistics of $r_n=s_n/s_{n-1}$. In particular, in Ref.~\cite{atas2013}, they obtained a Wigner-like surmise for the ratio distribution function, which reads $P(r)=1/(1+r^2)$ for Poisson statistics (exhibiting level clustering as $r\to0$), while for Wigner-Dyson statistics it is 
\begin{equation}
    P(r)\propto\frac{(r+r^2)^\beta}{(1+r+r^2)^{1+3\beta/2}}
\end{equation}
(with level repulsion $P(r)\propto r^\beta$ as $r\to0$). In Ref.~\cite{atas2013long} the exact joint distribution of all ratios $r_n$ (which are $N-2$ for an $N$-level sequence) was computed for an arbitrary rotationally-invariant ensemble, in terms of an $(N-2)$-fold integral.

Generalizations of the consecutive spacing ratios were considered in Refs.~\cite{atas2013long,chavda2013,atas2013long,tekur2018pre,tekur2018prb,tekur2018c,bhosale2018,srivastava2018}. Ref.~\cite{atas2013long} studied the $k$-overlap ratio, defined as $(E_{n+k+1}-E_n)/(E_{n+k}-E_{n-1})$ (the $r_n$ above is recovered with $k=0$). Other higher-order spacing ratios were considered in Refs.~\cite{atas2013long,tekur2018pre,tekur2018prb,tekur2018c,bhosale2018}. Ref.~\cite{srivastava2018} considered the ratio of the nearest-neighbor (NN) spacing to the next-to-nearest-neighbor (NNN) spacing, which does \emph{not} coincide with $r_n$ defined above when both the NN and the NNN are to the same side of a given level. The transition between Poisson and GOE statistics at the level of ratios~\cite{chavda2013} has also been considered recently.

We finally consider complex spectra. The level spacing statistics is defined as in the real case, by the distance to the NN. By a direct generalization~\cite{grobe1988} of the Berry-Tabor and Bohigas-Giannoni-Schmit conjectures, we expect classically integrable systems and systems with a chaotic semiclassical limit to follow Poisson and Ginibre level statistics, respectively. Poisson statistics on the plane lead to $P(s)=2s$, i.e. linear level repulsion (level clustering is a particularity of a one-dimensional spectrum; in two dimensions the measure $s\d s$ leads to level repulsion; nonetheless, the level repulsion of Poisson processes is always weaker than that of random matrices). For random matrices from the Ginibre ensembles (the paradigmatic complex ensembles) one finds cubic level repulsion, $P(s)\propto s^3$ for small $s$. The behavior for all $s$ is given in terms of an infinite product~\cite{haake2013} or can also be described by a Wigner-like surmise, given in terms of modified Bessel functions~\cite{hamazaki2019}. Remarkably, all three Ginibre ensembles (GinOE, GinUE and GinSE) have the same cubic level repulsion~\cite{grobe1988,grobe1989,haake2013}, independently of the index $\beta$ (i.e. independently of the behavior under time-reversal symmetry). In Ref.~\cite{hamazaki2019} it was shown that different degrees of level repulsion can, nonetheless, exist in nonhermitian matrices, by considering the behavior of the systems under transposition instead of hermitian conjugation. Theoretical spacing distributions (on the unfolded scale) in nonhermitian ensembles are, similarly to their hermitian counterparts, well understood by now~\cite{grobe1988,grobe1989,akemann2009,haake2013,fyodorov1997,hamazaki2019}. The same cannot be said, however, for ratios of complex spacing ratios.

To compare the predictions of the Ginibre ensembles (whose fluctuations are again universal within the respective symmetry classes) with actual spectra, we again need unfolding. In the complex plane the situation is even worse: the unfolding is in principle ambiguous; even so, one can find a minimal prescription that guarantees uniform unfolded complex level density~\cite{akemann2019}. To bypass the difficult and unreliable unfolding procedure, it is particularly natural to consider ratios of spectral observables in the complex plane. In Ref.~\cite{sa2019CSR} and in Chapter~\ref{chapter:spacings} of this thesis, we introduce and study, for the first time, complex spacing ratios. 
\cleardoublepage


\chapter{Random Liouvillian operators}
\label{chapter:model}

In this chapter, we discuss in more detail the concept of a quantum superoperator (such as the Liouvillian) introduced in Section~\ref{subsection:lindbladian_dynamics}. We also look at the importance of its spectrum and eigenstates for the description of the dynamics of a quantum system (Section~\ref{section:superoperators}). We then describe the unbiased construction of two matrix representations of Liouvillian superoperators: the tensor-product representation in Liouville space (Section \ref{section:tensor_product}), which will form the basis of all analytic and numeric work carried out in this thesis, and the coherence-vector representation (Section ~\ref{section:coherence_vector}).
Finally, in Section \ref{section:random_ensembles}, we discuss some relevant RMT ensembles and how to construct a \emph{random} Liouvillian.

\section{Quantum superoperators}
\label{section:superoperators}

Let $\mathcal{H}$ be an Hilbert space (of pure states) of dimension $N$ with (orthonormal) basis $\ket{n}$. The Liouville space $\mathcal{K}$ of (bounded) operators on $\mathcal{H}$ (that is, the space of density matrices or mixed states) is also an Hilbert space of dimension $N^2$ with inner product $\langle A,B\rangle=\mathrm{Tr}(A^\dagger B)=\langle B,A\rangle^*$. A superoperator~$\mathcal{S}$ is a linear map $\mathcal{S}:\mathcal{K}\to\mathcal{K}$. As usual, $K\in\mathcal{K}$ is an eigenoperator of $\mathcal{S}$ if $\mathcal{S}[K]=\Lambda K$, with $\Lambda\in\mathbb{C}$ the corresponding eigenvalue.

As discussed in Section~\ref{subsection:lindbladian_dynamics}, the continuous-time evolution of a Markovian $N$-level OQS is determined by a Liouvillian superoperator of Lindblad form. If we formally integrate the Lindblad equation, $\partial_t\rho(t)=\mathcal{L}[\rho]$, the state of a system at time $t$ is
\begin{equation}
    \rho(t)=e^{t\mathcal{L}}\rho(0)=\sum_{\alpha=0}^{N^2-1} \langle \tilde{\rho}_\alpha,\rho(0)\rangle e^{\Lambda_\alpha t}\rho_\alpha\,,
\end{equation}
where we have expanded the density matrix in the eigenbasis of the system. Here, the right eigenoperators
of $\mathcal{L}$, respecting $\mathcal{L}\left[\rho_{\alpha}\right]=\Lambda_{\alpha}\rho_{\alpha}$, with $\alpha=0,\dots,N^{2}-1$, are denoted by $\rho_{\alpha}$, with
$\Lambda_{\alpha}$ the respective eigenvalues, while $\tilde{\rho}_\alpha$ are left eigenvectors, satisfying $\mathcal{L}^{\dagger}[\tilde{\rho}_\alpha^{\dagger}]=\Lambda_\alpha\tilde{\rho}_\alpha^{\dagger}$. Because the Liouvillian is not a normal operator ($\mathcal{L}\neq\mathcal{L}^\dagger$), left and right eigenoperators do not coincide with each other, but, nonetheless, satisfy $\langle\tilde{\rho}_\alpha,\rho_\beta\rangle=\delta_{\alpha\beta}$. Since $\mathcal{L}[\rho]^\dagger=\mathcal{L}[\rho^\dagger]$, the eigenvalues of $\mathcal{L}$ are real or come in complex conjugated pairs. Furthermore, by construction, $\re\left(\Lambda_{\alpha}\right)\le\Lambda_{0}=0$.
The zero-mode $\rho_{0}$, if unique, is the asymptotic steady-state, left invariant
by the evolution, $\lim_{t\to\infty}\rho(t)=\rho_0$, since all other states decay in time with rate $\abs{\re\left(\Lambda_\alpha\right)}$. 

We thus see that the spectral and steady-state properties of an OQS can be determined by solving the eigenvalue problem of $\mathcal{L}$. Next, we introduce two matrix representations of $\mathcal{L}$ such that we can employ usual numerical and analytical techniques for solving eigenvalue problems.

\section{Tensor-product representation}
\label{section:tensor_product}

The first representation relies on the identification of an operator in $\mathcal{K}$ with a vector in $\mathcal{H}\otimes\mathcal{H}$.  In the basis $\ket{n}$ of $\mathcal{H}$, an operator $K=\sum_{nm}K_{nm}\ket{n}\bra{m}\in\mathcal{K}$ is \emph{vectorized} as
\begin{equation}\label{eq:vectorization}
    \kket{K}=\sum_{nm}K_{nm}\kket{nm}\equiv\sum_{nm}K_{nm}\ket{n}\otimes\bra{m}^{\sf T}\,.
\end{equation}
If the superoperator $\mathcal{S}_L$ acts on $K$ by left multiplication by $S\in\mathcal{K}$, i.e.\ $\mathcal{S}_L[K]=SK$, we have 
\begin{equation}
     SK=\sum_{nm}K_{nm}S\ket{n}\bra{m}\longrightarrow \sum_{nm}K_{nm}(S\ket{n})\otimes\bra{m}^{\sf T}=(S\otimes \mathbbm{1})\kket{K}\,.
\end{equation}
If $\mathcal{S}_R$ acts by right multiplication, i.e $\mathcal{S}_R[K]=KS$, then 
\begin{equation}
  KS=\sum_{nm}K_{nm}\ket{n}\bra{m}S\longrightarrow\sum_{nm}K_{nm}\ket{n}\otimes S^{\sf T}\bra{m}^{\sf{T}}=(\mathbbm{1}\otimes S^{\sf T})\kket{K}\,.
\end{equation}
So, we have the identifications $\mathcal{S}_L\cong S\otimes \mathbbm{1}$ and $\mathcal{S}_R\cong \mathbbm{1}\otimes S^{\sf T}$. 

In this representation, operators in $\mathcal{K}$ are represented by $N^2\times 1$ (column-) vectors and superoperators by $N^2\times N^2$ complex matrices. One can readily check that the identification corresponds to taking all the rows of the matrix $K$ and juxtaposing them in a column vector $\kket{K}$.  

Specializing to the case of the Liouvillian superoperator, we see that the action of $\mathcal{L}$ on $\rho$ is given entirely by left and right multiplication by the Hamiltonian and by the jump operators. The superoperator $\mathcal{L}[\cdot]$ in Eq.~(\ref{eq:Liouvillian_diagonal}) is thus represented by the matrix 
\begin{equation}\label{eq:Liouvillian_tensor}
\mathcal{L}=-i\left(H\otimes\mathbbm{1}-\mathbbm{1}\otimes H^{\T}\right)+\sum_{\ell=1}^{r}\left[W_\ell\otimes W_\ell^*-\frac{1}{2}\left(W_\ell^{\dagger}W_\ell\otimes\mathbbm{1}+\mathbbm{1}\otimes W_\ell^{\sf T}W_\ell^*\right)\right]\,.
\end{equation}
The eigenvalues of the matrix representation of $\mathcal{L}$ are the $\Lambda_\alpha$ and the eigenoperators $\rho_\alpha$ can be reconstructed from the eigenvectors of the matrix representation by running Eq.~(\ref{eq:vectorization}) in reverse.

Before moving on to the second representation, we recast the Liouvillian into a canonical form~\cite{hall2014}. This new form is also the one used in the numerical implementation, since it expresses the matrix representation in a given basis. We define a complete orthogonal basis $\left\{G_{i}\right\}$ in $\mathcal{K}$, with $i=0,\dots,N^{2}-1$, respecting $\Tr\left[G_{i}^{\dagger}G_{j}\right]=\delta_{ij}$, with $G_{0}=\mathbbm{1}/\sqrt{N}$ proportional to the identity. A particularly useful operator-basis is discussed in Section \ref{section:coherence_vector}. The Hamiltonian is decomposed as $H=\sum_{j=0}^{N^2-1}h_jG_j$. Each jump operator is decomposed as $W_{\ell}=g\sum_{j=1}^{N^{2}-1}l_{j\ell}G_{j}$. Note that the $W_{\ell}$ are taken to be traceless, i.e.\ orthogonal to $G_{0}$, to ensure that the dissipative term in Eq.~(\ref{eq:Liouvillian_tensor}) does not contribute to the Hamiltonian dynamics. Recall that we denote by $r$ the number of jump operators, i.e.\ $\ell=1,\dots,r$. Furthermore, we assume that all $r$ decay channels have the same average strength, parameterized by the coupling constant $g>0$. Then, in the $\left\{ G_{i}\right\}$ basis, the tensor-product representation of the Liouvillian is given by
\begin{equation}\label{eq:Liouvillian_canonical}
\mathcal{L}=-i\sum_{j=1}^{N^2-1}h_j\left(G_j\otimes\mathbbm{1}-\mathbbm{1}\otimes G_j\right)+g^2\sum_{j,k=1}^{N^{2}-1}d_{jk}\left( G_{j}\otimes G_{k}^*-\frac{1}{2}\left(G_{k}^{\dagger}G_{j}\otimes \mathbbm{1}+\mathbbm{1}\otimes G_{k}^{\T}G_{j}^*\right)\right),
\end{equation}
where 
\begin{equation}
    d_{jk}=\sum_{\ell=1}^{r}l_{j\ell}l_{k\ell}^{*}=(ll^{\dagger})_{jk}
\end{equation} 
is an $(N^2-1)\times(N^2-1)$ positive-definite matrix. After a choice of basis, the Liouvillian is then fully specified by two matrices, the $N\times N$ Hamiltonian $H$ and the $(N^2-1)\times(N^2-1)$ dissipation matrix $d$. It is the matrix of Eq.~(\ref{eq:Liouvillian_canonical}) which is numerically diagonalized in Chapter~\ref{chapter:numerics}.

\section{Coherence-vector representation}
\label{section:coherence_vector}

In the second representation, which we present here for completeness, we decompose the operators in the Lindblad equation in a specific orthogonal operator-basis in $\mathcal{K}$ (the $\{G_i\}$ of the previous section) and rewrite the equation in terms of the coefficients of this decomposition. Since density matrices are hermitian and have unit trace, one can work with $N^2-1$ \emph{real} coefficients by choosing the $G_i$ hermitian and traceless. A particularly useful basis set is the set of generators of the algebra $\mathfrak{su}(N)$ in the defining representation. The $N^2-1$ expansion coefficients form a vector $\vec{v}\in\mathbb{R}^{N^2-1}$  (see Eq.~(\ref{eq:decomposition_rho}) below), called the coherence vector and, accordingly, this representation is dubbed the coherence-vector representation~\cite{alicki2007,kimura2003,meyerov2018}. $\vec{v}$ is the $N$-dimensional analog of the Bloch vector for a qubit~\cite{nielsen2002}.

The generators $G_i$ satisfy the commutation and anti-commutation relations 
\begin{subequations}\label{eq:comm_acomm_su(n)}
    \begin{alignat}{2}
        \label{eq:commG}
            \comm{G_i}{G_j}&=2\,if_{ijk}G_k\,,\\
        \label{eq:acommG}
            \acomm{G_i}{G_j}&=\frac{4}{N}\delta_{ij}\mathbbm{1}+2c_{ijk}G_k\,,
    \end{alignat}
\end{subequations}
where the real structure constants $f_{ijk}$ and $c_{ijk}$ form a totally antisymmetric tensor and a totally symmetric tensor, respectively. Using the commutation/anticommutation relations, we can express the product of two generators as
\begin{equation}\label{eq:product_two_generators}
    G_iG_j=\frac{2}{N}\delta_{ij}\mathbbm{1}+z_{ijk}G_k\,,
\end{equation}
with the complex structure constants $z_{ijk}=c_{ijk}+if_{ijk}$, which satisfy $z_{ijk}=z^*_{jik}=z^*_{ikj}$, i.e. interchanging any pair of indices amounts to replacing $z\to z^*$. We have adopted a convention where the generators are orthogonal but \emph{not} normalized, satisfying instead $\Tr[G_iG_j]=2\delta_{ij}$. As mentioned before, the generators are traceless, $\Tr[G_i]=0$. Trace identities with more generators which will be used in the derivation below can be found in Ref.~\cite{haber2017}.

We decompose $\rho$, $H$ and $W_\ell$ in the basis of the generators of $\mathfrak{su}(N)$ (note there are slightly different conventions from the previous representation),
\begin{subequations}\label{eq:decomposition}
    \begin{alignat}{3}
        \label{eq:decomposition_rho}
            \rho&=\frac{1}{N}\mathbbm{1}+\frac{1}{2}v_iG_i\,,\\
        \label{eq:decomposition_H}
            H&=\frac{1}{2}h_iG_i\,,\\
        \label{eq:decomposition_L}
            W_\ell&=\frac{g}{\sqrt 2} \gamma_\ell\,l_{i\ell}G_i\,,
    \end{alignat}
\end{subequations}
with $i,\ell\in\{1,\dots,N^2-1\}$ and a sum over repeated indices understood. The prefactor $1/N$ in the decomposition of $\rho$ is such that $\Tr\rho=1$ and the values of the remaining prefactors in $\rho$, $H$ and $W_\ell$ are chosen for later convenience. Note that in the decompositions of $W_\ell$ and $H$ there is no term proportional to $\mathbbm{1}$: for the jump operator we have imposed tracelessness as a condition; the Hamiltonian only enters the Liouvillian via a commutator hence the identity component does not contribute and can be dropped. Note also that since $\rho$ and $H$ are hermitian its decomposition coefficients $v_i$ and $h_i$ are real, while the $l_{i\ell}$ are complex because no hermiticity condition is imposed on the $W_\ell$. We further impose that $\sum_\ell \gamma_\ell^2=1$, such that the overall dissipation strength is set by $g$ and the $\gamma_\ell$ control the number of decay channels and their relative strength. The convention of the previous section is recovered by setting $\gamma_\ell=1/\sqrt{r}$ for $\ell=1,\dots r$ and $\gamma_\ell=0$ for $\ell>r$. The definition of the dissipation matrix $d$ in this representation includes the weights $\gamma_\ell$:
\begin{align}\label{eq:d_matrix}
    d_{ij}=\sum_{\ell=1}^{N^2-1} \gamma_\ell^2 l_{i\ell}l_{j\ell}^*\,.
\end{align}

Inserting Eqs.~(\ref{eq:decomposition_rho})--(\ref{eq:decomposition_L}) into Eq.~(\ref{eq:Liouvillian_diagonal}) and using Eqs.~(\ref{eq:comm_acomm_su(n)}) and (\ref{eq:product_two_generators}), we obtain
\begin{equation}\label{eq:dot_v_sigma}
\begin{split}
    \dot{v}_iG_i&=-\frac{i}{2}h_i v_j\comm{G_i}{G_j}+\frac{g^2}{2}d_{ij}\left(G_i(\mathbbm{1}+v_kG_k)G_j-\frac{1}{2}\acomm{G_jG_i}{\mathbbm{1}+v_kG_k}\right)\\
    &=f_{ijk}h_iv_jG_k+\frac{g^2}{4}d_{ij}v_k\left(2G_iG_kG_j-G_jG_iG_k-G_kG_jG_i\right)+\frac{g^2}{2}d_{ij}\comm{G_i}{G_j}\\
    &=-f_{ijk}h_kv_jG_i+i\frac{g^2}{2}d_{kl}\left(z_{ikm}f_{jlm}-z^*_{ilm}f_{jkm}\right)v_jG_i+ig^2 f_{ijm}d_{ij}G_k\,.
\end{split}
\end{equation}
To remove the basis operators from Eq.~(\ref{eq:dot_v_sigma}), we project it onto some basis element, i.e. we apply the superoperator $\Tr[G_a\,\cdot\,]$, and use the trace normalization $\Tr[G_iG_j]=2\delta_{ij}$, yielding
\begin{align}
    \dot{v}_i&=(A_{ik}+g^2 S_{ik})v_k+g^2 b_i\,,
\end{align}
where 
\begin{subequations}\label{eq:general_ASb}
    \begin{alignat}{3}
        \label{eq:general_A}
            A_{ij}&=-f_{ijk}\,h_k\,,\\
        \label{eq:general_S}
            S_{ij}&=\frac{i}{2}\left(z_{ikm}f_{jlm}-z^*_{ilm}f_{jkm}\right)d_{kl}\,,\\
        \label{eq:general_b}
            b_{i}&=if_{ijk}\,d_{jk}\,.&
    \end{alignat}
\end{subequations}
The matrix $A$ is antisymmetric, while $S$ is symmetric if $N=2$ (for which the structure constants are $f_{ijk}=\varepsilon_{ijk}$, the Levi-Civita symbol, and $c_{ijk}=0$) or if the jump operators are hermitian. In the latter case, the dissipation matrix is symmetric $d_{jk}=d_{kj}$, whence $S_{ij}=-f_{ikm}f_{jlm}d_{kl}=S_{ji}$, and furthermore $\vec{b}=0$ since $b_i$ is the contraction of a symmetric tensor with an antisymmetric one.

By setting $\dot{\vec{v}}=0$, we obtain the steady-state, $\vec{v}_0=-g^2(A+g^2 S)^{-1}\vec{b}$, and denoting $M_g\equiv A+g^2 S$, the evolution of the coherence vector is given by:
\begin{equation}\label{eq:coherence_vector_evolution}
    \dot{\vec{v}}=M_g(\vec{v}-\vec{v}_0)\,.
\end{equation}

Recall that $\sigma(\mathcal{L})=\{\Lambda_i\}_{i=0}^{N^2-1}$, where we denote the spectrum of an operator $\mathcal{O}$ by $\sigma(\mathcal{O})$. We now show that $\sigma(M_g)=\{\Lambda_i\}_{i=1}^{N^2-1}$, i.e.\ that the spectra of $\mathcal{L}$ and $M_g$ coincide modulo a zero-eigenvalue, $\sigma\left(\mathcal{L}\right)=\sigma\left(M_g\right)\cup\{0\}$. By the expansion of Eq.~(\ref{eq:decomposition_rho}), and again using the projection superoperator $\Tr[\sigma_a\,\cdot]$, we obtain $\dot{\vec{v}}=\Lambda \vec{v}$, or, by Eq.~(\ref{eq:coherence_vector_evolution}), $M_g(\vec{v}-\vec{v}_0)=\Lambda \vec{v}$. Differentiating this relation we arrive at
\begin{equation}
    (M_g-\Lambda\mathbbm{1})\dot{\vec{v}}=0\,,
\end{equation}
which implies that either $\dot{\vec{v}}=0$, whence $\Lambda=0$, or $\det(M_g-\Lambda\mathbbm{1})=0$, whence $\Lambda$ is an eigenvalue of $M_g$. This computation also proves that a steady-state always exists and that it is unique if $M_g$ is diagonalizable. Since the set of nondiagonalizable matrices has measure zero in the space of all matrices, if we choose a random $M_g$, then the steady-state is unique.


\section{Random matrix ensembles}
\label{section:random_ensembles}

Having identified the relevant superoperator whose eigenvalue problem we seek to solve and having constructed matrix representations which can be diagonalized, we now introduce randomness into the model by specifying the random ensembles from which to draw the Hamiltonian $H$ and the jump operators coefficient matrices $l$. We consider two cases: real matrices, $H_{ij}=H_{ji},l_{ij}\in\mathbb{R}$ (labeled by $\beta=1$), and complex matrices, $H_{ij}=H_{ji}^{*},l_{ij}\in\mathbb{C}$ (labeled by $\beta=2$).

The Hamiltonian is restricted to be hermitian, hence we draw $H$ from a Gaussian ensemble with unit variance,
\begin{equation}\label{eq:weight_Gaussian}
    P_{N\times N}\left(H\right)\propto e^{-\frac{1}{2}\Tr\left(H^{2}\right)}\,.
\end{equation}
Because of the of the Gaussian character of the weight, all moments of $H$ can be related to the propagator $\langle H_{ij}H_{kl} \rangle=\delta_{il}\delta_{jk}$ by Wick contraction. 

The spectral density of the eigenvalues of $H$ (denote them by $E$) is the Wigner semicircle law, which can be derived in several ways (see, for instance, Refs.~\cite{haake2013,livan2018,jurkiewicz2008}):
\begin{equation}
    \varrho_{\mathrm{W}}(E)=\frac{2}{\pi \omega^2}\sqrt{\omega^2-E^2}\,,\quad\quad -\omega<E<\omega\,,
\end{equation}
where $\omega>0$ sets the energy scale. The choice of weight of Eq.~(\ref{eq:weight_Gaussian}) fixes $\omega^2=2\beta N$. Since $\varrho(E)$ is a symmetric function, the eigenvalues $E$ have zero mean, $\mu_\mathrm{W}=0$, and variance $\sigma_\mathrm{W}^2=\langle E^2\rangle=\omega^2/4$. With the scaling $\omega^2=2\beta N$, the standard variation of the semicircle law is $\sigma_\mathrm{W}=\sqrt{\beta N/2}$.

The jump operators have no constraints (besides tracelessness which has already been imposed \textit{a priori}) and hence we draw the $(N^2-1)\times r$ matrices $l$ from a Ginibre ensemble, also with unit variance, 
\begin{equation}\label{eq:weight_Ginibre}
 P_{(N^{2}-1)\times r}\left(l\right)\propto e^{-\frac{1}{2}\Tr\left(l^{\dagger}l\right)}\,.   
\end{equation}
Once again, all moments of $l$ are related by Wick contractions to the second moment $\langle l_{j\ell}l_{k\ell'}^*\rangle=\delta_{jk}\delta_{\ell\ell'}$.

We now consider the scaling of the dissipative term. Let $W$ be an $M\times N$ matrix with $M/N\equiv m\geq1$ and consider the limit $N,M\to\infty$ with $m$ fixed. In Chapter~\ref{chapter:analytics} we show, Eq.~(\ref{eq:self_energy_Wishart}), that the self-energy of the Wishart ensemble (i.e.\ of matrices $W^\dagger W$) is $\Sigma_W[G(z)]=\lambda m/(z-\lambda G(z))$ ($\lambda$ sets the overall scale, analogously to $\omega$ above). The choice of weight of Eq.~(\ref{eq:weight_Ginibre}) sets $\lambda=\beta N$ (we set $\lambda=1$ in Chapter~\ref{chapter:analytics}). From $\Sigma_W$, one calculates the Green's function and then the spectral density (recall Section~\ref{section:free_probability} and see Section~\ref{subsection:diagrammatics_Wishart} for further details), which is the Marchenko-Pastur (MP) law,
\begin{equation}\label{eq:xi_MP}
    \varrho_\mathrm{MP}(x)=\frac{1}{2\pi\lambda x}\sqrt{(\xi_+-x)(x-\xi_-)}\,,\quad \xi_-<x<\xi_+\,,
\end{equation}
where $\xi_{\pm}=\lambda(1\pm\sqrt{m})^2$.
The mean of the MP law is $\mu_\mathrm{MP}=m\lambda$ and the standard deviation $\sigma_\mathrm{MP}=\lambda\sqrt{m}$. 

A related matrix which will also arise in this thesis is $\Gamma=\sum_{\ell=1}^rW_\ell^\dagger W_\ell$, with $W_\ell$ now square matrices ($m=1$). Following Ref.~\cite{can2019}, we show, using free-addition of random matrices, that the spectrum of $\Gamma$ also follows an MP law, but with modified endpoints $\xi_\pm$. From Chapter~\ref{chapter:state_of_the_art}, we know that the self-energy (or $\mathcal{R}$-transform) is additive. It then immediately follows that the self-energy of $\Gamma$ is
\begin{equation}
    \Sigma_\Gamma[G(z)]=\sum_{\ell=1}^r\Sigma_{W}[G(z)]=\frac{\lambda r}{1-\lambda G(z)}\,.
\end{equation}
If we now notice that this coincides exactly with the self-energy of the Wishart ensemble quoted above, upon changing $m\to r$, we immediately conclude that the eigenvalues of $\Gamma$ are also distributed according to the MP law, Eq.~(\ref{eq:xi_MP}), but with $\xi_{\pm}=(1\pm\sqrt{r})^2$. In conclusion, we see that the distributions for a single rectangular jump operator (parameterized by $m$) and for a sum of square jump operators (parameterized by $r$) are completely equivalent. This result will be heavily used in Chapter~\ref{chapter:analytics}. 
\cleardoublepage


\chapter{Numerical results}
\label{chapter:numerics}

The aim of this chapter is to numerically study the properties of the spectrum and the steady-state of general Liouvillian operators, Eq.~(\ref{eq:Liouvillian_canonical}), parameterized by $N$, $\beta$, $r$ and $g$ and drawn from the ensemble described in Section~\ref{section:random_ensembles}. The spectrum and eigenvectors of the Liouvillian were obtained by exact diagonalization of this matrix representation. We are interested in the large-$N$ (thermodynamic) limit, particularly in scaling properties. The numerical results of this chapter are complemented by analytic calculations in Chapter~\ref{chapter:analytics}.

The chapter is organized as follows. First, we study the global spectra of the Liouvillian, focusing on the spreading of decay rates, $X$, and on cuts of the complex spectrum along certain lines (Section~\ref{section:global_spectrum}). Next, we consider the spectral gap, $\Delta$, (Section~\ref{section:spectral_gap}). We also consider the steady-state and study its purity and spectral distribution (Section~\ref{section:steady_state}). Finally, we separately analyze the case of a single decay channel, which is qualitatively different (Section~\ref{section:r=1}).

This chapter is based on our results in Ref.~\cite{sa2019}.

\section{Global spectral features}
\label{section:global_spectrum}

\begin{figure}[htbp]
\centering 
\includegraphics[width=0.99\columnwidth]{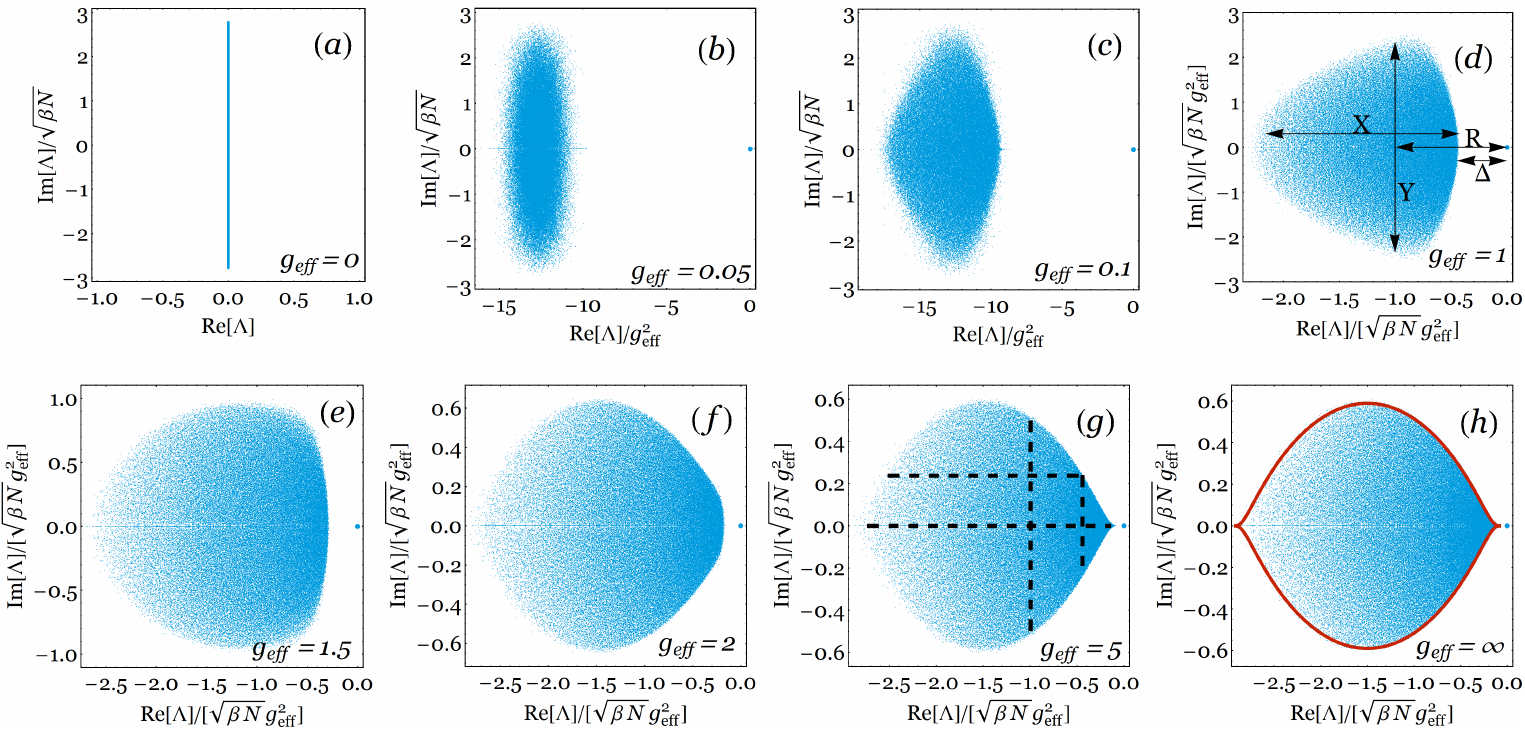}
\caption{Spectrum of a random Liouvillian operator for different values of $g_{\mathrm{eff}}=\left(2r\beta N\right)^{1/4}g$, computed for $N=80$, $\beta=2$ and $r=2$. $(d)$ depicts $R$, $X$, $Y$ and $\Delta$ which are, respectively, the center of mass of the spectrum, the standard deviation along the real and imaginary axes, and the spectral gap. The horizontal (vertical) dashed lines in $(g)$ correspond to $0$, $Y$ ($R$, $R+X$). The solid line in $(h)$ is the (shifted and rescaled) boundary explicitly computed in Ref.~\cite{denisov2018}.}
\label{fig:spectrum}
\end{figure}

We first consider the shape of the spectrum of random Liouvillians in the complex plane for various values of the dissipation strength $g$, Fig.~\ref{fig:spectrum}. The boundaries of the spectrum evolve from a straight line along the imaginary axis, for zero dissipation, to an ellipsoid, for small $g$, to a lemon-like shape, at large $g$.

In Fig.~\ref{fig:spectrum}~$(h)$ we plot (solid line) the spectral boundary, explicitly computed in Ref.~\cite{denisov2018} for the case $r=N^{2}-1$, given by the solutions $z\in\mathbb{C}$ to the equation
\begin{equation}
    \mathrm{Im}\left[z+G(z)\right]=0\,,
\end{equation}
where
\begin{equation}
    G(z)=2z\left\{1-\frac{1}{3\pi}\left[(4+z^2)\,E\!\left(\frac{4}{z^2}\right)+(4-z^2)\,K\!\left(\frac{4}{z^2}\right)\right]\right\}\,,
\end{equation}
and $E(\cdot)$, $K(\cdot)$ are elliptic integrals of the first and second kind, respectively. The spectral boundary thus obtained is centered at the origin and has to be displaced by a shift $-(1+r)/(2\sqrt{r})$ and then rescaled by $\sqrt{2}$.\footnote{The reasoning for this shift and rescaling is as follows. In Ref.~\cite{denisov2018} the spectrum is centered at zero by subtracting its mean value (center of mass), which, for $r=N^2-1$ coincides with the center of the spectral support. For finite $r$, this is no longer the case, and the center of the support is $-(\xi_++\xi_-)/2$, where $\xi_\pm=(1\pm\sqrt{r})^2$ are defined in Eq.~(\ref{eq:xi_MP}), see also the discussion at the end of Section~\ref{subsection:holomorphic_G_Liouvillian}. Furthermore, since the result of Ref.~\cite{denisov2018} is obtained in the infinite-$r$ limit, and no $r$-scaling is taken into account, an overall $\sqrt{r}$ has to be factored out, see below for a discussion of this scaling. Hence, the shift is $-(1+r)/(2\sqrt{r})$ as stated. Finally, the overall rescaling of $1/\sqrt{2}$ is due to a difference in normalizations.} 
These results, obtained here for $r=2$, indicate that the lemon-shaped spectral boundary is ubiquitous in the strong dissipation regime. However, contrarily to the $r=N^{2}-1$ case, the spectral distribution is not symmetric (along the real axis) with respect to the center of the spectral support.

In Fig.~\ref{fig:spectrum}~$(d)$, we represent several relevant energy- (or inverse time-) scales. 
The spectral gap, $\Delta=\min_{\alpha>0}\re\left(-\Lambda_{\alpha}\right)$ describes the typical time it takes the system to reach the steady-state. 
The variance along the real axis, $X^{2}=\sum_{\alpha}\left[\re\left(\Lambda_{\alpha}-R\right)\right]^{2}/N^{2}$, sets the spread of decay rates and also the typical minimum time to observe the onset of dissipation and decoherence. 
The variance along the imaginary axis, $Y^{2}=\sum_{\alpha}\left(\im\Lambda_{\alpha}\right)^{2}/N^{2}$, gives the timescale for the oscillations of the states' phases.
We also depict the center of mass of the spectrum, $R=\sum_{\alpha}\Lambda_{\alpha}/N^{2}$, which can usually be trivially shifted away.

\begin{figure}[tbp]
\centering 
\includegraphics[width=0.99\columnwidth]{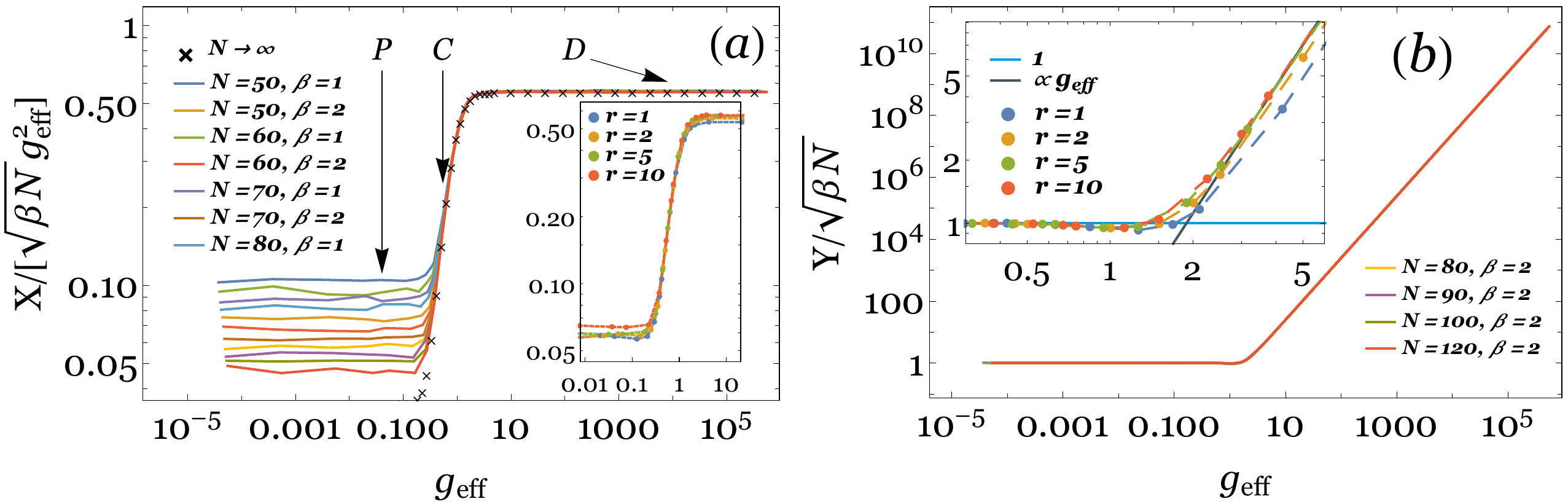}
\caption{Scaling of $X$ and $Y$ for different values of $N$ and $r$ as a function of $g_{\mathrm{eff}}=\left(2r\beta N\right)^{1/4}g$.}
\label{fig:spectrum_XY}
\end{figure}

It is worthwhile to note that the axes of Figs.~\ref{fig:spectrum} $(a)$--$(c)$ are scaled differently from Figs.~\ref{fig:spectrum} $(d)$--$(h)$, which signals that the finite-size scaling properties of the system vary as a function of dissipation strength. This fact can be motivated by the scaling of the two terms in Eq.~(\ref{eq:Liouvillian_canonical}) as follows.
The spectral density of $H$ is given by the Wigner semicircle law, with zero mean and standard deviation $\sqrt{\beta N/2}$ (recall Section~\ref{section:random_ensembles}). The dissipation term follows a Marcheko-Pastur law, with mean $\beta N r g^2$ and standard deviation $\beta N \sqrt{r}g^2$. Since the eigenvalues of $H$ and $d$ are real, when considered separately, i.e.\ at the non-dissipative ($g=0$) or at the fully dissipative ($g=\infty$) limits, respectively, they only contribute to the imaginary or to the real parts of $\Lambda$, respectively. Although at finite $g$ these considerations are no longer exact, we expect them to yield the leading scaling behavior. Thus, at large $g$ the main contribution to $X$ comes from the dissipative term and $X\propto \beta N \sqrt{r}g^2$ (see Fig.~\ref{fig:spectrum_XY}~$(a)$). For weak dissipation the $Y$ scaling is dominated by the Hamiltonian term, hence $Y\propto\sqrt{\beta N}$ (see Fig.~\ref{fig:spectrum_XY}~$(b)$). The passage from the weak dissipation scalings to the strong dissipation scalings should occur when the two terms in the Liouvillian are of the same order, $\sqrt{\beta N/2}\simeq \beta N \sqrt{r} g^2$, i.e.\ $g_\mathrm{eff}\simeq 1$, where we define
\begin{equation}\label{eq:geff_def}
g_{\mathrm{eff}}=\left(2r\beta N\right)^{1/4}g\,.
\end{equation}

\begin{figure}[tbp]
\centering
    \includegraphics[width=0.99\columnwidth]{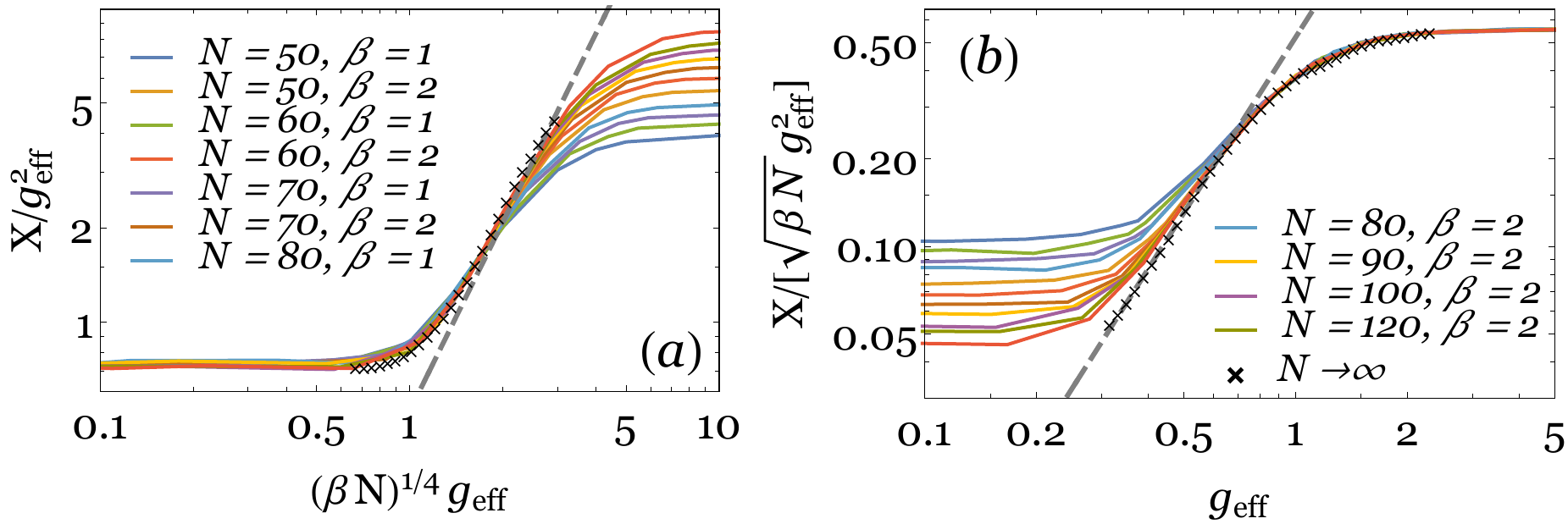}
    \caption{$X$ as a function of $g_\mathrm{eff}$ for $r=2$ and various $N$ and $\beta$. The black crosses give the extrapolation of the data for $N\to\infty$ and the dashed gray line is $\propto g_\mathrm{eff}^{2}$. Collapse to the universal curve for small-$g_\mathrm{eff}$ in $(a)$ and for large-$g_\mathrm{eff}$ in $(b)$.}
    \label{fig:SM_X_ScalingFunc} 
\end{figure}

The above heuristic reasoning for the scalings of Fig.~\ref{fig:spectrum_XY} can be put on firmer grounds by scaling function arguments. We first analyze the specific case of $X$ and then, in Section~\ref{subsection:compatibility}, we give the general procedure.
The three representative points (P, C, D) of Fig.~\ref{fig:spectrum_XY}~$(a)$ correspond to regimes for which $X$ depicts a different qualitative behavior. The observed scaling collapse (see also Fig.~\ref{fig:SM_X_ScalingFunc}~$(a)$) shows that, for large $g_{\mathrm{eff}}$, the standard deviation along the real axis behaves as $X\propto\left[(\beta N)^{1/4}g_{\mathrm{eff}}\right]^{2}f_{X}^{>}(g_{\mathrm{eff}})$, with $f_{X}^{>}$ an unknown scaling function satisfying $f_{X}^{>}(x)\propto x^{0}$ for $x\!\to\!\infty$ and $f_{X}^{>}(x)\!\propto\!x^{2}$ for $x\to0$. A similar scaling collapse can be obtained for small $g_{\mathrm{eff}}$ (see Fig.~\ref{fig:SM_X_ScalingFunc}~$(b)$) yielding $X\simeq g_{\mathrm{eff}}^{2}f_{X}^{<}\left[(\beta N)^{1/4}g_{\mathrm{eff}}\right]$, for which the scaling function satisfies $f_{X}^{<}(x)\propto x^{0}$ when $x\!\to\!0$ and $f_{X}^{<}(x)\propto x^{2}$ when $x\!\to\!\infty$.
We can thus identify the three regimes: P, for $g_{\mathrm{eff}}\lesssim\left(\beta N\right){}^{-1/4}$; C, for $(\beta N)^{-1/4}\lesssim g_{\mathrm{eff}}\lesssim(\beta N)^{0}$; and D, for $(\beta N)^{0}\lesssim g_{\mathrm{eff}}$, corresponding to each representative point in Fig.~\ref{fig:spectrum_XY}. 

Note that the rescaled quantities plotted in Fig.~\ref{fig:spectrum_XY} do not seem to depend on the index $\beta$. There is a small $r$-dependence, which converges rapidly for increasing $r$ (see insets) and is most likely due to the reduced sizes possible to attain numerically.

The asymptotic power law behavior of the scaling functions is hard to determine from the available values of $N\leq120$. However, it is fully compatible with large-$N$ extrapolation (see black crosses in Fig.~\ref{fig:spectrum_XY}~$(a)$ and Fig.~\ref{fig:SM_X_ScalingFunc}). For both large-$g_\mathrm{eff}$ (Fig.~\ref{fig:SM_X_ScalingFunc}~$(a)$) and small-$g_\mathrm{eff}$ (Fig.~\ref{fig:SM_X_ScalingFunc}~$(b)$), the extrapolated data to $N\to\infty$, taken as the $y$-intercept of a linear fit in $(\beta N)^{-1}$, agrees very well with the power-law $g_\mathrm{eff}^{2}$ (gray line) in regime C. Upon entering regimes P and D, the extrapolation points naturally start deviating from the power-law curve.
This result is further corroborated by the compatibility conditions $\lim_{g_{\mathrm{eff}}\to0}(\beta N)^{1/2}f_{X}^{>}(g_{\mathrm{eff}})\simeq\lim_{(\beta N)^{1/4}g_{\mathrm{eff}}\to\infty}f_{X}^{<}\left[(\beta N)^{1/4}g_{\mathrm{eff}}\right]$ (see also Section~\ref{subsection:compatibility}).

For the standard deviation along the imaginary axis we find, $Y\simeq\sqrt{\beta N}$ for regimes P and C, and $Y\propto\sqrt{\beta N}g_{\mathrm{eff}}^{2}$ within regime D. The variance in regime P can be explained by a perturbative treatment of the dissipative term--the value of $Y$ corresponds to the unit variance of the random Hamiltonian as discussed above.

\begin{figure}[htbp]
\centering
    \includegraphics[width=0.5\columnwidth]{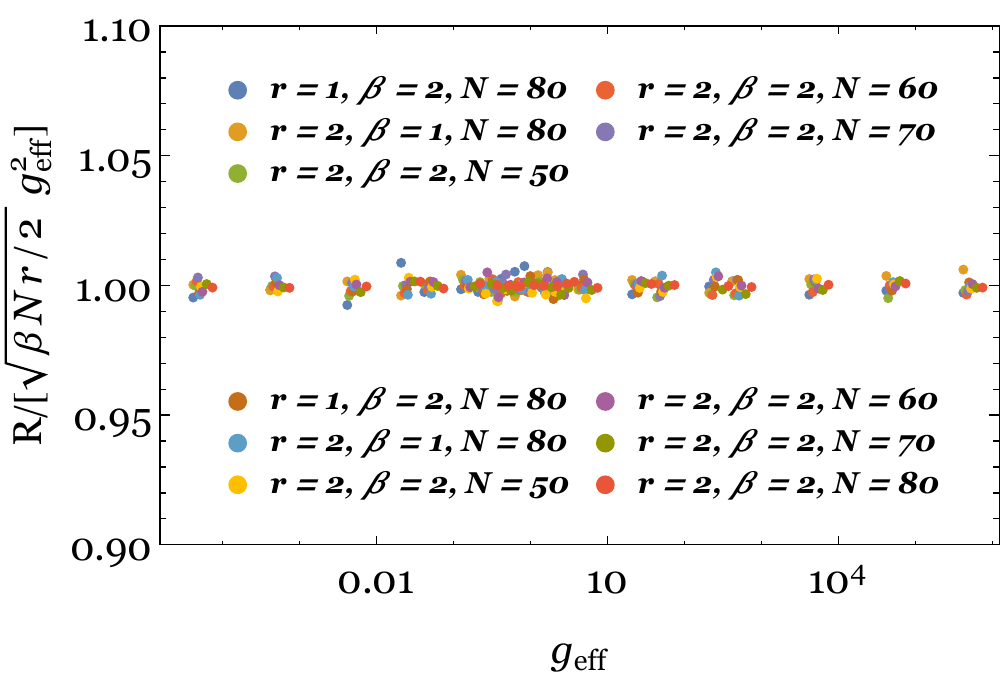}
    \caption{Center of mass $R$ of the spectrum for different combinations of $N$, $\beta$, $r$ and $g$.}
    \label{fig:SM_R}
\end{figure}

For completeness, we compute the exact value of $R$ for arbitrary $N$. From its definition, we have $R=(1/N^2)\sum_{\alpha}\Lambda_\alpha=(1/N^2)\left\langle\Tr\mathcal{L}\right\rangle$. Inserting the expression for $\mathcal{L}$ from Eq.~(\ref{eq:Liouvillian_canonical}), recalling that the operator-basis $\{G_i\}$ is traceless and orthonormalized, noting that the Hamiltonian contribution is a commutator and hence traceless, and using $\Tr[A\otimes B]=\Tr A\,\Tr B$ for any matrices $A$ and $B$, we obtain
\begin{equation}
    R=-\frac{1}{N}g^2\sum_{j=1}^{N^2-1}\langle d_{jj}\rangle=-\frac{1}{N}g^2\sum_{j=1}^{N^2-1}\sum_{\ell=1}^r\langle l_j^\ell {l_j^\ell}^*\rangle=-\frac{1}{N}\beta g^2\sum_{j=1}^{N^2-1}\delta_{jj}\sum_{\ell=1}^r\delta_{\ell\ell}=-\beta N r g^2 \left(1-\frac{1}{N^2}\right).
\end{equation}
In the limit of large $N$, the center of mass thus scales as
$R=-\beta Nrg^{2}$. The numerical results for different combinations of $N$, $\beta$ and $r$ are given in Fig.~\ref{fig:SM_R}, with perfect agreement. Note that by taking traces of powers of $\mathcal{L}$ and $\mathcal{L}^\dagger$ we cannot access higher-order moments because the operator $\mathcal{L}$ is not normal (i.e.\ it does not commute with its hermitian conjugate). We can only constrain $\Tr\mathcal{L}^2=X^2-Y^2$ but since $\Tr\mathcal{L}^\dagger\mathcal{L}\neq X^2+Y^2$, we cannot obtain $X$ and $Y$ individually this way.

\subsection{Compatibility of scaling functions}
\label{subsection:compatibility}
The compatibility of scaling functions arguments given for $X$ can be generalized for any quantity which has the same qualitative behavior. This procedure also allows us to systematize the scaling exponents found, by defining three types of exponents: $\nu$, $\kappa$ and $\lambda$, related to the finite-size scaling of the spectral and steady-state quantities (say, $X$), the finite-size scaling of the boundaries of the multiple regimes (say, P, C, D), and the $g_\mathrm{eff}$ scaling of the quantities in the crossover regime (say, of $X$ in C), respectively.

We define three regimes with $N$-dependent boundaries in which some quantity $Q$ has qualitatively distinct behaviors: P$_Q$, for $g_\mathrm{eff}\lesssim(\beta N)^{\kappa_Q^<}$; C$_Q$ for $(\beta N)^{\kappa_Q^<}\lesssim g_\mathrm{eff}\lesssim(\beta N)^{\kappa_Q^>}$; and D$_Q$ for $(\beta N)^{\kappa_Q^>}\lesssim g_\mathrm{eff}$. In P$_Q$, $Q$ behaves as 
\begin{equation}
    Q\propto g_\mathrm{eff}^2(\beta N)^{\nu_Q^P}f_Q^<[(\beta N)^{-\kappa_Q^<}g_\mathrm{eff}]\,,
\end{equation} while in D$_Q$,
\begin{equation}
    Q\propto g_\mathrm{eff}^2(\beta N)^{\nu_Q^D}f_Q^>[(\beta N)^{-\kappa_Q^>}g_\mathrm{eff}]\,.
\end{equation}
Note that some extra factors of $g_\mathrm{eff}$ may exist (such as the extra $g_\mathrm{eff}^2$ for $X$) but they do not modify the argument, as long as they are the same in regimes P$_Q$ and D$_Q$. If they are different, a straightforward modification of Eq.~(\ref{eq:exponents_constraint}) below is required, but this issue did not arise for the quantities studied in this work. We further assume that the scaling functions $f_Q$ have asymptotic power-law behaviors, that is
\begin{equation}
    f_Q^<(x) \propto
    \begin{cases}
        x^0, & \mathrm{if}\ x\to0\\
        x^{\lambda_Q^<}, & \mathrm{if}\ x\to\infty
    \end{cases}
    \quad\quad \mathrm{and} \quad\quad
    f_Q^>(x) \propto
    \begin{cases}
        x^{\lambda_Q^>}, & \mathrm{if}\ x\to0\\
        x^0, & \mathrm{if}\ x\to\infty
    \end{cases}\,.
\end{equation}
The two limiting behaviors of $Q$ should match in the intermediate regime C$_Q$, that is,
\begin{equation}
        \lim_{g_\mathrm{eff}(\beta N)^{-\kappa_Q^<}\to0}(\beta N)^{\nu_Q^P}f_Q^<\left[(\beta N)^{-\kappa_Q^<}g_\mathrm{eff}\right]\simeq
        \lim_{g_\mathrm{eff}(\beta N)^{-\kappa_Q^>}\to\infty}(\beta N)^{\nu_Q^D}f_Q^>\left[(\beta N)^{-\kappa_Q^>}g_\mathrm{eff}\right]\,,
\end{equation}
whence the equality
\begin{equation}
    g_\mathrm{eff}^{2+\lambda_Q^<}\left(\beta N\right)^{\nu_Q^P-\kappa_Q^<\lambda_Q^<}=g_\mathrm{eff}^{2+\lambda_Q^>}\left(\beta N\right)^{\nu_Q^D-\kappa_Q^>\lambda_Q^>}
\end{equation}
follows. This equality implies that $\lambda_Q^<=\lambda_Q^>\equiv\lambda_Q$ and establishes a relation between $\nu$, $\kappa$ and $\lambda$ exponents, which are thus not all independent but constrained by
\begin{equation}\label{eq:exponents_constraint}
    \lambda_Q\left(\kappa_Q^>-\kappa_Q^<\right)=\nu_Q^D-\nu_Q^P\,.
\end{equation}

Note that the exponent $\nu_Q^C$ (giving the collapse of curves of different $\beta N$ in regime $C_Q$) is not defined in the above argument and thus does not enter the constraint. However for all cases of interest we either found $\nu_Q^C=\nu_Q^P$ or $\nu_Q^C=\nu_Q^D$.

In Table~\ref{tab:exponents}, at the end of this chapter, we summarize the values of all exponents for $r=1$ and $r>1$, as defined above. It can be checked that they all satisfy Eq.~(\ref{eq:exponents_constraint}).

\subsection{Cuts of the spectral density along representative lines}

\begin{figure}[htbp]
\centering \includegraphics[width=0.99\columnwidth]{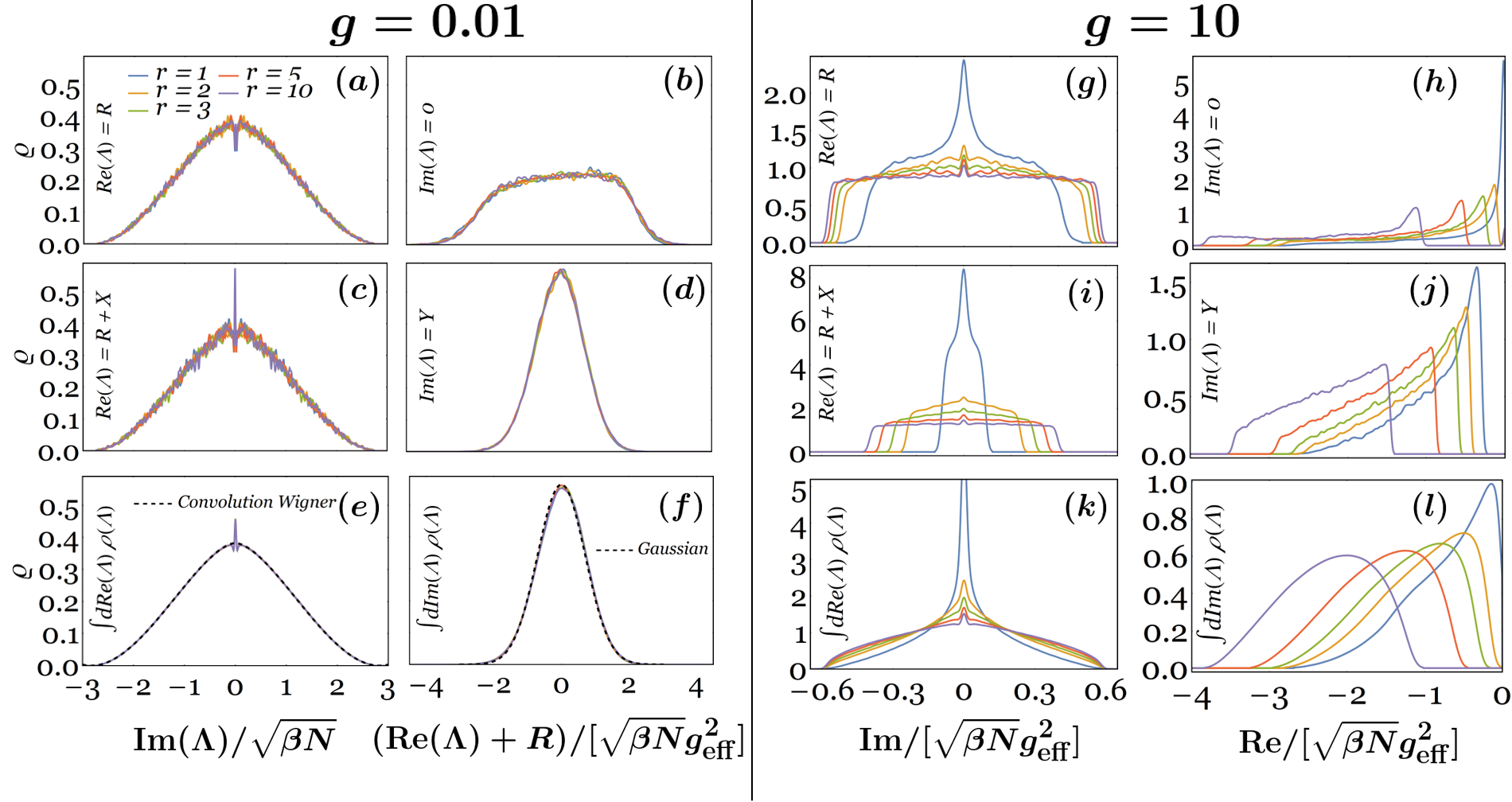}
\caption{Spectral density along the cuts $\re\left(\Lambda\right)=R$, $R+X$ and $\im\left(\Lambda\right)=0$, $Y$, computed for $N=80$, $\beta=2$, several values of $r$ and two values of $g$, such that the system is in regime P: $(a)-(d)$; and in regime D: $(g)-(j)$. $(e)$, $(f)$: integrated density along the real and imaginary axes, $\varrho_{\mathrm{I}}\left(\Lambda_{\mathrm{I}}\right)$ and $\varrho_{\mathrm{R}}\left(\Lambda_{\mathrm{R}}\right)$, for regime P. $(k)$ and $(l)$: same quantities for regime D. In $(b)$, $(d)$, $(f)$, $(h)$, $(j)$ and $(l)$ the zero eigenvalue is omitted.}
\label{fig:spectrum_cuts}
\end{figure}

Figures~\ref{fig:spectrum_cuts} $(a)$ to $(f)$ and $(g)$ to $(l)$ show the spectral density, $\varrho\left(\Lambda\right)=\sum_{\alpha}\delta^{2}\left(\Lambda-\Lambda_{\alpha}\right)$, along the cuts depicted in Figs.~\ref{fig:spectrum} $(g)$ for regimes P and D, respectively, for various numbers of jump operators. Results for $\beta=1$, i.e.\ real matrices, show the same limiting behavior in the large-$N$ limit, after proper rescaling.

For P, the density along $\re\left(\Lambda\right)=R$ and $R+X$, does not depend on the number of jump operators $r$, since we are in the regime dominated by unitary evolution. The integrated distribution of the imaginary parts, $\varrho_{\mathrm{I}}\left(\Lambda_{\mathrm{I}}\right)=\int\d^{2}\Lambda\,\delta\!\left(\Lambda_{\mathrm{I}}-\im\Lambda\right)\varrho\left(\Lambda\right)$, is well described by a convolution of Wigner's semicircle laws, $\varrho_{\mathrm{W}}(E)=(1/\pi)\sqrt{2-E^2}$, i.e.\
\begin{equation}
    \varrho_{\mathrm{I}}\left(\Lambda_{I}\right)\simeq\int\d E_{1}\d E_{2}\,\delta\!\left(\Lambda_{I}-E_{1}+E_{2}\right)\varrho_{\mathrm{W}}\left(E_{1}\right)\varrho_{\mathrm{W}}\left(E_{2}\right)\,,
\end{equation}
except at $\Lambda_{I}=0$ where there is an increase of spectral weight depleted from the immediate vicinity of this point. The spectral weights along the cuts $\im\left(\Lambda\right)=0$, $Y$ scale with $\sqrt{r}$ and the integrated distribution of the real parts, $\varrho_{\mathrm{R}}\left(\Lambda_{\mathrm{R}}\right)=\int\d^{2}\Lambda\,\delta\!\left(\Lambda_{\mathrm{R}}-\re\Lambda\right)\varrho\left(\Lambda\right)$, is well approximated by a Gaussian. As was the case for the variances in the P regime, these results follow from a perturbative small-$g$ expansion. 

In the strongly dissipative regime, D, the spectral density along $\re\left(\Lambda\right)=R$, $R+X$ depends on $r$ but converges rapidly to the $r\to\infty$ limit. The case $r=1$ is qualitatively different from $r>1$ (it will be studied in Section~\ref{section:r=1}). This can also be observed in the $\im\left(\Lambda\right)=0$, $Y$ cuts, and in $\varrho_{\mathrm{R}}$, where for $r=1$ the spectral weight is finite for $\Lambda_{\mathrm{R}}\to0^{-}$, in contrast with the $r>1$ results that develop a spectral gap.

\section{Spectral gap}
\label{section:spectral_gap}
\begin{figure}[htbp]
\centering 
\includegraphics[width=0.99\columnwidth]{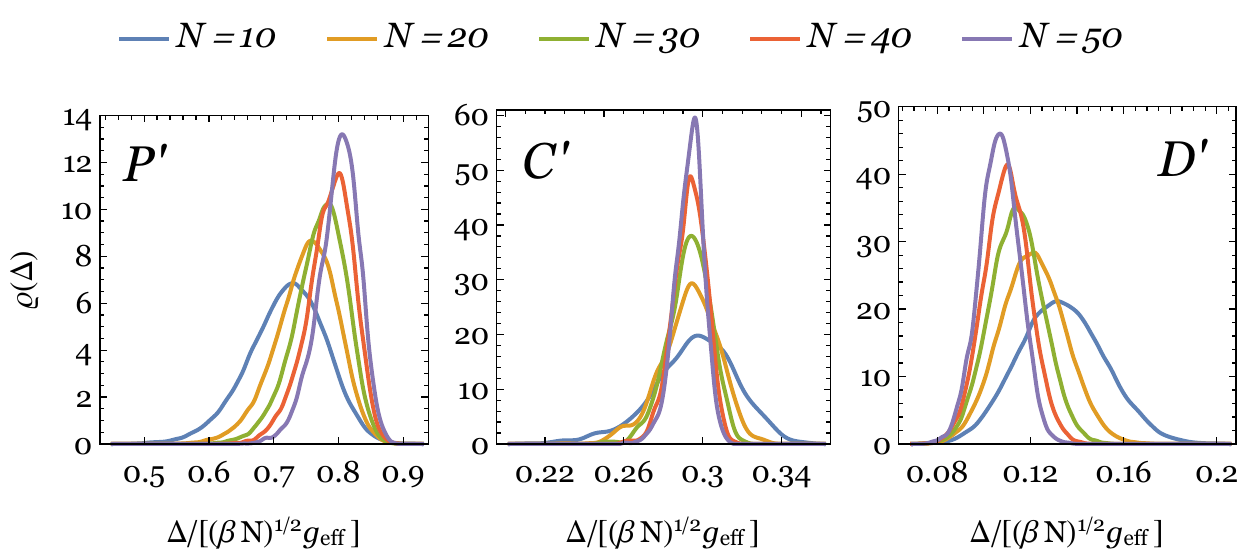}
\caption{\label{fig:SM_gap_distribution_function}Scaled gap distribution function for each of the regimes P$'$, C$'$, D$'$ (regimes defined below).}
\end{figure}

We now turn to the study of $\Delta$, which is a particularly important spectral feature since it determines the long-time relaxation asymptotics. For all dissipation regimes, the variance of the gap decreases with $N$ and the value of $\Delta$ becomes sharply peaked around its mean, see the gap distribution functions depicted in Fig.~\ref{fig:SM_gap_distribution_function}; hence, the mean gap accurately describes the long-time dynamics. The mean of the different distributions in Fig.~\ref{fig:SM_gap_distribution_function} is not constant due to finite-size effects to be discussed shortly.

\begin{figure}[tbp]
\centering \includegraphics[width=0.99\columnwidth]{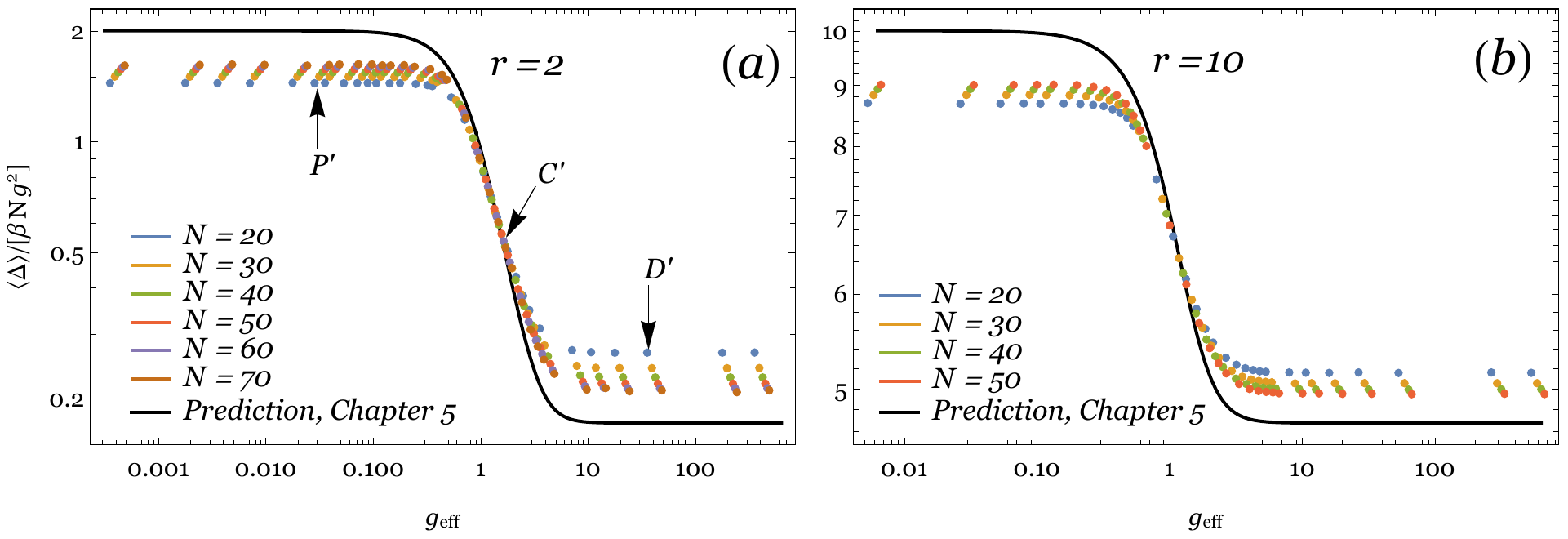}
\caption{Average spectral gap as a function of $g_{\mathrm{eff}}$ plotted for different values of $N$, $\beta=2$ and $r=2$, $(a)$, and $r=10$, $(b)$. Dots: exact diagonalization; solid line: solution to Eq.~(\ref{eq:K_boundary}), see also Section~\ref{subsection:comparison_gap}.}
\label{fig:gap}
\end{figure}

Figure~\ref{fig:gap} shows the average spectral gap, $\langle\Delta\rangle$ as a function of $g_{\mathrm{eff}}$, for $\beta=2$, $r=2$, $10$ and different values of $N$. We also find three qualitatively different regimes (P$'$, C$'$, D$'$) whose boundaries \emph{do not} coincide with those in Figs.~\ref{fig:spectrum_cuts}. Here, the boundaries $g_{\mathrm{eff}}\simeq g_{\mathrm{P'C'}}$ and $g_{\mathrm{eff}}\simeq g_{\mathrm{C'D'}}$, separating the P$'$ and C$'$, and the C$'$ and D$'$ regimes, respectively, are independent of $N$ for $N\to\infty$. For regime P$'$, the average gap behaves as $\langle\Delta\rangle\propto(\beta N)^{1/2}g_{\mathrm{eff}}^{2}$; for C$'$, the gap varies as $\langle\Delta\rangle\propto(\beta N)^{1/2}g_{\mathrm{eff}}$; for D$'$, we observe again (like for P$'$) $\langle\Delta\rangle\propto(\beta N)^{1/2}g_{\mathrm{eff}}^{2}$. The values of the exponents (as defined in Section~\ref{subsection:compatibility}) thus obtained are again tabulated in Table~\ref{tab:exponents}.
The equalities $\nu_P'=\nu_D'$ and $\kappa_\Delta^<=\kappa_\Delta^>$ imply that the exponent $\lambda_\Delta$ is not defined since the arguments of Section~\ref{subsection:compatibility} do not determine it. Said in a different way, for the gap, different scaling functions for small and large dissipation, which asymptotically match in an intermediate regime of size growing with $N$, do not exist. Instead, the gap is described by a single function for all $g$.

As before, these results are only possible to establish by performing a large-$N$ extrapolation of the available data, due to the presence of large finite-size corrections. Fitting directly the numerical average spectral gap for several $N$ and multiple $r>1$ seems to indicate that there is an $r$-dependence of the scaling exponents $\nu_{P'}$, $\nu_{C'}$, $\nu_{D'}$. In particular, $\nu_{P'}=\nu_{C'}=\nu_{D'}$ only for large $r$, but $\nu_{P'}>\nu_{C'}>\nu_{D'}$ for small $r>1$. However, this is incompatible with both the picture drawn in Section~\ref{subsection:compatibility} and with the finiteness of the gap in the thermodynamic limit for $r>1$, alluded to in the previous section and discussed in detail below. Since we can only access relatively small $N$, this apparent $r$-dependence is likely a finite-size effect suppressed by $r$. Indeed, values of the exponents obtained by data extrapolation to large-$N$ are compatible with $\nu_{P'}=\nu_{C'}=\nu_{D'}=1/2$, independently of $r$. 

\subsection{Comparison with the analytic expression}
\label{subsection:comparison_gap}
In Chapter~\ref{chapter:analytics}, we analytically compute the spectral gap as a function of dissipation strength and number of decay channels, using holomorphic Green's functions methods. In the limit $N\to\infty$, the spectral gap is given by the solution to Eq.~(\ref{eq:K_boundary}). A comparison of this prediction with the numerical results for $r=2$, $10$ is given by the solid lines in Fig.~\ref{fig:gap}, showing a good qualitative description for all $g_\mathrm{eff}$ and also a quantitatively accurate result in regime C$'$. However, for the asymptotic values as $g_\mathrm{eff}\to0,\infty$ there is a discrepancy, with the theoretical result overestimating the correct result at small dissipation and underestimating it at strong dissipation. This deviation has, at least, two sources. First, it is due to finite-size effects, which we have already seen are quite significant here (recall, for instance, Fig.~\ref{fig:SM_gap_distribution_function}). Indeed, in the center of regime C$'$, the finite-size effects are negligible and the agreement with the theoretical curve is very good; for small (large) $g_\mathrm{eff}$ the overestimation (underestimation) of the numerical results is consistent with the finite-size scaling, since $\Delta$ increases (decreases) with $N$. Second, the agreement improves as $r$ increases: the relative deviation between the numerical and analytical results at, say, small dissipation, is $\sim25\%$ for $r=2$, but only $\sim10\%$ for $r=10$ and $\sim5\%$ at $r=50$ (not shown). Combining these two observations we conclude that a perfect match between the numerical and analytical results only occurs in the double scaling limit $N\to\infty$, $r\to\infty$.

Closed-form expressions for the solution of Eq.~(\ref{eq:K_boundary}) exist but are rather involved. In contrast, the limiting values for small- and large-$g_{\mathrm{eff}}$ can be stated succinctly. Both limits can also be derived independently from Eq.~(\ref{eq:K_boundary}): using degenerate perturbation theory for small-$g_\mathrm{eff}$ and using holomorphic Green's functions for the Wishart ensemble for large-$g_\mathrm{eff}$ (both computations are also performed explicitly in Chapter~\ref{chapter:analytics}).
For large $g_{\mathrm{eff}}$, we find that 
\begin{equation}
    \langle\Delta\rangle=\beta Ng^{2}(1-\sqrt{r})^{2}=\sqrt{\beta N}g_\mathrm{eff}\frac{(1-\sqrt{r})^2}{\sqrt{2r}}\,,
\end{equation} 
which is compatible with Ref.~\cite{can2019}.
For small $g_\mathrm{eff}$, we find that 
\begin{equation}
    \langle\Delta\rangle=\beta Ng^{2}r=\frac{1}{2}g_\mathrm{eff}^2\,.
\end{equation}

\begin{figure}[htbp]
\centering
    \includegraphics[width=0.99\columnwidth]{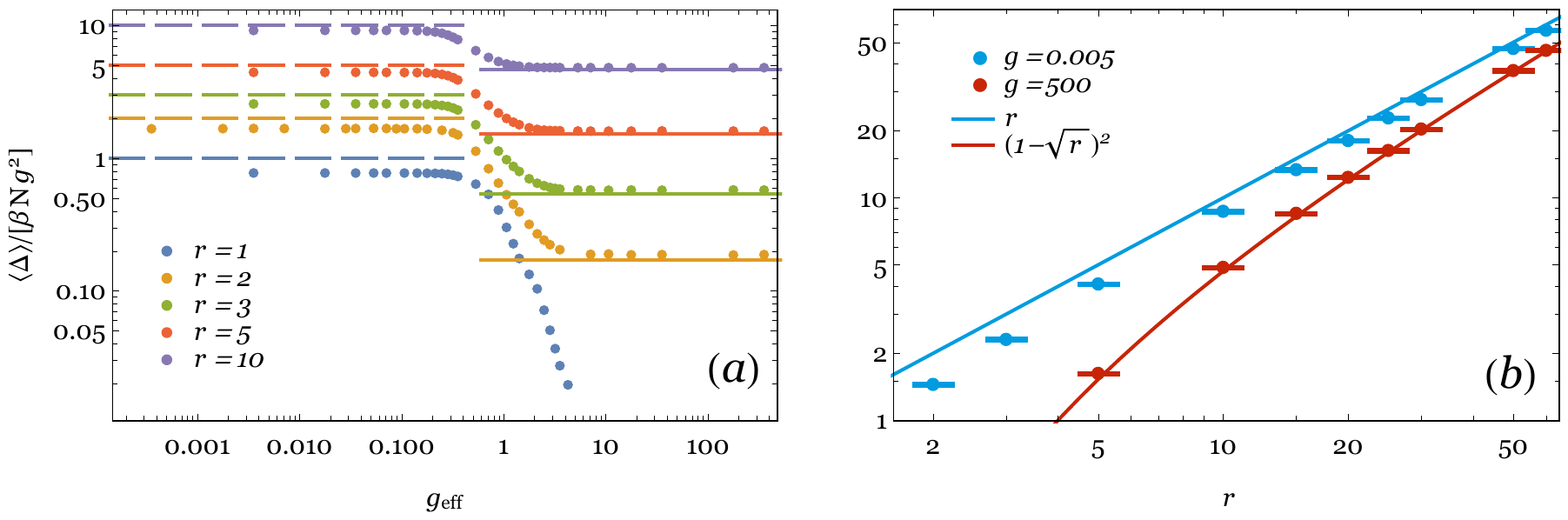}
    \caption{$(a)$: Evolution of the extrapolated spectral gap at $N\to\infty$ with $g_\mathrm{eff}$ for $r=2$, $3$, $5$ and $10$. The the dashed lines are the perturbative results $\langle\Delta\rangle=\beta N g^2 r$ for small $g_\mathrm{eff}$, the full lines the analytic predictions $\langle\Delta\rangle=\beta N g^2(1-\sqrt{r})^2$ for large $g_\mathrm{eff}$. $(b)$: Evolution of the spectral gap with the number of jump operators $r$, for $N=60$, $\beta=1,2$ and $g=0.005$ (blue) and $g=500$ (red); the full lines correspond to the analytic predictions.}
    \label{fig:SM_gap_r}
\end{figure}

These predictions are depicted in Fig.~\ref{fig:SM_gap_r}~$(a)$, together with extrapolated data to $N\to\infty$ for several values of $r$ up to $10$, and in Fig.~\ref{fig:SM_gap_r}~$(b)$, for fixed $N=60$ and $r$ up to $60$. We see that they describe the gap increasingly well for growing $r$; for $r=2$, although not exact, they give a good estimate, see Fig.~\ref{fig:SM_gap_r}~$(b)$. Notwithstanding that the above two results are derived in the limits $g_{\mathrm{eff}}\to0$ and $g_{\mathrm{eff}}\to\infty$, they provide a remarkable description for the whole P$'$ and D$'$ regimes, respectively.

For the special case $r=1$, although three regimes are also present, see Section~\ref{section:r=1}, the scaling of the gap with $N$ changes in the strongly dissipative regime and $\langle\Delta\rangle\to0$ as $g_\mathrm{eff}\to\infty$, in the thermodynamic limit.

\section{Steady-state}
\label{section:steady_state}

\subsection{Steady-state purity}
\label{subsection:SS_purity}
We now turn to the characterization of the steady-state $\rho_{0}$. Because of the normalization of probabilities, we have $\Tr\rho_0=\langle p_0\rangle=1$ identically, where $p_0$ is a steady-state probability (eigenvalue of $\rho_0$). Therefore, we consider instead the variance, $\sigma_{\rho_{0}}^{2}$, of the eigenvalues of $\rho_{0}$. This quantity is related to the difference between the purity of the steady-state, $\mathcal{P}_{0}=\Tr\left[\rho_{0}^{2}\right]$, which quantifies the degree of mixing of $\rho_{0}$, and that of a fully-mixed state $\mathcal{P}_{\mathrm{FM}}=1/N$, $\mathcal{P}_{0}-\mathcal{P}_{\mathrm{FM}}=N\sigma_{\rho_{0}}^{2}$.

\begin{figure}[tbp]
\centering 
\includegraphics[width=0.99\columnwidth]{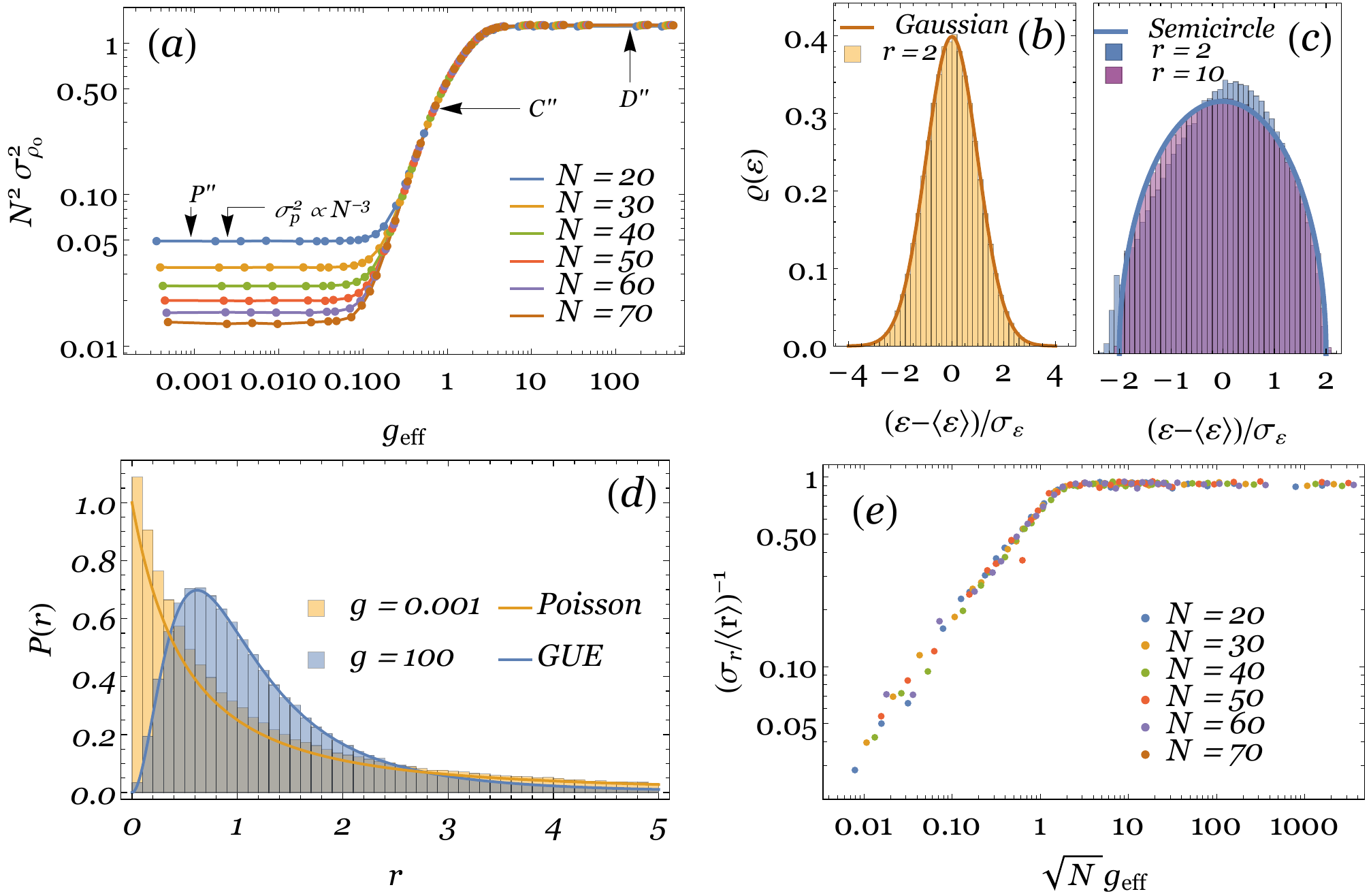}
\caption{$(a)$: Variance of steady-state probabilities as a function of $g_{\mathrm{eff}}$ plotted for different values of $N$, $\beta=2$ and $r=2$. $(b)-(c)$: spectral density for the effective Hamiltonian $\mathcal{H}$ for $g=0.001$ and $g=100$, respectively. $(d)$: Statistics of level spacing ratios for weak and strong dissipation and comparison with approximate analytic predictions for Poisson and RMT statistics (solid lines). $(e)$: ratio of the first two moments of the distribution of $r$, $(\sigma_{r}/\langle r\rangle)^{-1}$, for $\beta=2$ and $r=2$.}
\label{fig:steady_state}
\end{figure}

\begin{figure}[tbp]
\centering
    \includegraphics[width=0.99\columnwidth]{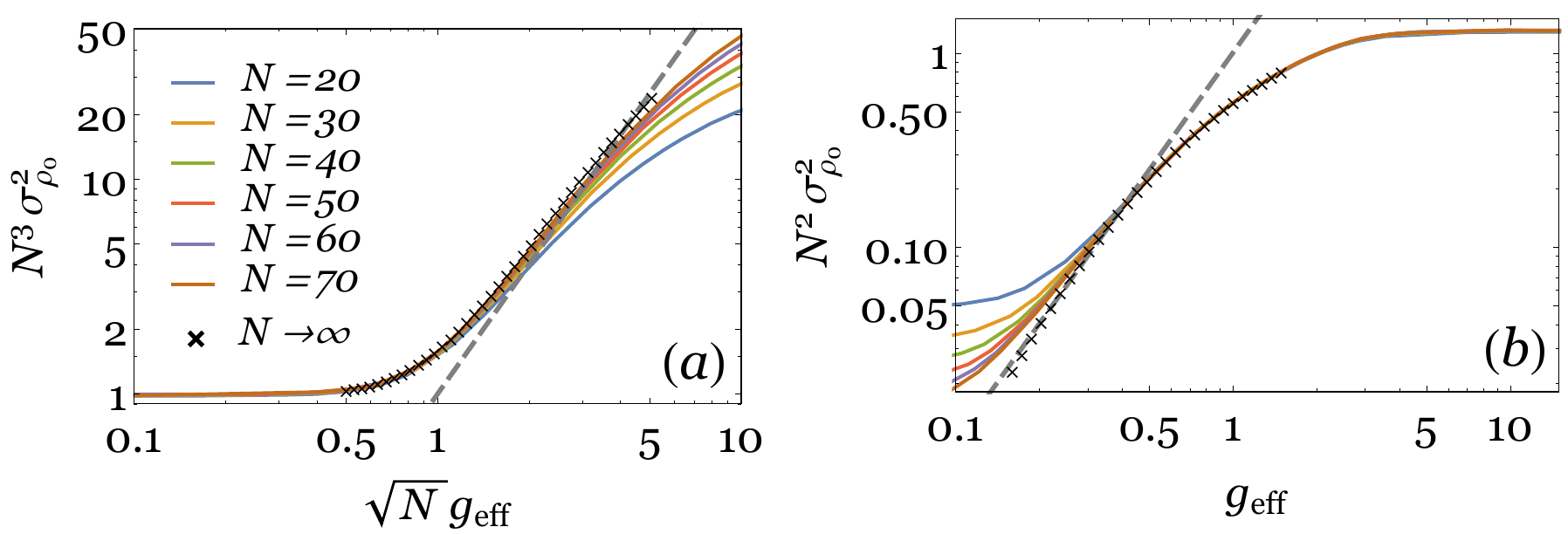}
    \caption{$\sigma_{\rho_0}^2$ as a function of $g_\mathrm{eff}$ for $r=2$ and various $N$. The black crosses give the extrapolation of the data for $N\to\infty$ and the dashed gray line is $\propto g_\mathrm{eff}^{\lambda_{\rho_0}}$. Note data collapse to the universal curve for small-$g_\mathrm{eff}$ in $(a)$ and for large-$g_\mathrm{eff}$ in $(b)$.}
    \label{fig:SM_purity_ScalingFunc}
\end{figure}

Figure~\ref{fig:steady_state}~$(a)$ shows the variance~$\sigma_{\rho_{0}}^{2}$ as a function of $g_{\mathrm{eff}}$ for $\beta=2$ and $r=2$. 
Here, again, three different regimes can be observed whose boundaries \emph{do not} coincide with those given for previous quantities. In regime P$''$, $g_{\mathrm{eff}}\lesssim (\beta N)^{-\frac{1}{2}}$, we observe $\sigma_{\rho_{0}}^{2}\propto N^{-3}f_{\rho_{0}}^{<}\left(N^{1/2}g_{\mathrm{eff}}\right)$, with $f_{\rho_{0}}^{<}(x)\propto x^{0}$, for $x\!\to\!0$, and $f_{\rho_{0}}^{<}(x)\!\propto\!x^{2}$ for $x\!\to\!\infty$. In regime D$''$, for $N^{0}\lesssim g_{\mathrm{eff}}$, we have $\sigma_{\rho_{0}}^{2}\propto N^{-2}f_{\rho_{0}}^{>}(g_{\mathrm{eff}})$, with $f_{\rho_{0}}^{>}(x)\!\propto\!x^{0}$ for $x\!\to\!\infty$ and $f_{\rho_{0}}^{>}(x)\!\propto\!x^{2}$ for $x\!\to\!0$. The crossover regime C$''$ can be accessed by both asymptotic expansions and corresponds to $\sigma_{\rho_{0}}^{2}\propto N^{-2}g_{\mathrm{eff}}^{2}$. The finite-size scaling exponents $\nu_{P''}=-3$ and $\nu_{C''}=\nu_{D''}=-2$ (see Table~\ref{tab:exponents} for the complete list) are in very good agreement with fits of the data to the respective power-laws, for various $r>1$ ($r=2$, $3$, $5$, $10$) and show no dependence on $r$ as expected. The exponents $\lambda_{\rho_0}$ and $\kappa_{\rho_0}^<$ are also compatible with data extrapolation, see Fig.~\ref{fig:SM_purity_ScalingFunc}, by proceeding in the same way as for $X$.

These scalings imply that, up to subleading $1/N$ corrections, the steady-state is fully mixed in regime P$''$, $\langle\mathcal{P}_{0}\rangle=\mathcal{P}_{\mathrm{FM}}+\mathcal{O}\left(1/N^{2}\right)$, while in regimes C$''$ and D$''$, $\mathcal{P}_{0}$ is only proportional to $\mathcal{P}_{\mathrm{FM}}$. 

The case with $r=1$ is again qualitatively different from $r>1$ at strong dissipation, the purity $\langle\mathcal{P}_0\rangle$ being of order $N^0$ (see Section~\ref{section:r=1}), signaling a steady-state closer to a pure state than a fully-mixed one. 

\subsection{Steady-state spectrum}
\label{subsection:SS_spectrum}
Next, we investigate the steady-state's spectrum. For that, it is useful to introduce the effective Hamiltonian $\mathcal{H}=-\log\rho_{0}$. We denote the eigenvalues of $\mathcal{H}$ by $\varepsilon_i$ and present their spectral density $\varrho(\varepsilon)$ in Figs.~\ref{fig:steady_state}~$(b)$ and $(c)$, corresponding to a point in regime P$''$ and D$''$, respectively. In the weak coupling regime, P$''$, $\varrho(\varepsilon)$ is well described by a Gaussian, while at strong coupling, D$''$, it acquires a non-Gaussian shape; for large $r$, $\varrho(\varepsilon)$ is increasingly well described by a Wigner semicircle distribution. A remarkable agreement can already be seen for $r=10$ in the example of Fig.~\ref{fig:steady_state}~$(c)$. For small $r$ (see $r=2$ in Fig.~\ref{fig:steady_state}~$(c)$) there is a systematic skewing of the spectrum to the right.

Figure~\ref{fig:steady_state}~$(d)$ presents the probability distribution, $P(r)$, of consecutive spacing ratios, $r_{i}=s_{i}/s_{i-1}$, with $s_{i}=\varepsilon_{i+1}-\varepsilon_{i}$, which automatically unfolds the spectrum of $\mathcal{H}$ (see the discussion in Chapter~\ref{chapter:state_of_the_art}). The analytic predictions for the GUE and for the Poisson distribution \cite{atas2013} (full lines) are given for comparison. The agreement of the numerical data in the P$''$ and D$''$ regimes with the Poisson and GUE predictions, respectively, is remarkable. Within regime C$''$, we observe a crossover between these two regimes. We thus arrive at the remarkable result that a fully-chaotic Liouvillian may support an integrable steady-state in an appropriate (weak enough) dissipation regime.

To further illustrate the changing of the spectral properties of $\mathcal{H}$ with $g_{\mathrm{eff}}$, we provide in Fig.~\ref{fig:steady_state}~$(e)$ the ratio of the first two moments of the distribution of $r$, $(\sigma_{r}/\langle r\rangle)^{-1}$, which can distinguish between Poisson and GUE statistics. Since in the Poissonian case $P(r)=1/(1+r)^{2}$, the $n$-th moment of the distribution diverges faster than the $(n-1)$-th and thus $\sigma_{r}/\langle r\rangle\to\infty$. On the other hand, for GUE or GOE this ratio is given by a finite number of order unity (e.g. $\sigma_{r}/\langle r\rangle=256\pi^{2}/(27\sqrt{3}-4\pi)^{2}-1\simeq1.160$ for GUE). Figure~\ref{fig:steady_state}~$(e)$ shows that the Poissonian statistics are only attained in the dissipationless limit $N^{1/2}g_{\mathrm{eff}}\to0$. On the other hand, the GUE (for $\beta=2$) or GOE (for $\beta=1$) values are attained for $g_{\mathrm{eff}}N^{1/2}\simeq1$. Thus, in the thermodynamic limit the effective Hamiltonian $\mathcal{H}$ is quantum chaotic for all finite values of $g_{\mathrm{eff}}$.

\section{$r=1$ is different}
\label{section:r=1}
For weak dissipation, the unitary contribution to the Liouvillian dominates and, therefore, the spectral and steady-state properties do not differ qualitatively between $r=1$ and $r>1$. In fact, in regime P$''$, the exponents are always the same in both cases, see Table~\ref{tab:exponents}. However, there are important differences at large $g_\mathrm{eff}$, where dissipation dominates, for both the spectral gap and the steady-state. 

\begin{figure}[htbp]
\centering
    \includegraphics[width=0.99\columnwidth]{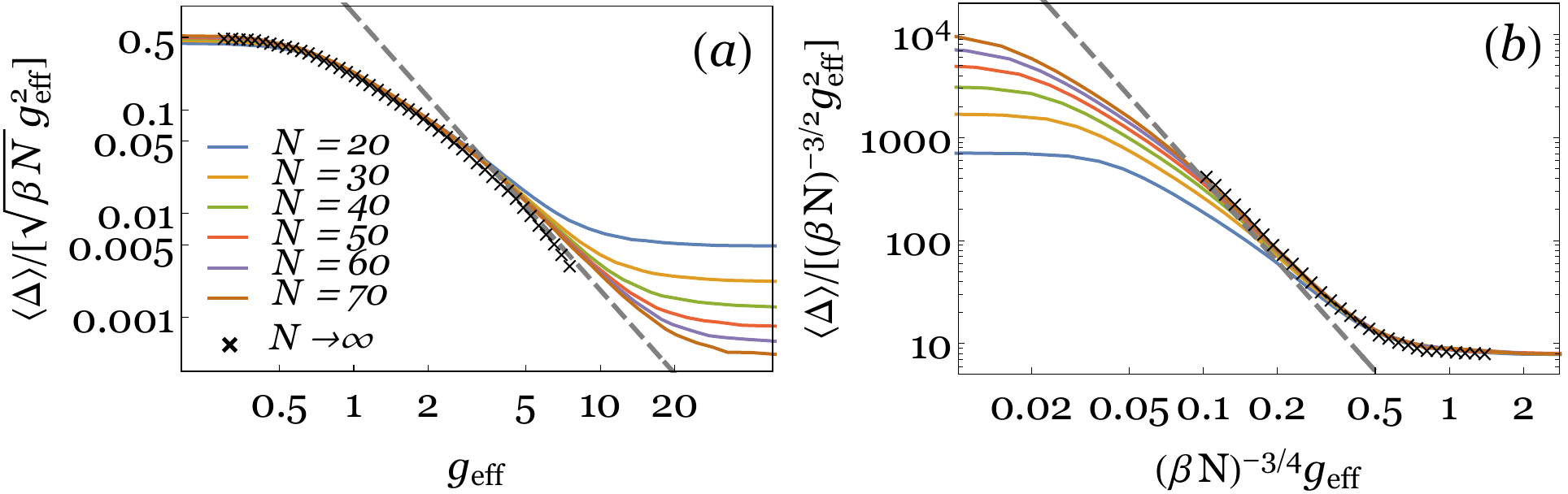}
    \caption{Average spectral gap as a function of $g_\mathrm{eff}$ for $r=1$ and various $N$. The black crosses give the extrapolation of the data for $N\to\infty$ and the dashed gray line is $\propto g_\mathrm{eff}^{\lambda_\Delta}$. Collapse to the universal curve for small-$g_\mathrm{eff}$ in $(a)$ and for large-$g_\mathrm{eff}$ in $(b)$.}
    \label{fig:SM_r1_gap_ScalingFunc}
\end{figure}

Contrary to $r>1$, for $r=1$ the scaling of the gap at large-$g_\mathrm{eff}$ and at small-$g_\mathrm{eff}$ is not the same. In particular, since $\nu_{D'}$, $\lambda_\Delta<0$, in the thermodynamic limit the gap starts to close as dissipation increases (strictly closing only for $g_\mathrm{eff}=+\infty$). This can be seen in Fig.~\ref{fig:SM_r1_gap_ScalingFunc}, where we plot the evolution of $\langle\Delta\rangle$ as a function of $g_\mathrm{eff}$: choosing the scaling of small-$g_\mathrm{eff}$ (which coincides with the scaling for all $g_\mathrm{eff}$ for $r>1$), at large-$g_\mathrm{eff}$ the gap follows the dashed line (power-law $g_\mathrm{eff}^{2-\lambda_\Delta}$) and hence closes at $g_\mathrm{eff}=\infty$. Once again, the exponents listed in Table~\ref{tab:exponents} are compatible with data extrapolation, see the black crosses in Fig.~\ref{fig:SM_r1_gap_ScalingFunc}. The procedure for data extrapolation is the same as before. 

Furthermore, contrary to $r>1$, the values of $g_\mathrm{eff}$ that define the boundaries between different regimes are $N$-dependent. Hence $\lambda_\Delta$ can be defined, for $r=1$. The gap has a nonzero exponent $\kappa_\Delta^>$ and the matching procedure of Section~\ref{subsection:compatibility} is applicable. 
In particular, with $\kappa_\Delta^>=3/4$ (see Fig.~\ref{fig:SM_r1_gap_ScalingFunc}~$(b)$), we can use Eq.~(\ref{eq:exponents_constraint}) to determine $\lambda_\Delta=-8/3$. Hence, we find, at large dissipation, $\langle\Delta\rangle\propto g_\mathrm{eff}^{-2/3}$ in agreement with Ref.~\cite{can2019}. The fact that (in the thermodynamic limit) $\langle\Delta\rangle\to0$ as $g_\mathrm{eff}\to\infty$ can be seen as a manifestation of the quantum Zeno effect~\cite{misra1977,itano1990}, which is a statement that although a small dissipative coupling enhances decay, too strong dissipation may suppress it.

\begin{figure}[t]
\centering
    \includegraphics[width=0.99\textwidth]{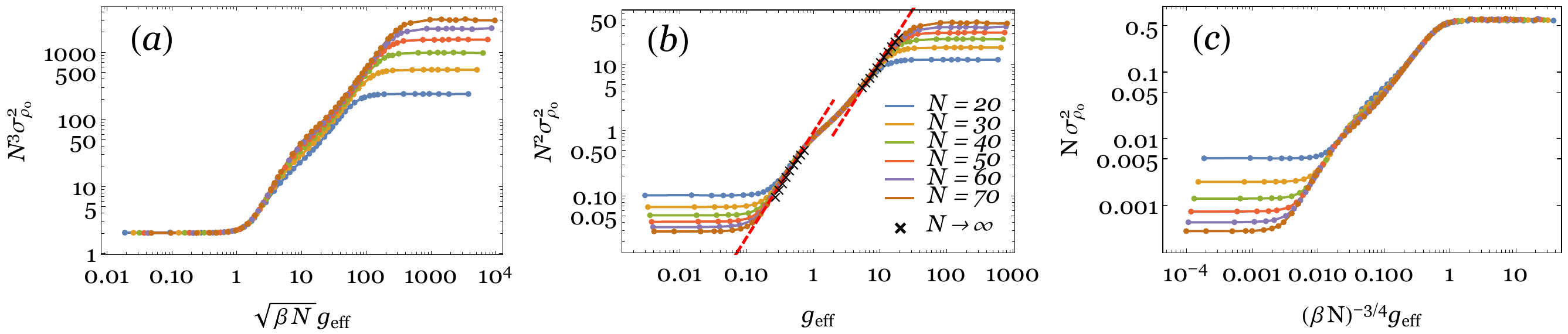}
    \caption{Variance of steady-state probabilities as a function of $g_{\mathrm{eff}}$, plotted for a single decay channel ($r=1$), different values of $N$ and $\beta=2$. The rescalings of $\sigma^2_{\rho_0}$ and of $g_\mathrm{eff}$ are those which collapse the different curves to a single curve in the $(a)$ P$''$ regime, $(b)$ C$''$ regime, and $(c)$ D$''$ regime. The dashed lines in $(b)$ are obtained by fitting extrapolated data for $N\to\infty$ (black crosses) to power-law behavior $g_\mathrm{eff}^{\lambda_{\rho_0}}$ near the boundaries of regime C$''$.}
    \label{fig:SM_r1_purity}
\end{figure}

Regarding the steady-state at large dissipation and $r=1$, in Fig.~\ref{fig:SM_r1_purity} we show the evolution of the variance of the steady-state eigenvalues as a function of $g_\mathrm{eff}$. For $r=1$, the condition  $\nu_{C''}=\nu_{D''}$ is no longer verified, instead $\nu_{C''}=-2$. Using the values of $\nu_{P''}$ and $\kappa_{\rho_0}^<$ obtained from Fig.~\ref{fig:SM_r1_purity}~$(a)$, and of $\nu_{D''}$ and $\kappa_{\rho_0}^>$ obtained from Fig.~\ref{fig:SM_r1_purity}~$(c)$,  Eq.~(\ref{eq:exponents_constraint}) gives the value $\lambda_{\rho_0}=8/5$. Note that in this case, both boundaries of regime C$''$ scale with $N$ and the scaling function does not have a single power-law behavior throughout the intermediate regime. 
However, the power-law $g_\mathrm{eff}^{\lambda_{\rho_0}}$ describes accurately the behavior of $\sigma_{\rho_0}^2$ both when entering regime C$''$ from regime P$''$ and when entering regime C$''$ from regime D$''$, see the dashed lines in Fig.~\ref{fig:SM_r1_purity} $(b)$. The two power-laws are shifted with respect to each other and are linked by a crossover regime in the center of regime C$''$.

Furthermore, we have $\sigma^2_{\rho_0}\propto1/N$ for strong dissipation, which gives a purity $\mathcal{P}_0\propto N^0$ in the large-$N$ limit. Thus, contrarily to the $r>1$ case, the steady-state's purity is not proportional to that of the maximally mixed state but remains finite for large $N$. 

\section{Summary of exponents}
\label{section:exponents}

We wrap up this chapter by summarizing the various exponents encountered in the study of random Liouvillians, see Table~\ref{tab:exponents}.

\begin{table*}[htbp]
\begin{center}
\caption{\label{tab:exponents}Scaling exponents for $r=1$ and $r>1$, as defined in the main text. $\nu$ exponents give the finite-size scaling of the spectral and steady-state quantities within one of the regimes P, C, D and their primed counterparts; $\lambda$ exponents give the extra $g_\mathrm{eff}$-scaling besides $g_\mathrm {eff}^2$ in regime C and its primed counterparts; $\kappa$ exponents give the finite-size scaling of the boundaries of the regimes.}
\vspace{+1em}
\begin{tabular}{r|rrrrrr|rrrrrr}
$r$ &
$\nu_P$ &
$\nu_C$ &
$\nu_{D}$ &
$\lambda_X$ &
$\kappa_X^<$ &
$\kappa_X^>$ &
$\nu_{P'}$ &
$\nu_{C'}$ &
$\nu_{D'}$ &
$\lambda_{\Delta}$ &
$\kappa_{\Delta}^<$ &
$\kappa_{\Delta}^>$ \\
\hline\hline
$=1$  & $0$ & $1/2$ & $1/2$ & $2$ & $-1/4$ & $0$ & $1/2$ & $1/2$ & $-3/2$ & $-8/3$ & 0 & $3/4$\\
$>1$  & $0$ & $1/2$ & $1/2$ & $2$ & $-1/4$ & $0$ & $1/2$ & $1/2$ & $1/2$  & ---    & $0$ & $0$\\
\end{tabular}
\vspace{+1em}
\begin{tabular}{r|rrrrrr}
$r$ &
$\nu_{P''}$ &
$\nu_{C''}$ &
$\nu_{D''}$ &
$\lambda_{\rho_0}$ &
$\kappa_{\rho_0}^<$ &
$\kappa_{\rho_0}^>$\\
\hline\hline
$=1$  & $-3$ & $-2$ & $-1$ & $8/5$ & $-1/2$ & $3/4$\\
$>1$  & $-3$ & $-2$ & $-2$ & $2$   & $-1/2$ & $0$   \\
\end{tabular}
\end{center}
\end{table*}

\vfill 
\cleardoublepage


\chapter{Analytic results for the spectral gap}
\label{chapter:analytics}

Having characterized random Liouvillians numerically in the preceding chapter, one may naturally ask which (if any) properties can be determined analytically. Regarding global spectral properties, a quantity of major interest would be to obtain the spectral density inside the support of the Liouvillian. This remains, however, an open question. In Ref.~\cite{denisov2018} the boundary of the spectral support was obtained using free probability for $H=0$. In contrast, in this chapter, we focus on the long-time asymptotics, namely on the spectral gap.

We start, in Section~\ref{section:holomorphic_G}, by employing holomorphic Green's function methods to obtain the spectral gap at infinite dissipation strength. We then generalize that procedure for arbitrary $g_\mathrm{eff}$ in Section~\ref{section:holomorphic_G_K}. Afterwards, in Section~\ref{section:degenerate_PT}, we note that, since the unitary evolution under the Hamiltonian $H$ can be solved exactly, we can apply degenerate perturbation theory in the small $g_\mathrm{eff}$ limit, obtaining the gap at weak dissipation.

\section{Holomorphic Green's functions I.\ Infinite dissipation}
\label{section:holomorphic_G}

As discussed in Chapter~\ref{chapter:state_of_the_art}, holomorphic Green's function methods do not allow one to study the spectral density of nonhermitian operators, as is the case of Liouvillian superoperators, since this method is only valid when the geometric series in $1/z$ is convergent. However, one can still use them to compute the spectral form factor (the trace of the propagator $\exp{t\mathcal{L}}$)~\cite{can2019}, from which one can then extract the spectral gap and the time evolution of ensemble-averaged autocorrelation functions. Remarkably the holomorphic Green's function itself contains the endpoints of the spectrum of $\mathcal{L}$ along the real direction in the complex plane (that is, the spectral gap and the left-most point of the lemon-shaped support, which is the gap plus some order-unity multiple of $X$). This indicates that the geometric series fails to converge at exactly the real endpoints. Unfortunately, the information along the imaginary axis (for instance $Y$) and the spectral density inside the support cannot be accessed by this method, and one needs to resort to the nonholomorphic (block) Green's function, as defined in Chapter~\ref{chapter:state_of_the_art}.

In this section, we will set up a diagrammatic expansion for the holomorphic Green's function (henceforth referred to simply as \emph{the} Green's function) for a Liouvillian superoperator (which carries four indices, in contrast to the two of an operator). From it, we can then extract $\Delta$ and $X$ as discussed above. During the preparation of this thesis, the novel diagrammatic expansion developed in Section~\ref{subsection:holomorphic_G_Liouvillian} was independently reported in Ref.~\cite{can2019} and applied to a wide range of Liouvillians.

\subsection{General remarks -- setup of the diagrammatic expansion}
We consider an ensemble of complex $M\times N$ random matrices $\phi_{ka}$ ($k=1,\dots,M$; $a=1,\dots, N$) with distribution function
\begin{equation}
    P(\phi)=\exp{-N\,\Tr V[\phi^\dagger\phi]}\,,
\end{equation}
where $V[\phi^\dagger\phi]$ is some polynomial in $\phi^\dagger\phi$. The spectral density $\varrho$ of some (possibly different) function $X[\phi^\dagger\phi]$ can be obtained from the one-point Green's function,
\begin{equation}
    G(z)=\frac{1}{N}\Tr\left\langle\frac{1}{z-X}\right\rangle=\frac{1}{N}\sum_{n=0}^\infty\frac{1}{z^{n+1}}\Tr\langle X^n\rangle.
\end{equation}
If the spectrum of $X$ is real, then, in the $N\to\infty$ limit, the spectral density and the Green's function are related by
\begin{equation}\label{eq:G_rho_relation_real}
    G(z)=\int\d\nu\,\varrho(\nu)\frac{1}{z-\nu}\Leftrightarrow \varrho(z)=\pm \frac{1}{\pi}\Im G(z\mp i\varepsilon)\,,
\end{equation}
where the limit $\varepsilon\to0^+$ is understood. If the spectrum of $X$ is complex, then, in the thermodynamic limit, the Green's function ceases to be an analytic function of $z$ and the method below fails, as discussed in Chapter~\ref{chapter:state_of_the_art}.

To relate the moments of $X$ and $\phi$, we introduce the $M\times N$ sources $J_{ka}$ and the generating functional
\begin{equation}
    Z[J]=\int \d \phi\, \exp{-N\,\Tr V[\phi^\dagger\phi]+\Tr[\phi^\dagger J]+\Tr[J^\dagger\phi]}\,,
\end{equation}
which can be differentiated to give all moments of $\phi$,
\begin{equation}
   \langle \phi^*_{k_1a_1}\cdots\phi^*_{k_pa_q}\phi^{}_{\ell_1b_1}\cdots\phi^{}_{\ell_qb_q}\rangle=\frac{\partial}{\partial J_{k_1a_1}}\cdots \frac{\partial}{\partial J_{k_pa_p}}\frac{\partial}{\partial J^*_{\ell_1b_1}}\cdots \frac{\partial}{\partial J^*_{b_q\ell_q}}Z[J]\,\biggr\rvert_{J=0}\,.
\end{equation}

We are interested in the case where the matrix $\phi$ is a jump operator $W$ (for the moment we consider a single dissipator, but we shall lift this condition later) and $X$ is the Liouvillian. Since we are in the strong (infinite) dissipation limit, we set $H=0$. Recall that the matrix representation of the superoperator is, according to Eq.~(\ref{eq:Liouvillian_tensor}),
\begin{equation}
    \mathcal{L}=W\otimes W^*-\frac{1}{2}\left(W^\dagger W\otimes\mathbbm{1}+\mathbbm{1}\otimes W^{\sf T}W^*\right).
\end{equation}
We choose the following index structure for a single Liouvillian, as well as for the product of Liouvillians: 
\begin{subequations}
\begin{align}
    &\mathcal{L}_{ijk\ell}=W_{ij}W^*_{k\ell}-\frac{1}{2}W^*_{mi}W_{mj}\delta_{k\ell}-\frac{1}{2}\delta_{ij}W_{mk}W^*_{m \ell}\label{eq:index_L}\\
    &(\mathcal{L}^n)_{ijk\ell}=(\mathcal{L}^{n-1})_{i\alpha k\beta}\,\mathcal{L}_{\alpha j \beta \ell}\,.\label{eq:index_L2}
\end{align}
\end{subequations}

If we draw the matrix $W$ from the Ginibre ensemble, then all moments $\langle\mathcal{L}^n\rangle$ will involve only sums of averages of the type $\langle W^{}_{ab}W^*_{cd}\rangle$ (with different index structures) which are moments of the Wishart ensemble. We will start by studying this ensemble, following Ref.~\cite{jurkiewicz2008}, obtaining its diagrammatic expansion and its Green's function, which will then be related to the Liouvillian case. 

\subsection{Green's function for the Wishart ensemble}
\label{subsection:diagrammatics_Wishart}
Following Ref.~\cite{jurkiewicz2008}, we take $V=X=\phi^\dagger \phi$ and fix $m\equiv M/N \geqslant1$ as $M,N\to \infty$. Since the probability distribution of $\phi$ is Gaussian, the generating functional can be computed exactly, $Z[J]=\exp{\frac{1}{N}\Tr J^\dagger J}=\exp{\frac{1}{N}J^*_{ka}J_{ka}}$. Also because of the Gaussian character of the weight, all moments of $\phi$ can be related to the propagator
\begin{equation}
    \langle\phi^*_{ka}\phi^{}_{\ell b}\rangle=\frac{1}{N}\delta_{k\ell}\delta_{ab}= 
    \begin{tikzpicture}[baseline=($0.4*(a)+0.6*(k)$)]
        \begin{feynman}[inline=(a)]
        \vertex[label=180:$a$]                      (a);
        \vertex[right=2cm of a,label=360:$b$]       (b);
        \vertex[below=0.5cm of a,label=180:$k$]     (k);
        \vertex[right=2cm of k,label=360:$\ell$]    (l);
        \diagram*{
            (a) -- (b) 
            (k) --[scalar] (l)
        };
        \end{feynman}
    \end{tikzpicture}
\end{equation}
by Wick contraction.

If we define $G(z)\equiv\frac{1}{N}\delta_{ab}G_{ab}(z)$, the expansion of the ``untraced'' Green's function $G_{ab}$ is 
\begin{equation}
\begin{split}
    G_{ab}(z)&=\sum_{n=0}^\infty \frac{\langle(X^n)\rangle_{ab}}{z^{n+1}}=\left\langle\frac{1}{z}\delta_{ab}+\frac{1}{z}X_{ab}\frac{1}{z}+\frac{1}{z}X_{ac}\frac{1}{z}X_{cb}\frac{1}{z}+\frac{1}{z}X_{ac}\frac{1}{z}X_{cd}\frac{1}{z}X_{db}\frac{1}{z}+\cdots\right\rangle\\
    &=\left\langle\frac{1}{z}\delta_{ab}+\frac{1}{z}\phi^*_{ka}\phi^{}_{kb}\frac{1}{z}+\frac{1}{z}\phi^*_{ka}\phi^{}_{kc}\frac{1}{z}\phi^*_{\ell c}\phi^{}_{\ell b}\frac{1}{z}+\frac{1}{z}\phi^*_{ka}\phi^{}_{kc}\frac{1}{z}\phi^*_{\ell c}\phi^{}_{\ell d}\frac{1}{z}\phi^*_{md}\phi^{}_{mb}\frac{1}{z}+\cdots\right\rangle\\
    &=\begin{tikzpicture}[baseline=(a)]
       \begin{feynman}[inline=(a)]
        \vertex[label=270:$a$]                        (a);
        \vertex[blob,scale=1.5,right=0.8cm of a]      (k) {};
        \vertex[right=1.1cm of k,label=270:$b$]       (b);
        \diagram*{(a) -- (k) -- (b)};
       \end{feynman}
       \end{tikzpicture}\,.
\end{split}
\end{equation}

We now evaluate this expansion in the large-$N$ limit. Each propagator $\langle\phi^*_{ka}\phi^{}_{\ell b}\rangle$ contributes with a factor $1/N$, while each dashed loop contributes with factor $\sim\delta_{kk}=M$ and a full-line loop with a factor $\sim\delta_{aa}=N$. The first four orders in the expansion of $G_{ab}$ give
\begingroup
\allowdisplaybreaks
\begin{subequations}\label{eq:Wishart_G_expansion}
\begin{align}
    G^{(0)}_{ab}=&
        \feynmandiagram [baseline=(a),horizontal=a to b]
        {a [label=270:$a$] --  b[label=270:$b$]};\\
    G^{(1)}_{ab}=&
    \hspace{-0.5cm}
        \begin{tikzpicture}[baseline=(a)]
        \begin{feynman}[inline=(a)]
        \vertex[label=270:$a$]                      (a);
        \vertex[right=0.6cm of a]                   (c);
        \vertex[dot,right=0.3cm of c,label=270:$k$] (k) {};
        \vertex[right=0.3cm of k]                   (d);
        \vertex[right=0.6cm of d,label=270:$b$]     (b);
        \diagram*{
            (a) -- (c) --[half left, out=135, in=45, loop, min distance=2cm] (d) -- (b)
        };
        \end{feynman}
        \draw [/tikzfeynman/scalar] (k) to[out=135, in=45, loop, min distance=1.5cm] (k);
        \end{tikzpicture}
        \\
    G^{(2)}_{ab}=&
    \hspace{-0.5cm}
        \begin{tikzpicture}[baseline=(a)]
        \begin{feynman}[inline=(a)]
        \vertex[label=270:$a$]                         (a);
        \vertex[right=0.6cm of a]                      (c);
        \vertex[dot,right=0.3cm of c,label=270:$k$]    (k) {};
        \vertex[right=0.3cm of k]                      (d);
        \vertex[right=1cm of d]                        (e);
        \vertex[dot,right=0.3cm of e,label=270:$\ell$] (l) {};
        \vertex[right=0.3cm of l]                      (f);
        \vertex[right=0.6cm of f,label=270:$b$]        (b);
        \diagram*{
            (a) -- (c) --[half left, out=135, in=45, loop, min distance=2cm] (d) -- (e) --[half left, out=135, in=45, loop, min distance=2cm] (f) -- (b)
        };
        \end{feynman}
        \draw [/tikzfeynman/scalar] (k) to[out=135, in=45, loop, min distance=1.5cm] (k);
        \draw [/tikzfeynman/scalar] (l) to[out=135, in=45, loop, min distance=1.5cm] (l);
    \end{tikzpicture}
    \hspace{-0.5cm}
    +
    \hspace{-0.5cm}
    \begin{tikzpicture}[baseline=(a)]
        \begin{feynman}[inline=(a)]
        \vertex[label=270:$a$]                         (a);
        \vertex[right=0.6cm of a]                      (c);
        \vertex[dot,right=0.3cm of c,label=270:$k$]    (k) {};
        \vertex[right=0.3cm of k]                      (d);
        \vertex[right=0.5cm of d]                      (e);
        \vertex[dot,right=0.3cm of e,label=270:$\ell$] (l) {};
        \vertex[right=0.3cm of l]                      (f);
        \vertex[right=0.6cm of f,label=270:$b$]        (b);
        \diagram*{
            (a) -- (c) --[half left, out=135, in=45, loop, min distance=2cm] (f) -- (b)
            (d) -- (e)
        };
        \end{feynman}
        \draw [/tikzfeynman/scalar] (k) to[out=135, in=45, loop, min distance=1.5cm] (l);
        \draw [/tikzfeynman/scalar] (k) to[out=110, in=70, loop, min distance=0.8cm] (l);
        \draw (d) to[out=110, in=70, loop, min distance=0.5cm] (e);
    \end{tikzpicture}
    \\
    \begin{split}
    G^{(3)}_{ab}=&
    \hspace{-0.5cm}
        \begin{tikzpicture}[baseline=(a)]
        \begin{feynman}[inline=(a)]
        \vertex[label=270:$a$]                         (a);
        \vertex[right=0.6cm of a]                      (c);
        \vertex[dot,right=0.3cm of c,label=270:$k$]    (k) {};
        \vertex[right=0.3cm of k]                      (d);
        \vertex[right=1cm of d]                        (e);
        \vertex[dot,right=0.3cm of e,label=270:$\ell$] (l) {};
        \vertex[right=0.3cm of l]                      (f);
        \vertex[right=1cm of f]                        (g);
        \vertex[dot,right=0.3cm of g,label=270:$m$]    (m) {};
        \vertex[right=0.3cm of m]                      (h);
        \vertex[right=0.6cm of h,label=270:$b$]        (b);
        \diagram*{
            (a) -- (c) --[half left, out=135, in=45, loop, min distance=2cm] (d) -- (e) --[half left, out=135, in=45, loop, min distance=2cm] (f) -- (g) --[half left, out=135, in=45, loop, min distance=2cm] (h) -- (b)
        };
        \end{feynman}
        \draw [/tikzfeynman/scalar] (k) to[out=135, in=45, loop, min distance=1.5cm] (k);
        \draw [/tikzfeynman/scalar] (l) to[out=135, in=45, loop, min distance=1.5cm] (l);
        \draw [/tikzfeynman/scalar] (m) to[out=135, in=45, loop, min distance=1.5cm] (m);
    \end{tikzpicture}
    \hspace{-0.5cm}
    +
    \hspace{-0.5cm}
    \begin{tikzpicture}[baseline=(a)]
        \begin{feynman}[inline=(a)]
        \vertex[label=270:$a$]                         (a);
        \vertex[right=0.6cm of a]                      (c);
        \vertex[dot,right=0.3cm of c,label=270:$k$]    (k) {};
        \vertex[right=0.3cm of k]                      (d);
        \vertex[right=0.5cm of d]                      (e);
        \vertex[dot,right=0.3cm of e,label=270:$\ell$] (l) {};
        \vertex[right=0.3cm of l]                      (f);
        \vertex[right=1cm of f]                        (g);
        \vertex[dot,right=0.3cm of g,label=270:$m$]    (m) {};
        \vertex[right=0.3cm of m]                      (h);
        \vertex[right=0.6cm of h,label=270:$b$]        (b);
        \diagram*{
            (a) -- (c) --[half left, out=135, in=45, loop, min distance=2cm] (f) -- (g) --[half left, out=135, in=45, loop, min distance=2cm] (h) -- (b)
            (d) -- (e)
        };
        \end{feynman}
        \draw [/tikzfeynman/scalar] (k) to[out=135, in=45, loop, min distance=1.5cm] (l);
        \draw [/tikzfeynman/scalar] (k) to[out=110, in=70, loop, min distance=0.8cm] (l);
        \draw [/tikzfeynman/scalar] (m) to[out=135, in=45, loop, min distance=1.5cm] (m);
        \draw (d) to[out=110, in=70, loop, min distance=0.5cm] (e);
    \end{tikzpicture}
    \\
    +&
    \hspace{-0.5cm}
        \begin{tikzpicture}[baseline=(a)]
        \begin{feynman}[inline=(a)]
        \vertex[label=270:$a$]                         (a);
        \vertex[right=0.6cm of a]                      (c);
        \vertex[dot,right=0.3cm of c,label=270:$k$]    (k) {};
        \vertex[right=0.3cm of k]                      (d);
        \vertex[right=1cm of d]                        (e);
        \vertex[dot,right=0.3cm of e,label=270:$\ell$] (l) {};
        \vertex[right=0.3cm of l]                      (f);
        \vertex[right=0.5cm of f]                      (g);
        \vertex[dot,right=0.3cm of g,label=270:$m$]    (m) {};
        \vertex[right=0.3cm of m]                      (h);
        \vertex[right=0.6cm of h,label=270:$b$]        (b);
        \diagram*{
            (a) -- (c) --[half left, out=135, in=45, loop, min distance=2cm] (d) -- (e) --[half left, out=135, in=45, loop, min distance=2cm] (h) -- (b)
            (f) -- (g)
        };
        \end{feynman}
        \draw [/tikzfeynman/scalar] (k) to[out=135, in=45, loop, min distance=1.5cm] (k);
        \draw [/tikzfeynman/scalar] (l) to[out=135, in=45, loop, min distance=1.5cm] (m);
        \draw [/tikzfeynman/scalar] (l) to[out=110, in=70, loop, min distance=0.8cm] (m);
        \draw (f) to[out=110, in=70, loop, min distance=0.5cm] (g);
        \end{tikzpicture}
    \hspace{-0.5cm}
    +
    \hspace{-1cm}
    \begin{tikzpicture}[baseline=(a)]
        \begin{feynman}[inline=(a)]
        \vertex[label=270:$a$]                         (a);
        \vertex[right=0.6cm of a]                      (c);
        \vertex[dot,right=0.3cm of c,label=270:$k$]    (k) {};
        \vertex[right=0.3cm of k]                      (d);
        \vertex[right=0.3cm of d]                      (e);
        \vertex[dot,right=0.3cm of e,label=270:$\ell$] (l) {};
        \vertex[right=0.3cm of l]                      (f);
        \vertex[right=0.3cm of f]                      (g);
        \vertex[dot,right=0.3cm of g,label=270:$m$]    (m) {};
        \vertex[right=0.3cm of m]                      (h);
        \vertex[right=0.6cm of h,label=270:$b$]        (b);
        \diagram*{
            (a) -- (c) --[half left, out=135, in=45, loop, min distance=3cm] (h) -- (b)
            (d) -- (e) --[half left, out=110, in=70, loop, min distance=1cm] (f) -- (g)
        };
        \end{feynman}
        \draw [/tikzfeynman/scalar] (l) to[out=135, in=45, loop, min distance=0.9cm] (l);
        \draw [/tikzfeynman/scalar] (k) to[out=135, in=45, loop, min distance=2.5cm] (m);
        \draw [/tikzfeynman/scalar] (k) to[out=110, in=70, loop, min distance=1.6cm] (m);
        \draw (d) to[out=120, in=60, loop, min distance=1.4cm] (g);
    \end{tikzpicture}
    \\
    +&
    \hspace{-0.8cm}
    \begin{tikzpicture}[baseline=(a)]
        \begin{feynman}[inline=(a)]
        \vertex[label=270:$a$]                         (a);
        \vertex[right=0.6cm of a]                      (c);
        \vertex[dot,right=0.3cm of c,label=270:$k$]    (k) {};
        \vertex[right=0.3cm of k]                      (d);
        \vertex[right=0.5cm of d]                      (e);
        \vertex[dot,right=0.3cm of e,label=270:$\ell$] (l) {};
        \vertex[right=0.3cm of l]                      (f);
        \vertex[right=0.5cm of f]                      (g);
        \vertex[dot,right=0.3cm of g,label=270:$m$]    (m) {};
        \vertex[right=0.3cm of m]                      (h);
        \vertex[right=0.6cm of h,label=270:$b$]        (b);
        \diagram*{
            (a) -- (c) --[half left, out=135, in=45, loop, min distance=2.5cm] (h) -- (b)
            (d) -- (e) 
            (f) -- (g)
        };
        \end{feynman}
        \draw [/tikzfeynman/scalar] (k) to[out=135, in=45, loop, min distance=2cm] (m);
        \draw [/tikzfeynman/scalar] (k) to[out=110, in=90, loop, min distance=1cm] (l);
        \draw [/tikzfeynman/scalar] (l) to[out=90, in=70, loop, min distance=1cm] (m);
        \draw (d) to[out=110, in=70, loop, min distance=0.5cm] (e);
        \draw (f) to[out=110, in=70, loop, min distance=0.5cm] (g);
    \end{tikzpicture}
    \hspace{-0.8cm}
    +
    \begin{tikzpicture}[baseline=(a)]
        \begin{feynman}[inline=(a)]
        \vertex[label=270:$a$]                         (a);
        \vertex[right=0.6cm of a]                      (c);
        \vertex[dot,right=0.3cm of c,label=270:$k$]    (k) {};
        \vertex[right=0.3cm of k]                      (d);
        \vertex[right=0.5cm of d]                      (e);
        \vertex[dot,right=0.3cm of e,label=270:$\ell$] (l) {};
        \vertex[right=0.3cm of l]                      (f);
        \vertex[right=0.5cm of f]                      (g);
        \vertex[dot,right=0.3cm of g,label=270:$m$]    (m) {};
        \vertex[right=0.3cm of m]                      (h);
        \vertex[right=0.6cm of h,label=270:$b$]        (b);
        \diagram*{
            (a) -- (c) --[half left, out=250, in=290, loop, min distance=1.5cm] (g) -- (f) -- [half left, out=110, in=70, loop, min distance=1cm] (e) -- (d) --[half left, out=250, in=290, loop, min distance=1.5cm] (h) -- (b)
        };
        \end{feynman}
        \draw [/tikzfeynman/scalar] (k) to[out=135, in=45, loop, min distance=2cm] (m);
        \draw [/tikzfeynman/scalar] (k) to[out=110, in=90, loop, min distance=1cm] (l);
        \draw [/tikzfeynman/scalar] (l) to[out=90, in=70, loop, min distance=1cm] (m);
    \end{tikzpicture}
    \end{split}
\end{align}
\end{subequations}
\endgroup

The zeroth, first and second-order diagrams all have the same number of double-line propagators and of loops, hence they are of $\mathcal{O}(1)$. More concretely, the zeroth-order diagram is $\sim 1$, the first order diagram is $\sim M/N=m$, the first of the two second-order diagrams is $\sim M^2/N^2=m^2$ and the second one is $\sim M\cdot N/N^2=m$. The third-order expansion is the first to have non-planar diagrams (i.e.\ diagrams with crossings of lines), which are suppressed by some power of $N$ (because they necessarily have more propagators than loops) and hence a vanishing contribution in the large-$N$ limit. In particular, the last one of those diagrams has only one (dashed) loop and hence is $\sim M/N^3=m/N^2$.

The self-energy $\Sigma(z)\equiv\frac{1}{N}\delta_{ab}\Sigma_{ab}(z)$ is defined as the sum of all one-particle irreducible (1PI) diagrams, such that
\begin{equation}\label{eq:Wishart_G_Sigma}
\begin{split}
    G(z)=\frac{1}{z}+\frac{1}{z}\Sigma(z)\frac{1}{z}+\frac{1}{z}\Sigma(z)\frac{1}{z}\Sigma(z)\frac{1}{z}+\cdots=\frac{1}{z-\Sigma(z)}\,.
\end{split}
\end{equation}
Collecting all planar 1PI diagrams from the expansion of Eq.~(\ref{eq:Wishart_G_expansion}), we obtain the the large-$N$ expansion of the self-energy,
\begingroup
\allowdisplaybreaks
\begin{align*}
    \Sigma_{ab}&=
    \hspace{-1cm}
        \begin{tikzpicture}[baseline=($0.2*(current bounding box.north)+0.8*(current bounding box.south)$)]
        \begin{feynman}[inline=(a)]
        \vertex                       (a);
        \vertex[dot,right=0.3cm of a] (k) {};
        \vertex[right=0.3cm of k]     (b);
        \diagram*{
            (a) --[half left, out=135, in=45, loop, min distance=2cm] (b)
        };
        \end{feynman}
        \draw [/tikzfeynman/scalar] (k) to[out=135, in=45, loop, min distance=1.5cm] (k);
        \end{tikzpicture}
    \hspace{-1cm}
    +
    \hspace{-1cm}
        \begin{tikzpicture}[baseline=($0.2*(current bounding box.north)+0.8*(current bounding box.south)$)]
        \begin{feynman}[inline=(a)]
        \vertex                          (c);
        \vertex[dot,right=0.3cm of c]    (k) {};
        \vertex[right=0.3cm of k]        (d);
        \vertex[right=0.5cm of d]        (e);
        \vertex[dot,right=0.3cm of e]    (l) {};
        \vertex[right=0.3cm of l]        (f);
        \diagram*{
            (c) --[half left, out=135, in=45, loop, min distance=2cm] (f)
            (d) -- (e)
        };
        \end{feynman}
        \draw [/tikzfeynman/scalar] (k) to[out=135, in=45, loop, min distance=1.5cm] (l);
        \draw [/tikzfeynman/scalar] (k) to[out=110, in=70, loop, min distance=0.8cm] (l);
        \draw (d) to[out=110, in=70, loop, min distance=0.5cm] (e);
        \end{tikzpicture}
    \hspace{-1cm}
    +
    \Bigg(
    \hspace{-1.5cm}
        \begin{tikzpicture}[baseline=($0.25*(current bounding box.north)+0.75*(current bounding box.south)$)]
        \begin{feynman}[inline=(a)]
        \vertex[right=0.6cm of a]        (c);
        \vertex[dot,right=0.3cm of c]    (k) {};
        \vertex[right=0.3cm of k]        (d);
        \vertex[right=0.3cm of d]        (e);
        \vertex[dot,right=0.3cm of e]    (l) {};
        \vertex[right=0.3cm of l]        (f);
        \vertex[right=0.3cm of f]        (g);
        \vertex[dot,right=0.3cm of g]    (m) {};
        \vertex[right=0.3cm of m]        (h);
        \diagram*{
            (c) --[half left, out=135, in=45, loop, min distance=3cm] (h)
            (d) -- (e) --[half left, out=110, in=70, loop, min distance=1cm] (f) -- (g)
        };
        \end{feynman}
        \draw [/tikzfeynman/scalar] (l) to[out=135, in=45, loop, min distance=0.9cm] (l);
        \draw [/tikzfeynman/scalar] (k) to[out=135, in=45, loop, min distance=2.5cm] (m);
        \draw [/tikzfeynman/scalar] (k) to[out=110, in=70, loop, min distance=1.6cm] (m);
        \draw (d) to[out=120, in=60, loop, min distance=1.4cm] (g);
        \end{tikzpicture}
    \hspace{-1.5cm}
    +
    \hspace{-1.2cm}
        \begin{tikzpicture}[baseline=($0.25*(current bounding box.north)+0.75*(current bounding box.south)$)]
        \begin{feynman}[inline=(a)]
        \vertex                         (c);
        \vertex[dot,right=0.3cm of c]   (k) {};
        \vertex[right=0.3cm of k]       (d);
        \vertex[right=0.5cm of d]       (e);
        \vertex[dot,right=0.3cm of e]   (l) {};
        \vertex[right=0.3cm of l]       (f);
        \vertex[right=0.5cm of f]       (g);
        \vertex[dot,right=0.3cm of g]   (m) {};
        \vertex[right=0.3cm of m]       (h);
        \diagram*{
            (c) --[half left, out=135, in=45, loop, min distance=2.5cm] (h)
            (d) -- (e) 
            (f) -- (g)
        };
        \end{feynman}
        \draw [/tikzfeynman/scalar] (k) to[out=135, in=45, loop, min distance=2cm] (m);
        \draw [/tikzfeynman/scalar] (k) to[out=110, in=90, loop, min distance=1cm] (l);
        \draw [/tikzfeynman/scalar] (l) to[out=90, in=70, loop, min distance=1cm] (m);
        \draw (d) to[out=110, in=70, loop, min distance=0.5cm] (e);
        \draw (f) to[out=110, in=70, loop, min distance=0.5cm] (g);
        \end{tikzpicture}
    \hspace{-1.2cm}
    \Bigg)\\
    &+\Bigg(
    \hspace{-1.7cm}
        \begin{tikzpicture}[baseline=($0.25*(current bounding box.north)+0.75*(current bounding box.south)$)]
        \begin{feynman}[inline=(a)]
        \vertex[right=0.6cm of a]        (c);
        \vertex[dot,right=0.3cm of c]    (k) {};
        \vertex[right=0.3cm of k]        (d);
        \vertex[right=0.3cm of d]        (e);
        \vertex[dot,right=0.3cm of e]    (l) {};
        \vertex[right=0.3cm of l]        (f);
        \vertex[right=0.3cm of f]        (g);
        \vertex[dot,right=0.3cm of g]    (m) {};
        \vertex[right=0.3cm of m]        (h);
        \vertex[right=0.3cm of h]        (i);
        \vertex[dot,right=0.3cm of i]    (n) {};
        \vertex[right=0.3cm of n]        (j);
        \diagram*{
            (c) --[half left, out=135, in=45, loop, min distance=3cm] (j)
            (d) -- (e)
            (f) -- (g) --[half left, out=110, in=70, loop, min distance=1cm] (h) -- (i)
        };
        \end{feynman}
        \draw [/tikzfeynman/scalar] (m) to[out=135, in=45, loop, min distance=0.9cm] (m);
        \draw [/tikzfeynman/scalar] (k) to[out=135, in=45, loop, min distance=2.5cm] (n);
        \draw [/tikzfeynman/scalar] (k) to[out=110, in=90, loop, min distance=1cm] (l);
        \draw [/tikzfeynman/scalar] (l) to[out=100, in=70, loop, min distance=1.5cm] (n);
        \draw (d) to[out=120, in=60, loop, min distance=0.5cm] (e);
        \draw (f) to[out=120, in=60, loop, min distance=1.4cm] (i);
        \end{tikzpicture}
    \hspace{-1.7cm}
    +
    \hspace{-1.7cm}
        \begin{tikzpicture}[baseline=($0.25*(current bounding box.north)+0.75*(current bounding box.south)$)]
        \begin{feynman}[inline=(a)]
        \vertex[right=0.6cm of a]        (c);
        \vertex[dot,right=0.3cm of c]    (k) {};
        \vertex[right=0.3cm of k]        (d);
        \vertex[right=0.3cm of d]        (e);
        \vertex[dot,right=0.3cm of e]    (l) {};
        \vertex[right=0.3cm of l]        (f);
        \vertex[right=0.3cm of f]        (g);
        \vertex[dot,right=0.3cm of g]    (m) {};
        \vertex[right=0.3cm of m]        (h);
        \vertex[right=0.3cm of h]        (i);
        \vertex[dot,right=0.3cm of i]    (n) {};
        \vertex[right=0.3cm of n]        (j);
        \diagram*{
            (c) --[half left, out=135, in=45, loop, min distance=3cm] (j)
            (h) -- (i)
            (d) -- (e) --[half left, out=110, in=70, loop, min distance=1cm] (f) -- (g)
        };
        \end{feynman}
        \draw [/tikzfeynman/scalar] (l) to[out=135, in=45, loop, min distance=0.9cm] (l);
        \draw [/tikzfeynman/scalar] (k) to[out=135, in=45, loop, min distance=2.5cm] (n);
        \draw [/tikzfeynman/scalar] (k) to[out=110, in=80, loop, min distance=1.5cm] (m);
        \draw [/tikzfeynman/scalar] (m) to[out=90, in=70, loop, min distance=1cm] (n);
        \draw (h) to[out=120, in=60, loop, min distance=0.5cm] (i);
        \draw (d) to[out=120, in=60, loop, min distance=1.4cm] (g);
        \end{tikzpicture}
    \hspace{-1.7cm}
    +
    \hspace{-1.4cm}
    \begin{tikzpicture}[baseline=($0.25*(current bounding box.north)+0.75*(current bounding box.south)$)]
        \begin{feynman}[inline=(a)]
        \vertex                         (c);
        \vertex[dot,right=0.3cm of c]   (k) {};
        \vertex[right=0.3cm of k]       (d);
        \vertex[right=0.5cm of d]       (e);
        \vertex[dot,right=0.3cm of e]   (l) {};
        \vertex[right=0.3cm of l]       (f);
        \vertex[right=0.5cm of f]       (g);
        \vertex[dot,right=0.3cm of g]   (m) {};
        \vertex[right=0.3cm of m]       (h);
        \vertex[right=0.5cm of h]       (i);
        \vertex[dot,right=0.3cm of i]   (n) {};
        \vertex[right=0.3cm of n]       (j);
        \diagram*{
            (c) --[half left, out=135, in=45, loop, min distance=2.5cm] (j)
            (d) -- (e) 
            (f) -- (g)
            (h) -- (i)
        };
        \end{feynman}
        \draw [/tikzfeynman/scalar] (k) to[out=135, in=45, loop, min distance=2cm] (n);
        \draw [/tikzfeynman/scalar] (k) to[out=90, in=90, loop, min distance=1cm] (l);
        \draw [/tikzfeynman/scalar] (l) to[out=90, in=90, loop, min distance=1cm] (m);
        \draw [/tikzfeynman/scalar] (m) to[out=90, in=90, loop, min distance=1cm] (n);
        \draw (d) to[out=110, in=70, loop, min distance=0.5cm] (e);
        \draw (f) to[out=110, in=70, loop, min distance=0.5cm] (g);
        \draw (h) to[out=110, in=70, loop, min distance=0.5cm] (i);
        \end{tikzpicture}
    \hspace{-1.4cm}
    \\
    &+
    \hspace{-1.7cm}
    \begin{tikzpicture}[baseline=($0.25*(current bounding box.north)+0.75*(current bounding box.south)$)]
        \begin{feynman}[inline=(a)]
        \vertex                         (c);
        \vertex[dot,right=0.3cm of c]   (k) {};
        \vertex[right=0.3cm of k]       (d);
        \vertex[right=0.5cm of d]       (e);
        \vertex[dot,right=0.3cm of e]   (l) {};
        \vertex[right=0.3cm of l]       (f);
        \vertex[right=0.5cm of f]       (g);
        \vertex[dot,right=0.3cm of g]   (m) {};
        \vertex[right=0.3cm of m]       (h);
        \vertex[right=0.5cm of h]       (i);
        \vertex[dot,right=0.3cm of i]   (n) {};
        \vertex[right=0.3cm of n]       (j);
        \diagram*{
            (c) --[half left, out=135, in=45, loop, min distance=3cm] (j)
            (d) -- (e) --[half left, out=135, in=45, loop, min distance=1.2cm] (f) -- (g) --[half left, out=135, in=45, loop, min distance=1.2cm] (h) -- (i)
        };
        \end{feynman}
        \draw [/tikzfeynman/scalar] (k) to[out=135, in=45, loop, min distance=2.5cm] (n);
        \draw [/tikzfeynman/scalar] (k) to[out=110, in=70, loop, min distance=1.6cm] (n);
        \draw [/tikzfeynman/scalar] (l) to[out=135, in=45, loop, min distance=0.8cm] (l);
        \draw [/tikzfeynman/scalar] (m) to[out=135, in=45, loop, min distance=0.8cm] (m);
        \draw (d) to[out=110, in=70, loop, min distance=1.3cm] (i);
        \end{tikzpicture}
    \hspace{-1.7cm}
    +
    \hspace{-1.7cm}
        \begin{tikzpicture}[baseline=($0.25*(current bounding box.north)+0.75*(current bounding box.south)$)]
        \begin{feynman}[inline=(a)]
        \vertex                         (c);
        \vertex[dot,right=0.3cm of c]   (k) {};
        \vertex[right=0.3cm of k]       (d);
        \vertex[right=0.5cm of d]       (e);
        \vertex[dot,right=0.3cm of e]   (l) {};
        \vertex[right=0.3cm of l]       (f);
        \vertex[right=0.5cm of f]       (g);
        \vertex[dot,right=0.3cm of g]   (m) {};
        \vertex[right=0.3cm of m]       (h);
        \vertex[right=0.5cm of h]       (i);
        \vertex[dot,right=0.3cm of i]   (n) {};
        \vertex[right=0.3cm of n]       (j);
        \diagram*{
            (c) --[half left, out=135, in=45, loop, min distance=3.2cm] (j)
            (d) -- (e) --[half left, out=135, in=45, loop, min distance=1.5cm] (h) -- (i)
            (f) -- (g)
        };
        \end{feynman}
        \draw [/tikzfeynman/scalar] (k) to[out=135, in=45, loop, min distance=2.7cm] (n);
        \draw [/tikzfeynman/scalar] (k) to[out=110, in=70, loop, min distance=1.8cm] (n);
        \draw [/tikzfeynman/scalar] (l) to[out=135, in=45, loop, min distance=1.2cm] (m);
        \draw [/tikzfeynman/scalar] (l) to[out=90, in=90, loop, min distance=0.7cm] (m);
        \draw (d) to[out=110, in=70, loop, min distance=1.5cm] (i);
        \draw (f) to[out=110, in=70, loop, min distance=0.5cm] (g);
        \end{tikzpicture}
    \hspace{-1.7cm}
    \Bigg)+\cdots
    \stepcounter{equation}\tag{\theequation}\label{eq:Wishart_Sigma}\\
    &=\hspace{-1cm}
        \begin{tikzpicture}[baseline=($0.25*(current bounding box.north)+0.75*(current bounding box.south)$)]
        \begin{feynman}[inline=(a)]
        \vertex                        (a);
        \vertex[dot,right=0.3cm of a]  (k) {};
        \vertex[right=0.3cm of k]      (b);
        \diagram*{
            (a) --[half left, out=135, in=45, loop, min distance=2cm] (b)
        };
        \end{feynman}
        \draw [/tikzfeynman/scalar] (k) to[out=135, in=45, loop, min distance=1.5cm] (k);
        \end{tikzpicture}
    \hspace{-1cm}
    +
    \hspace{-1cm}
        \begin{tikzpicture}[baseline=($0.25*(current bounding box.north)+0.75*(current bounding box.south)$)]
        \begin{feynman}[inline=(a)]
        \vertex                          (c);
        \vertex[dot,right=0.3cm of c]    (k) {};
        \vertex[right=0.3cm of k]        (d);
        \vertex[blob,right=0.3cm of d]   (z) {};
        \vertex[right=0.5cm of z]        (e);
        \vertex[dot,right=0.3cm of e]    (l) {};
        \vertex[right=0.3cm of l]        (f);
        \diagram*{
            (c) --[half left, out=135, in=45, loop, min distance=2cm] (f)
            (d) -- (z)
            (z) -- (e)
        };
        \end{feynman}
        \draw [/tikzfeynman/scalar] (k) to[out=135, in=45, loop, min distance=1.6cm] (l);
        \draw [/tikzfeynman/scalar] (k) to[out=90, in=90, loop, min distance=0.9cm] (l);
        \draw (d) to[out=90, in=90, loop, min distance=0.7cm] (e);
        \end{tikzpicture}
    \hspace{-1cm}
    +
    \hspace{-1.2cm}
        \begin{tikzpicture}[baseline=($0.25*(current bounding box.north)+0.75*(current bounding box.south)$)]
        \begin{feynman}[inline=(a)]
        \vertex                         (c);
        \vertex[dot,right=0.3cm of c]   (k) {};
        \vertex[right=0.3cm of k]       (d);
        \vertex[blob,right=0.3cm of d]  (z) {};
        \vertex[right=0.5cm of z]       (e);
        \vertex[dot,right=0.3cm of e]   (l) {};
        \vertex[right=0.3cm of l]       (f);
        \vertex[blob,right=0.3cm of f]  (y) {};
        \vertex[right=0.5cm of y]       (g);
        \vertex[dot,right=0.3cm of g]   (m) {};
        \vertex[right=0.3cm of m]       (h);
        \diagram*{
            (c) --[half left, out=135, in=45, loop, min distance=2.5cm] (h)
            (d) -- (z)
            (z) -- (e) 
            (f) -- (y)
            (y) -- (g)
        };
        \end{feynman}
        \draw [/tikzfeynman/scalar] (k) to[out=135, in=45, loop, min distance=2cm] (m);
        \draw [/tikzfeynman/scalar] (k) to[out=90, in=90, loop, min distance=1cm] (l);
        \draw [/tikzfeynman/scalar] (l) to[out=90, in=90, loop, min distance=1cm] (m);
        \draw (d) to[out=90, in=90, loop, min distance=0.8cm] (e);
        \draw (f) to[out=90, in=90, loop, min distance=0.8cm] (g);
        \end{tikzpicture}
    \hspace{-1.2cm}
    +\cdots\\
    &\equiv
    \hspace{-1cm}
        \begin{tikzpicture}[baseline=($0.47*(current bounding box.north)+0.53*(current bounding box.south)$)]
        \begin{feynman}[inline=(a)]
        \vertex[label=270:$a$]                         (a);
        \vertex[dot,right=0.3cm of a,label=270:$k$]    (k) {};
        \vertex[dot,right=1cm of k,label=270:$\ell$] (l) {};
        \vertex[right=0.3cm of l,label=270:$b$]        (b);
        \diagram*{
            (a) --[half left, out=135, in=45, loop, min distance=2cm] (b)
            (k) --[photon] (l)
        };
        \end{feynman}
        \draw [/tikzfeynman/scalar] (k) to[out=135, in=45, loop, min distance=1.5cm] (l);
        \end{tikzpicture}
    \hspace{-1cm}
    =\frac{1}{N}\delta_{ab}\delta_{k\ell}F_{k\ell}\equiv\frac{M}{N}\delta_{ab}F\,.
\end{align*}
\endgroup
In Eq.~(\ref{eq:Wishart_Sigma}), in the first equality we have collected all 1PI diagrams with amputated external legs at each order of the expansion of the Green's function; in the second equality we noted that each diagram was composed of an overarching double propagator with insertions of the expansion terms of the Green's function of Eq.~(\ref{eq:Wishart_G_expansion}), with the diagrams reordered with respect to the number of those insertions; in the final three equalities we define the function $F$ which is the sum of all those reordered diagrams excluding the overarching double propagator and is represented by the wiggly line. From the last expression of Eq.~(\ref{eq:Wishart_Sigma}), it follows that $\Sigma(z)=mF(z)$. 

The function $F$ is evaluated as follows:
\begin{equation}\label{eq:Wishart_F}
\begin{split}
    F_{k\ell}&=
   \begin{tikzpicture}[baseline=(k)]
        \begin{feynman}[inline=(a)]
        \vertex[dot,label=270:$k$]                        (k) {};
        \vertex[dot,right=2cm of k,label=270:$\ell$]    (l) {};
        \diagram*{ (k) --[photon] (l) };
        \end{feynman}
    \end{tikzpicture}
    =\\
    &=
    \begin{tikzpicture}[baseline=(k)]
        \begin{feynman}[inline=(a)]
        \vertex[dot,label=270:$k$]                        (k) {};
        \vertex[dot,right=2cm of k,label=270:$\ell$]    (l) {};
        \diagram*{ (k) --[scalar] (l) };
        \end{feynman}
    \end{tikzpicture}
    +
    \begin{tikzpicture}[baseline=(k)]
        \begin{feynman}[inline=(a)]
        \vertex[dot,label=270:$k$]                        (k) {};
        \vertex[right=0.4cm of k]                         (d);
        \vertex[blob,right=0.3cm of d]                    (z) {};
        \vertex[right=0.5cm of z]                         (e);
        \vertex[dot,right=0.4cm of e,label=270:$\ell$]    (l) {};
        \diagram*{
            (d) -- (z)
            (z) -- (e)
        };
        \end{feynman}
        \draw [/tikzfeynman/scalar] (k) to[out=90, in=90, loop, min distance=1.2cm] (l);
        \draw (d) to[out=90, in=90, loop, min distance=0.8cm] (e);
    \end{tikzpicture}
    +
        \begin{tikzpicture}[baseline=(k)]
        \begin{feynman}[inline=(a)]
        \vertex[dot,label=270:$k$]                       (k) {};
        \vertex[right=0.4cm of k]                        (d);
        \vertex[blob,right=0.3cm of d]                   (z) {};
        \vertex[right=0.5cm of z]                        (e);
        \vertex[dot,right=0.4cm of e,label=270:$\ell$]   (l) {};
        \vertex[right=0.4cm of l]                        (f);
        \vertex[blob,right=0.3cm of f]                   (y) {};
        \vertex[right=0.5cm of y]                        (g);
        \vertex[dot,right=0.3cm of g,label=270:$m$]      (m) {};
        \diagram*{
            (d) -- (z)
            (z) -- (e) 
            (f) -- (y)
            (y) -- (g)
        };
        \end{feynman}
        \draw [/tikzfeynman/scalar] (k) to[out=90, in=90, loop, min distance=1.2cm] (l);
        \draw [/tikzfeynman/scalar] (l) to[out=90, in=90, loop, min distance=1.2cm] (m);
        \draw (d) to[out=90, in=90, loop, min distance=0.8cm] (e);
        \draw (f) to[out=90, in=90, loop, min distance=0.8cm] (g);
    \end{tikzpicture}
    +\cdots\\
    &=
    \begin{tikzpicture}[baseline=(k)]
        \begin{feynman}[inline=(a)]
        \vertex[dot,label=270:$k$]                        (k) {};
        \vertex[dot,right=2cm of k,label=270:$\ell$]      (l) {};
        \diagram*{ (k) --[scalar] (l) };
        \end{feynman}
    \end{tikzpicture}
    +
    \begin{tikzpicture}[baseline=(k)]
        \begin{feynman}[inline=(a)]
        \vertex[dot,label=270:$k$]                        (k) {};
        \vertex[right=0.4cm of k]                         (d);
        \vertex[blob,right=0.3cm of d]                    (z) {};
        \vertex[right=0.5cm of z]                         (e);
        \vertex[dot,right=0.4cm of e,label=270:$m$]       (l) {};
        \diagram*{
            (d) -- (z)
            (z) -- (e)
        };
        \end{feynman}
        \draw [/tikzfeynman/scalar] (k) to[out=90, in=90, loop, min distance=1.2cm] (l);
        \draw (d) to[out=90, in=90, loop, min distance=0.8cm] (e);
    \end{tikzpicture}
    \Bigg(
    \begin{tikzpicture}[baseline=(k)]
        \begin{feynman}[inline=(a)]
        \vertex[dot,label=270:$m$]                        (k) {};
        \vertex[dot,right=2cm of k,label=270:$\ell$]      (l) {};
        \diagram*{ (k) --[scalar] (l) };
        \end{feynman}
    \end{tikzpicture}
    +
    \begin{tikzpicture}[baseline=(k)]
        \begin{feynman}[inline=(a)]
        \vertex[dot,label=270:$m$]                        (k) {};
        \vertex[right=0.4cm of k]                         (d);
        \vertex[blob,right=0.3cm of d]                    (z) {};
        \vertex[right=0.5cm of z]                         (e);
        \vertex[dot,right=0.4cm of e,label=270:$\ell$]    (l) {};
        \diagram*{
            (d) -- (z)
            (z) -- (e)
        };
        \end{feynman}
        \draw [/tikzfeynman/scalar] (k) to[out=90, in=90, loop, min distance=1.2cm] (l);
        \draw (d) to[out=90, in=90, loop, min distance=0.8cm] (e);
    \end{tikzpicture}
    +\cdots\Bigg)\\
    &=
    \begin{tikzpicture}[baseline=(k)]
        \begin{feynman}[inline=(a)]
        \vertex[dot,label=270:$k$]                        (k) {};
        \vertex[dot,right=2cm of k,label=270:$\ell$]      (l) {};
        \diagram*{ (k) --[scalar] (l) };
        \end{feynman}
    \end{tikzpicture}
    +
    \begin{tikzpicture}[baseline=(k)]
        \begin{feynman}[inline=(a)]
        \vertex[dot,label=270:$k$]                        (k) {};
        \vertex[right=0.4cm of k]                         (d);
        \vertex[blob,right=0.3cm of d]                    (z) {};
        \vertex[right=0.5cm of z]                         (e);
        \vertex[dot,right=0.4cm of e,label=270:$m$]       (l) {};
        \vertex[dot,right=2cm of l,label=270:$\ell$]      (m) {};
        \diagram*{
            (d) -- (z)
            (z) -- (e)
            (l) --[photon] (m)
        };
        \end{feynman}
        \draw [/tikzfeynman/scalar] (k) to[out=90, in=90, loop, min distance=1.2cm] (l);
        \draw (d) to[out=90, in=90, loop, min distance=0.8cm] (e);
    \end{tikzpicture}\\
    &=\delta_{k\ell}+\frac{1}{N}\,\delta_{ab}\,\delta_{km}\,G_{ab}\, F_{m\ell}\,.
\end{split}
\end{equation}
From Eq.~(\ref{eq:Wishart_F}), it then follows that $F(z)=1+G(z)F(z)$, which in combination with Eq.~(\ref{eq:Wishart_Sigma}), $\Sigma(z)=mF(z)$, gives the self-energy of the Wishart ensemble,
\begin{equation}\label{eq:self_energy_Wishart}
    \Sigma(z)=\frac{m}{1-G(z)}\,.
\end{equation}
Using now Eq.~(\ref{eq:Wishart_G_Sigma}), $G(z)=1/(z-\Sigma(z))$, we are led to the algebraic equation $zG^2+(m-1-z)G+1=0$, which is solved by the Wishart ensemble's Green's function,
\begin{equation}\label{eq:Wishart_Greens_function_expr}
    G(z)=\frac{1}{2z}\left(z-(m-1)-\sqrt{\left(z-(m-1)\right)^2-4z}\right).
\end{equation}
From the two possible signs of the square root, we have chosen the minus sign because of the large-$z$ behavior of $G(z)$, $G(z)\to1/z$ as $z\to\infty$. Using Eq.~(\ref{eq:G_rho_relation_real}) and setting $m=1$ ($W$ is a square matrix), we obtain the spectral density of the Wishart ensemble, the MP law:
\begin{equation}
    \varrho_{\mathrm{MP}}(x)=\frac{1}{2\pi}\frac{\sqrt{x(4-x)}}{x},\quad 0<x<4\,.
\end{equation}

To obtain the number of planar diagrams which contribute to the Green's function at a given order, we write its moment expansion as 
\begin{equation}\label{eq:Wishart_G_series}
    G(z)=\frac{1}{z}\sum_{n=0}^\infty \frac{C_n}{z^n}\,,
\end{equation}
where $C_n=\langle \Tr X^n\rangle/N$ precisely counts the number of Wishart planar diagrams at $n^\mathrm{th}$ order. Using the definition of the expansion coefficients, and since we know $G(z)$ explicitly from Eq.~(\ref{eq:Wishart_Greens_function_expr}), we have
\begin{equation}\label{eq:c_n_def}
    C_n=\frac{1}{n!}\partial_{1/z}^n zG(z)\Big\rvert_{z=\infty}=\frac{1}{2\pi i}\oint_\infty \d\!\left(\frac{1}{z}\right)\frac{zG(z)}{(1/z)^{n+1}}=-\frac{1}{2\pi i}\oint_\infty\d z\, z^nG(z)\,,
\end{equation}
where the contour of integration in $\mathbb{C}\cup\{\infty\}$ winds counterclockwise around $\infty$. The integrand $z^nG(z)$ has a branch cut at $[0,4]$ on the real line hence the integral can be taken on a contour clockwise around that branch cut. Parameterizing $z=x\pm iy$, $x\in[0,4]$, $y\to0^+$, the $C_n$ can be expressed as a real integral,
\begin{equation}\label{eq:c_n_expr}
    C_n=\frac{1}{2\pi}\int_0^4\d x\,x^{n-1}\sqrt{4x-x^2}=\frac{1}{n+1}\binom{2n}{n}\,,
\end{equation}
and are identified as the Catalan numbers~\cite{weisstein2002,oeis2019}. By inserting the relation between the Green's function and the spectral density, Eq.~(\ref{eq:G_rho_relation_real}), into Eq.~(\ref{eq:c_n_def}) we obtain $C_n$ as the $n^\mathrm{th}$ moment of the MP distribution~$\varrho_\mathrm{MP}$:
\begin{equation}\label{eq:c_n_moments_MP}
    C_n=-\frac{1}{2\pi i}\oint_\infty\d z \int\d\nu\, \varrho_\mathrm{MP}(\nu)\frac{z^n}{z-\nu}=+\int\d\nu\,\varrho_\mathrm{MP}(\nu)\underset{z=\nu}{\Res}\frac{z^n}{z-\nu}=\int \d\nu\,\nu^n\varrho_\mathrm{MP}(\nu)\,.
\end{equation}

\subsection{Holomorphic Green's function for the Liouvillian}
\label{subsection:holomorphic_G_Liouvillian}
As noted before, all terms in the moment expansion of the Liouvillian Green's function can be reduced, using Wick's theorem, to products of double propagators of the Wishart ensemble. The ``untraced'' Green's function has in this case four external legs, $G(z)\equiv\frac{1}{N^2}\delta_{ij}\delta_{k\ell}G_{ijk\ell}$, and it is expanded as
\begin{equation}
    G_{ijk\ell}=\left\langle\frac{1}{z}\delta_{ij}\delta_{k\ell}+\frac{1}{z^2}\mathcal{L}_{ijk\ell}+\frac{1}{z^3}\mathcal{L}_{i\alpha k\beta}\mathcal{L}_{\alpha j \beta \ell}+\cdots\right\rangle\,.
\end{equation}
As previously, the only relevant diagrams in the large-$N$ limit are planar, which, at a given order $n$ of the expansion, have the same number $n$ of double propagators and loops. The planar diagrams in the first three orders in the expansion are given by
\begin{subequations}\label{eq:Liouvillian_G_expansion}
\begin{align}
    &G^{(0)}_{ijk\ell}=
    \begin{tikzpicture}[baseline=($0.5*(i)+0.5*(k)$)]
        \begin{feynman}[inline=(i)]
        \vertex[label=180:$i$]                      (i);
        \vertex[right=1.6cm of i,label=360:$j$]       (j);
        \vertex[below=0.5cm of i,label=180:$k$]     (k);
        \vertex[right=1.6cm of k,label=360:$\ell$]    (l);
        \diagram*{
            (i) -- (j) 
            (k) -- (l)
        };
        \end{feynman}
    \end{tikzpicture}\\
    &G^{(1)}_{ijk\ell}=-\frac{1}{2}\Bigg(
    \begin{tikzpicture}[baseline=($0.5*(i)+0.5*(k)$)]
        \begin{feynman}[inline=(i)]
        \vertex[label=180:$i$]                          (i);
        \vertex[right=0.6cm of i]                       (c);
        \vertex[dot,right=0.2cm of c,label=270:\scriptsize{$\mu$}]   (m) {};
        \vertex[right=0.2cm of m]                       (d);
        \vertex[right=0.6cm of d,label=360:$j$]         (j);
        \vertex[below=0.5cm of i,label=180:$k$]         (k);
        \vertex[right=1.6cm of k,label=360:$\ell$]        (l);
        \diagram*{
            (i) -- (c) --[half left, out=135, in=45, loop, min distance=1.2cm] (d) -- (j)
            (k) -- (l)
        };
        \end{feynman}
        \draw [/tikzfeynman/scalar] (m) to[out=135, in=45, loop, min distance=0.8cm] (m);
    \end{tikzpicture}
    +
    \begin{tikzpicture}[baseline=($0.5*(i)+0.5*(k)$)]
        \begin{feynman}[inline=(i)]
        \vertex[label=180:$i$]                          (i);
        \vertex[right=1.6cm of i,label=360:$j$]           (j);
        \vertex[below=0.5cm of i,label=180:$k$]         (k);
        \vertex[right=0.6cm of k]                       (c);
        \vertex[dot,right=0.2cm of c,label=90:\scriptsize{$\mu$}]   (m) {};
        \vertex[right=0.2cm of m]                       (d);
        \vertex[right=0.6cm of d,label=360:$\ell$]      (l);
        \diagram*{
            (i) -- (j)
            (k) -- (c) --[half left, out=225, in=315, loop, min distance=1.2cm] (d) -- (l)
        };
        \end{feynman}
        \draw [/tikzfeynman/scalar] (m) to[out=225, in=315, loop, min distance=0.8cm] (m);
    \end{tikzpicture}
    \Bigg)\\
    \begin{split}
        &G^{(2)}_{ijk\ell}=\frac{1}{4}\Bigg(
        \begin{tikzpicture}[baseline=($0.5*(i)+0.5*(k)$)]
        \begin{feynman}[inline=(i)]
        \vertex[label=180:$i$]                                     (i);
        \vertex[right=0.6cm of i]                                  (c);
        \vertex[dot,right=0.2cm of c,label=270:\scriptsize{$\mu$}] (m) {};
        \vertex[right=0.2cm of m]                                  (d);
        \vertex[right=0.6cm of d]                                  (e);
        \vertex[dot,right=0.2cm of e,label=270:\scriptsize{$\nu$}] (n) {};
        \vertex[right=0.2cm of n]                                  (f);
        \vertex[right=0.6cm of f,label=360:$j$]                    (j);
        \vertex[below=0.5cm of i,label=180:$k$]                    (k);
        \vertex[right=2.6cm of k,label=360:$\ell$]                 (l);
        \diagram*{
            (i) -- (c) --[half left, out=135, in=45, loop, min distance=1.2cm] (d) -- (e) --[half left, out=135, in=45, loop, min distance=1.2cm] (f) -- (j)
            (k) -- (l)
        };
        \end{feynman}
        \draw [/tikzfeynman/scalar] (m) to[out=135, in=45, loop, min distance=0.8cm] (m);
        \draw [/tikzfeynman/scalar] (n) to[out=135, in=45, loop, min distance=0.8cm] (n);
        \end{tikzpicture}
        +
        \begin{tikzpicture}[baseline=($0.5*(i)+0.5*(k)$)]
        \begin{feynman}[inline=(i)]
        \vertex[label=180:$i$]                                     (i);
        \vertex[right=0.6cm of i]                                  (c);
        \vertex[dot,right=0.2cm of c,label=270:\scriptsize{$\mu$}] (m) {};
        \vertex[right=0.2cm of m]                                  (d);
        \vertex[right=0.2cm of d]                                  (e);
        \vertex[dot,right=0.2cm of e,label=270:\scriptsize{$\nu$}] (n) {};
        \vertex[right=0.2cm of n]                                  (f);
        \vertex[right=0.6cm of f,label=360:$j$]                    (j);
        \vertex[below=0.5cm of i,label=180:$k$]                    (k);
        \vertex[right=2.2cm of k,label=360:$\ell$]                 (l);
        \diagram*{
            (i) -- (c) --[half left, out=135, in=45, loop, min distance=1.2cm] (f) -- (j)
            (d) -- (e)
            (k) -- (l)
        };
        \end{feynman}
        \draw [/tikzfeynman/scalar] (m) to[out=135, in=45, loop, min distance=0.9cm] (n);
        \draw [/tikzfeynman/scalar] (m) to[out=110, in=70, loop, min distance=0.5cm] (n);
        \draw (d) to[out=135, in=45, loop, min distance=0.5cm] (e);
        \end{tikzpicture}
        +
        \begin{tikzpicture}[baseline=($0.5*(i)+0.5*(k)$)]
        \begin{feynman}[inline=(i)]
        \vertex[label=180:$i$]                                     (i);
        \vertex[right=0.6cm of i]                                  (c);
        \vertex[dot,right=0.2cm of c,label=270:\scriptsize{$\mu$}] (m) {};
        \vertex[right=0.2cm of m]                                  (d);
        \vertex[right=1.6cm of d,label=360:$j$]                    (j);
        \vertex[below=0.5cm of i,label=180:$k$]                    (k);
        \vertex[right=1.6cm of k]                                  (e);
        \vertex[dot,right=0.2cm of e,label=90:\scriptsize{$\nu$}]  (n) {};
        \vertex[right=0.2cm of n]                                  (f);
        \vertex[right=0.6cm of f,label=360:$\ell$]                 (l);
        \diagram*{
            (i) -- (c) --[half left, out=135, in=45, loop, min distance=1.2cm] (d) -- (j)
            (k) -- (e) --[half left, out=225, in=315, loop, min distance=1.2cm] (f) -- (l)
        };
        \end{feynman}
        \draw [/tikzfeynman/scalar] (m) to[out=135, in=45, loop, min distance=0.8cm] (m);
        \draw [/tikzfeynman/scalar] (n) to[out=225, in=315, loop, min distance=0.8cm] (n);
        \end{tikzpicture}\\
        &\quad\quad\quad\quad\quad 
        \begin{tikzpicture}[baseline=($0.5*(i)+0.5*(k)$)]
        \begin{feynman}[inline=(i)]
        \vertex[label=180:$i$]                                     (i);
        \vertex[right=1.6cm of i]                                  (e);
        \vertex[dot,right=0.2cm of e,label=270:\scriptsize{$\nu$}]  (n) {};
        \vertex[right=0.2cm of n]                                  (f);
        \vertex[right=0.6cm of f,label=360:$j$]                    (j);
        \vertex[below=0.5cm of i,label=180:$k$]                    (k);
        \vertex[right=0.6cm of k]                                  (c);
        \vertex[dot,right=0.2cm of c,label=90:\scriptsize{$\mu$}]  (m) {};
        \vertex[right=0.2cm of m]                                  (d);
        \vertex[right=1.6cm of d,label=360:$\ell$]                 (l);
        \diagram*{
            (i) -- (e) --[half left, out=135, in=45, loop, min distance=1.2cm] (f) -- (j)
            (k) -- (c) --[half left, out=225, in=315, loop, min distance=1.2cm] (d) -- (l)
        };
        \end{feynman}
        \draw [/tikzfeynman/scalar] (m) to[out=225, in=315, loop, min distance=0.8cm] (m);
        \draw [/tikzfeynman/scalar] (n) to[out=135, in=45, loop, min distance=0.8cm] (n);
        \end{tikzpicture}
        +
        \begin{tikzpicture}[baseline=($0.5*(i)+0.5*(k)$)]
        \begin{feynman}[inline=(i)]
        \vertex[label=180:$i$]                                     (i);
        \vertex[right=2.6cm of i,label=360:$j$]                    (j);
        \vertex[below=0.5cm of i,label=180:$k$]                    (k);
        \vertex[right=0.6cm of k]                                  (c);
        \vertex[dot,right=0.2cm of c,label=90:\scriptsize{$\mu$}] (m) {};
        \vertex[right=0.2cm of m]                                  (d);
        \vertex[right=0.6cm of d]                                  (e);
        \vertex[dot,right=0.2cm of e,label=90:\scriptsize{$\nu$}] (n) {};
        \vertex[right=0.2cm of n]                                  (f);
        \vertex[right=0.6cm of f,label=360:$\ell$]                 (l);
        \diagram*{
            (i) -- (j)
            (k) -- (c) --[half left, out=225, in=315, loop, min distance=1.2cm] (d) -- (e) --[half left, out=225, in=315, loop, min distance=1.2cm] (f) -- (l)
        };
        \end{feynman}
        \draw [/tikzfeynman/scalar] (m) to[out=225, in=315, loop, min distance=0.8cm] (m);
        \draw [/tikzfeynman/scalar] (n) to[out=225, in=315, loop, min distance=0.8cm] (n);
        \end{tikzpicture}
        +
        \begin{tikzpicture}[baseline=($0.5*(i)+0.5*(k)$)]
        \begin{feynman}[inline=(i)]
        \vertex[label=180:$i$]                                     (i);
        \vertex[right=2.2cm of i,label=360:$j$]                    (j);
        \vertex[below=0.5cm of i,label=180:$k$]                    (k);
        \vertex[right=0.6cm of k]                                  (c);
        \vertex[dot,right=0.2cm of c,label=90:\scriptsize{$\mu$}] (m) {};
        \vertex[right=0.2cm of m]                                  (d);
        \vertex[right=0.2cm of d]                                  (e);
        \vertex[dot,right=0.2cm of e,label=90:\scriptsize{$\nu$}] (n) {};
        \vertex[right=0.2cm of n]                                  (f);
        \vertex[right=0.6cm of f,label=360:$\ell$]                 (l);
        \diagram*{
            (i) -- (j)
            (k) -- (c) --[half left, out=225, in=315, loop, min distance=1.2cm] (f) -- (l)
            (d) -- (e)
        };
        \end{feynman}
        \draw [/tikzfeynman/scalar] (m) to[out=225, in=315, loop, min distance=0.9cm] (n);
        \draw [/tikzfeynman/scalar] (m) to[out=250, in=290, loop, min distance=0.5cm] (n);
        \draw (d) to[out=225, in=315, loop, min distance=0.5cm] (e);
        \end{tikzpicture}
        \Bigg)
    \end{split}
\end{align}
\end{subequations}

From the first three orders we can infer the general structure of the planar diagrams at arbitrary order $n$: there are $n$ contractions of indices (loops) corresponding to dashed lines (represented by the dots $\mu$, $\nu$, $\cdots$ in Eq.~(\ref{eq:Liouvillian_G_expansion})) which are divided in all possible ways between the top ($i\to j$) and the bottom ($k\to\ell$) lines, resulting in $2^n$ different divisions. Note that, in organizing the diagrams in this way, the third and fourth diagrams of $G^{(2)}$ in Eq.~(\ref{eq:Liouvillian_G_expansion}) are considered different, i.e.\ a point up followed by one down is different from one down followed by one up, and similarly for more points. Then, for a given ordered division of points, say $n_1$ points in the upper line and $n_2$ in the bottom one (such that $n_1+n_2=n$), we draw all possible combinations of $n_1^\mathrm{th}$ order planar diagrams for the Wishart Green's function in the upper line and of $n_2^\mathrm{th}$ order planar Wishart diagrams in the lower line, and add them all up with a multiplicative factor of $(-1/2)^n$. 

It is relevant to note that all the diagrams in Eq.~(\ref{eq:Liouvillian_G_expansion}) arise from powers of the $-\frac{1}{2}W^*_{mi}W_{mj}\delta_{k\ell}-\frac{1}{2}\delta_{ij}W_{mk}W^*_{m \ell}$ term (called the `non-crossing term' in Ref.~\cite{can2019}) of the Liouvillian only. This happens because in the $W_{ij}W^*_{k\ell}$ term (called the `recycling term') there are no contractions and hence it is suppressed by $1/N$. Then, at order $n$ in the expansion, diagrams of $\mathcal{O}(1)$ arise exclusively from powers of the non-crossing term and using planar Wishart diagrams. Diagrams of $\mathcal{O}(1/N)$ arise from the product of one recycling term with $n-1$ non-crossing terms, using planar Wishart diagrams. For $1/N^2$ corrections we need either products of two recycling terms with $n-2$ non-crossing terms and planar Wishart diagrams or the $n^\mathrm{th}$ power of non-crossing terms with non-planar, suppressed by $1/N^2$, Wishart diagrams. Considering even higher-order correction becomes increasingly difficult because of the increasing number of ways to generate Liouvillian diagrams out of the two kinds of terms and highly suppressed Wishart diagrams. When considering the nonholomorphic Green's function inside the complex spectral support, the recycling term does contribute and the above reasoning fails. We can also easily see why the computation becomes extremely involved in that case (and remains an open problem): one the one hand, the two sectors of the tensor product become connected; on the other hand, instead of having to calculate averages of powers of $\mathcal{L}$ only, $\langle\mathcal{L}^n\rangle$, we would have to compute all possible powers of $\mathcal{L}$ and $\mathcal{L^\dagger}$ (which, recall, do not commute and, e.g.\ $\langle\mathcal{L}\mathcal{L}^\dagger\mathcal{L}\mathcal{L}^\dagger\rangle\neq\langle\mathcal{L}^2{\mathcal{L}^\dagger}^2\rangle)$).

We are now able to relate the series expansion of the Wishart and the Liouvillian Green's functions. To avoid confusion, we denote the Wishart Green's function by $G$ and the Liouvillian Green's function by $G_\mathcal{L}$. Using the counting of diagrams of the preceding paragraphs, we can immediately write down its expansion in terms of the Catalan numbers $C_n$:
\begin{equation}\label{eq:Liouvillian_G_series}
    G_\mathcal{L}(z)=\sum_{n=0}^\infty\frac{(-1)^n}{2^n}\frac{1}{z^{n+1}}\sum_{k=0}^n\binom{n}{k}C_k\,C_{n-k}\,.
\end{equation}

Inserting the form of $C_n$ from Eq.~(\ref{eq:c_n_moments_MP}) into Eq.~(\ref{eq:Liouvillian_G_series}), we obtain
\begin{equation}\label{eq:G_L_aux}
\begin{split}
    G_\mathcal{L}(z)=&\int\d\nu\d\nu^\prime\,\varrho_\mathrm{MP}(\nu)\varrho_\mathrm{MP}(\nu^\prime)\sum_{n=0}^\infty\sum_{k=0}^n\binom{n}{k}\left(\frac{-1}{2}\right)^{n}\frac{\nu^k{(\nu^\prime)}^{n-k}}{z^{n+1}}\\
    =&\int\d\nu\d\nu^\prime\, \frac{\varrho_\mathrm{MP}(\nu)\varrho_\mathrm{MP}(\nu^\prime)}{z+\frac{1}{2}(\nu+\nu^\prime)}\,.
\end{split}
\end{equation}
Introducing the spectral density of the Liouvillian, $\varrho_\mathcal{L}$, again by its relation with the Green's function, $G_\mathcal{L}(z)=\int\d x\,\varrho_\mathcal{L}(x)/(z-x)$, and comparing it with Eq.~(\ref{eq:G_L_aux}), we obtain the spectral density of $\mathcal{L}$ in terms of the spectral density of the Wishart ensemble:
\begin{equation}\label{eq:Liouvillian_rho_integral}
    \varrho_\mathcal{L}(x)=\int\d\nu\d\nu^\prime\, \varrho_\mathrm{MP}(\nu)\varrho_\mathrm{MP}(\nu^\prime)\dirac{x+\frac{1}{2}(\nu+\nu')}=
    2\int_{\nu_m}^{\nu_M}\d\nu\,\varrho_{\mathrm{MP}}(\nu)\varrho_{\mathrm{MP}}(-\nu-2x)\,,
\end{equation}
with $\nu_m=\max\{0,-2x-4\}$ and $\nu_M=\min\{4,-2x\}$, whence it follows that $-4<x<0$. Hence, the computation of the holomorphic Green's function has lead us to a convolution of two Marchenko-Pastur laws, which is the spectral density of half the sum of two \emph{uncorrelated} Wishart matrices. Although the spectral density $\varrho_\mathcal{L}$ does not correctly describe the spectrum of the Liouvillian, it does predict its endpoints and the interval $[-4,0]$ in which $\re(\Lambda)$ is supported. Note, in particular, that the spectrum is gapless for a single decay channel and infinite dissipation.

To generalize the above discussion to more than one jump operator, we briefly summarize the key steps above. When computing the holomorphic Green's function, the terms $W_\ell\otimes W^*_\ell$ do not contribute in the large-$N$ limit. The Liouvillian is then of the form $\mathcal{L} \simeq -\frac{1}{2}(\Gamma\otimes\mathbbm{1}+\mathbbm{1}\otimes\Gamma^{\T}$), where $\Gamma=W^\dagger W$ is a ($m=1$) Wishart matrix. Since the two sectors of the Liouvillian tensor-product representation (i.e. the top and bottom lines in the diagrammatic expansion) do not interact, we can do independent expansions for each Wishart matrix. It follows that the spectral density of the holomorphic Green's function is given by the convolution of the spectral densities of the uncorrelated matrices $\Gamma$. The endpoints of the convolution, $\nu_m$ and $\nu_M$, define the spectral support of $\re(\Lambda)$.

For more than one jump operator, $\Gamma$ is a sum of $r$ identical (i.e.\ independent but sampled from the same distribution) Wishart matrices. We saw in Section~\ref{section:random_ensembles} that the spectral density of $\Gamma$ is again a Marchenko-Pastur distribution, $\varrho_\Gamma(x)$, $\xi_-<x<\xi_+$, with endpoints dependent on $r$, $\xi_{\pm}=(1\pm\sqrt{r})^2$. By the same arguments as above, it immediately follows that $\mathcal{L}$ is supported in $[-\xi_+,-\xi_-]$. We thus see, in agreement with Ref.~\cite{can2019}, that
\begin{equation}\label{eq:gap_strong_dissipation}
    \langle\Delta\rangle=\beta N g^2\xi_-=\beta N r g^2=\sqrt{\beta N}g_\mathrm{eff}^2\frac{(1-\sqrt{r})^2}{\sqrt{2r}}\,,
\end{equation}
while 
\begin{equation}
    X\propto\beta N g^2(\xi_+-\xi_-)=4\beta N \sqrt{r} g^2\propto\sqrt{\beta N}g_\mathrm{eff}^2\,,
\end{equation}
with the proportionality constant of order unity, and where we have reintroduced the variances $\beta N g^2$. These expressions were verified numerically in Chapter \ref{chapter:numerics}, see Figs.~\ref{fig:spectrum_XY}~and~\ref{fig:SM_gap_r}.

\section{Holomorphic Green's functions II.\ Arbitrary dissipation}
\label{section:holomorphic_G_K}

The most important lesson to take from the previous section is that the recycling term $\sum_\ell W_\ell\otimes W_\ell^*$ does not contribute to the holomorphic Green's function in the large-$N$ limit. Calculations are then substantially simplified by the independence of the two sectors of the product sector. Since the Hamiltonian contributes to the Liouvillian only via a commutator, that independence is not spoiled by considering an arbitrary finite $g$. The ideas behind the results of this section were first brought to our attention by~\cite{can_private} (see also Ref.~\cite{can2019}).

At finite $g$ (or, equivalently, $H\neq0$), and neglecting the recycling term, the Liouvillian reads
\begin{equation}\label{eq:eff_Liouvillian}
    \mathcal{L}=-i\left\{\left(H-i\frac{g^2}{2}\sum_\ell W_\ell^\dagger W_\ell \right)\otimes\mathbbm{1}-\mathbbm{1}\otimes\left(H+i\frac{g^2}{2}\sum_\ell W_\ell^\dagger W_\ell \right)^{\sf{T}}\right\}\equiv -i\left[K\otimes\mathbbm{1}-\mathbbm{1}\otimes\left(K^\dagger\right)^{\T} \right]\,,
\end{equation}
where the last equality defines the nonhermitian Hamiltonian $K=H-i(g^2/2)\Gamma$, with $\Gamma=\sum_{\ell=1}^r W_\ell^\dagger W_\ell$. We thus see that in the large-$N$ limit, the (holomorphic version of the) Liouvillian acts on $\rho$ as $\mathcal L[\rho]=-i(K\rho-\rho K^\dagger)$. From Eq.~(\ref{eq:eff_Liouvillian}), we see that indeed the two sectors of the tensor product do not communicate and we can consider them independently. The eigenvalues of $\mathcal L$ can be immediately written down
\begin{equation}\label{eq:Lambda_kappa}
    \Lambda_{jk}=-i(\kappa_j-\kappa^*_k),\quad\quad j,k=1,\dots,N\,,
\end{equation}
with $\kappa_j$ the eigenvalues of $K$ (and $\kappa^*_k$ the eigenvalues of $K^\dagger$). If we denote $y_\mathrm{min}=\min\left(-\Im\kappa_{j}\right)$ (note that, because $\Gamma$ is positive-definite, all $\kappa$ have negative imaginary part), then it follows immediately from Eqs.~(\ref{eq:eff_Liouvillian})~and~(\ref{eq:Lambda_kappa}) that $\Delta=\min\left(-\Re\Lambda_{jk}\right)=2\min\left(\Re i\kappa_{j}\right)=2\min\left(-\Im\kappa_j\right)=2y_\mathrm{min}$. The problem of computing the gap of the Liouvillian superoperator is thus reduced to finding the gap of a nonhermitian Hamiltonian operator.

Nonhermitian Hamiltonians were extensively studied in the past~\cite{lehmann1995,haake1992,janik1997a,janik1997b} (recall also the discussion of Section~\ref{section:nonhermitian_RMT}), and their spectral density and boundary (and therefore the gap) were explicitly obtained by different methods. Although these authors considered dissipative terms $\Gamma$ build out of (in the language of this thesis) a single $M\times N$ jump operator with $m=M/N>1$ (this $m$ is the one introduced at the start of Section~\ref{subsection:diagrammatics_Wishart}), and we are instead interested in $r$ identically distributed $N\times N$ jump operators, we have seen in Section~\ref{section:random_ensembles} that two cases are formally equivalent (under the interchange $m\leftrightarrow r$). The results of these authors can, therefore, be directly imported into our work, after some straightforward rescalings.

First, notice that, for instance, Ref.~\cite{janik1997b} studies the nonhermitian Hamiltonian $\Tilde{K}=\Tilde{H}+i\Tilde{g}\Tilde{\Gamma}$, where $\Tilde{H},\Tilde{\Gamma}$ have spectra independent of $N$, i.e.\ $H=\sqrt{\beta N}\Tilde{H}$ and $\Gamma=\beta N\Tilde{\Gamma}$. We thus rewrite 
\begin{equation}
    K=\sqrt{\beta N}\left(\Tilde{H}+i\frac{\sqrt{\beta N}(-g)^2}{2}\Tilde{\Gamma}\right)\,,
\end{equation} 
and we identify $\Tilde{g}=-\sqrt{\beta N}g^2/2$. The prefactor $\sqrt{\beta N}$ is just an overall factor by which we must multiply our final results when working with these new scalings. Setting $\Tilde{\kappa}=\Tilde{x}+i\Tilde{y}$ (with $\Tilde{\kappa}$ the eigenvalues of $\Tilde{K}$ and $\Tilde{y}>0$), the boundary of the spectrum of $\Tilde{K}$ is given by the solutions to the following equation~\cite{janik1997b}:
\begin{equation}\label{eq:K_boundary_janik}
\Tilde{x}^2=\frac{4m}{\Tilde{g}\Tilde{y}}-\left(\frac{m}{\Tilde{y}}-\frac{2\Tilde{g}}{1+\Tilde{g}\Tilde{y}}+\frac{1}{\Tilde{g}}\right)^2\,.
\end{equation}

The minimal imaginary part, $\Tilde{y}_\mathrm{min}>0$, and the maximal imaginary part $\Tilde{y}_\mathrm{max}>0$, are attained at $\Tilde{x}=0$. Replacing $m$ by $r$, we arrive at the equation for the endpoints of the spectrum,
\begin{equation}\label{eq:K_boundary}
\frac{4r}{\Tilde{g} \Tilde{y}}=\left(\frac{r}{\Tilde{y}}-\frac{\Tilde{g}}{1+\Tilde{g}\Tilde{y}}+\frac{1}{\Tilde{g}}\right)^2\,.   
\end{equation}
This equation is a fourth order equation for $\Tilde{y}$ with two real and two complex solutions (which may degenerate into multiple real roots for certain values of $\Tilde{g}$). The two real solutions correspond to the real endpoints of the spectrum, $\Tilde{y}_\mathrm{min},\Tilde{y}_\mathrm{max}$. This equation can be solved explicitly for $\Tilde{y}$ and a plot of the solution as a function of $(2r)^{1/4}\sqrt{\Tilde{g}}$ was given in Fig.~\ref{fig:gap}, in comparison with the numerical results. We now consider the limiting cases $\Tilde{g}\to0,+\infty$.

For large $\Tilde{g}$, we can replace the second term inside brackets in Eq.~(\ref{eq:K_boundary}) by its limiting value $1/\Tilde{y}$. Although it would seem that we could also drop the last term, since $1/\Tilde{g}\to0$, we might have $\Tilde{y}\propto\Tilde{g}$ (as we know is the case, in hindsight), which would render the first and second terms of the same order of the third. The two solutions to resulting approximate equation are $\Tilde{y}=\Tilde{g}(1\pm\sqrt{r})^2$.

For small $\Tilde{g}$, we can drop the second term inside brackets. The equation now has a unique solution (single endpoint) $\Tilde{y}=\Tilde{g} r$. This means that for infinitesimal dissipation the spectrum gets shifted away from the imaginary axis into the complex plane to $\Tilde{g} r$ but does not spread, as we had discussed before.

Reintroducing our scalings, $\Delta=2y_\mathrm{min}=-2\sqrt{\beta N}\Tilde{y}_\mathrm{min}$ and $\Tilde{g}=-\sqrt{\beta N} g^2/2$, we obtain the asymptotic values of the spectral gap
\begin{equation}\label{eq:asymptotics_gap}
    \Delta=
    \begin{cases}
        \beta N g^2 r\,, &\mathrm{when}\ g\to0\,,\\
        \beta N g^2 (1-\sqrt{r})^2, &\mathrm{when}\ g\to\infty\,,\\
    \end{cases}
\end{equation}
in agreement with the numerical results of Chapter~\ref{chapter:numerics} and with the analytic results of the previous section and of Ref.~\cite{can2019}.

\section{Degenerate perturbation theory -- weak dissipation}
\label{section:degenerate_PT}

We now focus on the weakly dissipative limit, $g\to0$, by using degenerate perturbation theory around the Hamiltonian limit $g=0$, which we know how to solve exactly. This way, we will recover the asymptotic limiting result of the last section from an independent calculation.

We denote $\mathcal{L}=\mathcal{L}_\mathrm{H}+g^2\mathcal{L}_\mathrm{D}$, with $\mathcal{L}_\mathrm{H}\left(\rho\right)=-i\left[H,\rho\right]$ and $\mathcal{L}_\mathrm{D}\left(\rho\right)=\sum_\ell\left(W_{\ell}\rho W_{\ell}^{\dagger}-\frac{1}{2}W_{\ell}^{\dagger}W_{\ell}\rho-\frac{1}{2}\rho W_{\ell}^{\dagger}W_{\ell}\right)$. If $H$ is diagonal in the basis $\ket{k}$, $H=\sum_k\varepsilon_k\ket{k}\bra{k}$, then the eigenstates of $\mathcal{L}_\mathrm{H}$ are $\rho_{kk^\prime}=\ket{k}\bra{k^\prime}$ with eigenvalue $\Lambda_{kk^\prime}=-i(\varepsilon_k-\varepsilon_{k^\prime})$ ($k,k^\prime=1,\dots,N$). We thus have an $N$-fold degeneracy of stationary eigenstates, while the other eigenvalues form $N(N-1)/2$ complex conjugate pairs. Any amount of dissipation lifts the degeneracy of the zero-eigenvalue subspace. Nonetheless, for small enough dissipation the two sectors (of zero and nonzero eigenvalues at $g=0$) will not mix and we expect that the long-time dynamics (the gap) and the (now unique) steady-state are determined from the states in the (previously-) degenerate subspace.  We lift the degeneracy by diagonalizing the perturbation $\mathcal{L}_\mathrm{D}$ in that subspace, obtaining, to first order,
\begin{equation}\label{eq:A_from_LD}
\begin{split}
    A_{nm}\equiv&\,\frac{1}{g^2}\bbra{nn}\mathcal{L}_\mathrm{D}\kket{mm}\\
    =&\bbra{nn}\sum_{\ell=1}^r\left[W^{(\ell)}\otimes {W^{(\ell)}}^*-\frac{1}{2}\left({W^{(\ell)}}^\dagger W^{(\ell)}\otimes \mathbbm{1}+\mathbbm{1}\otimes {W^{(\ell)}}^{\sf T}{W^{(\ell)}}^*\right)\right]\kket{mm}\\
    =&\sum_{\ell=1}^r\Big(\bra{n}W^{(\ell)}\ket{m}\bra{m} {W^{(\ell)}}^\dagger\ket{n}-\delta_{nm}\bra{n} {W^{(\ell)}}^\dagger W^{(\ell)}\ket{m}\Big)\\
    =&\sum_{\ell=1}^r\Big(W^{(\ell)}_{nm}{W^{(\ell)}_{nm}}^*-\delta_{nm}\sum_k {W^{(\ell)}_{kn}}^*W^{(\ell)}_{kn}\Big)\\
    =&\begin{cases}
        \sum_{\ell}W^{(\ell)}_{nm}{W^{(\ell)}_{nm}}^*\,,\quad\quad\quad\quad\ \ \ \mathrm{if}\ m\neq n \\
        -\sum_{\ell}\sum_{k\neq m}W^{(\ell)}_{km}{W^{(\ell)}_{km}}^*\,,\quad \mathrm{if}\ m= n 
    \end{cases}\,.
\end{split}
\end{equation}

From Eq.~(\ref{eq:A_from_LD}) it follows that
\begin{subequations}
\begin{align}
    &A_{nm}\in\mathbb{R}\quad (\mathrm{for}\ \mathrm{all}\ n)\,,\\
    &A_{nm}>0\quad\ (n\neq m)\,,\\
    \sum_{n=1}^N &A_{nm}=0\quad\ (\mathrm{for}\ \mathrm{all}\ n)\,.\label{eq:restriction_A}
\end{align}
\end{subequations}
These three conditions imply that $A$ is the generator of a classical stochastic equation~\cite{vankampen1992,denisov2018,timm2009}, $\partial_t \vec{P}=A\vec{P}$, where $\vec P$ is a probability vector. The exponential of $A$, $T(t)=\exp{t A}$, is a stochastic matrix, meaning that $T_{nm}\geq0$ and $\sum_m T_{nm}=1$. In particular, we can see Eq.~(\ref{eq:restriction_A}) as a restriction on $A$ ensuring conservation of probability. Note also that $A$ is symmetric if the $W^{(\ell)}$ are hermitian.

We now wish to obtain spectral properties of $A$. In particular, we wish to study the low laying eigenvalues that determine the long time asymptotics. The diagonal entries $A_{mm}=-\sum_{k\neq m}A_{km}$ (by Eq.~(\ref{eq:restriction_A})) are determine by the off-diagonal entries, hence we must only determined the distribution of the latter. For $n\neq m$, the distribution of the entry $A_{nm}$ depends on the nature of the jump operator as follows. $W^{(\ell)}_{nm}$ is a Gaussian random variable with variance $\sigma^2$ and $\beta$ real degrees of freedom (in our numerical implementation we have $\sigma^2=1$). Then, $A_{nm}=\sum_\ell {W^{(\ell)}_{nm}}^*W^{(\ell)}_{nm}=\sum_\ell\abs{W^{(\ell)}_{nm}}^2$ is (in terms of real degrees of freedom) a sum of $r\beta$ squared Gaussian iid variables and hence follows a $\chi^2$-distribution with $k=r\beta$ degrees of freedom. It follows that the distribution function of the entries $A_{nm}$ is 
\begin{equation}\label{eq:distribution_A}
    P_k(A_{nm})=\frac{(A_{nm})^{\frac{k}{2}-1}\exp{-\frac{A_{nm}}{2\sigma^2}}}{(2\sigma^2)^\frac{k}{2}\,\Gamma\left(\frac{k}{2}\right)}\,.
\end{equation}

Ensembles of real exponentially-distributed matrices (corresponding to $k=2$), both symmetric and asymmetric, with the diagonal constraint~(\ref{eq:restriction_A}) were studied in Ref.~\cite{timm2009}. In particular, it was shown that, in the large $N$-limit, the spectral density depends only on the second moment of the distribution, which in our case is $\tau^2_k\equiv\langle A_{nm}^2\rangle-\langle A_{nm}\rangle^2=2k\sigma^4$. It can, therefore, be calculated from any distribution of entries having this variance, particularly from a Gaussian distribution, which was done in Ref.~\cite{staring2003} for symmetric $A$ (recall that symmetric $A$ corresponds to hermitian jump operators here). The spectral density obtained this way is not, however, the Wigner-semicircle law (for hermitian jump operators) or the Ginibre disk (for nonhermitian jump operators) because of the constraint on diagonal entries (if one proceeds diagrammatically, the constraint modifies the Feynman rules). Furthermore, Ref.~\cite{timm2009} numerically found that the average smallest nonzero eigenvalue of $A$ (the symmetric value of the spectral gap here) is $N\langle R\rangle$ for large $N$, where $\langle R\rangle$ is the mean of the exponential ($k=2$) distribution of $A_{nm}$. Generalizing to our $\chi^2$-distribution, with mean $\mu_k=k\sigma^2$, we finally arrive at the result quoted in Chapter~\ref{chapter:numerics} and computed in Eq.~(\ref{eq:asymptotics_gap}),
\begin{equation}\label{eq:gap_weak_dissipation}
    \langle\Delta\rangle=N\mu_kg^2=\beta N r g^2=\sqrt{\beta N r/2}g_\mathrm{eff}^2\,,
\end{equation}
where we have reintroduced the factor $g^2$ from Eq.~(\ref{eq:A_from_LD}).

\cleardoublepage


\chapter{Complex spacing ratios: a signature of dissipative quantum chaos}
\label{chapter:spacings}
In previous chapters, we studied the spectral and steady-state properties of a random Liouvillian ensemble. It models some expected characteristic features of generic open quantum systems, like the spectral gap, typical relaxation rates, or the steady-state purity. By construction, the spectrum of the random Liouvillian is fully chaotic. In this chapter, we tackle the problem from a different angle by looking at the universal spectral fluctuations in open quantum systems. We introduce a new signature of dissipative quantum chaos based on the \emph{complex} distance between levels, which is easily computed for numerical spectra. Besides giving the usual measure of level repulsion, complex spacing ratios contain additional information on the isotropy (or lack thereof) of the spectral correlations. Furthermore, by considering the ratios of complex spacings, we are automatically unfolding the spectrum, bypassing the need to define a specific unfolding procedure for each system, which is nontrivial.

Two comments are in order regarding complex spacing ratios. First, level sequences are usually assumed ordered~\cite{atas2013,atas2013long,srivastava2018,tekur2018pre,tekur2018prb,tekur2018c,chavda2013,bhosale2018} (recall the discussion of Section~\ref{section:review_spacing_distributions}). There is, however, no global order in the complex plane, and hence all ratios which relied on the ordering have to be abandoned. The only remaining spacing ratio is the NN by NNN ratio introduced in Ref.~\cite{srivastava2018} (and $k$th-nearest-neighbor generalizations). Second, all the aforementioned works considered \emph{undirected} spacings, i.e.\ the distance, $s>0$, between neighboring levels, without a direction). In contrast, we consider \emph{directed} spacings. In real spectra, this only introduces a sign (for a given level, the spacing is positive if both its NN and NNN are on the same side, negative if they are on opposite sides), which may seem a minor difference. For complex spectra, however, we get an angular dependence, which is markedly different for Poisson and Ginibre spectra and hence constitutes a new signature of dissipative quantum chaos.

The chapter is organized as follows. In Section~\ref{section:ratio_definitions_overview} we give the definition of the new signature, analyze its finite-size dependence and discuss the expected differences between integrable and chaotic spectra. Then, in Section~\ref{section:ratio_Louvillian}, we apply the complex spacing ratio to Liouvillian spectra, by numerically studying random Liouvillians and several different boundary-driven spin chains. The final three sections provide a complete analytic characterization of the ratio distribution: in Section~\ref{section:ratio_Poisson} for Poisson processes in $d$ dimensions, in Section~\ref{section:ratio_hermitian} for hermitian random matrix ensembles, and in Section~\ref{section:ratio_nonhermitian} for nonhermitian random matrix ensembles.

The results of this Chapter are also discussed in Ref.~\cite{sa2019CSR}.

\section{Definitions and overview}
\label{section:ratio_definitions_overview}
\begin{figure}[htbp]
    \centering
    \includegraphics[width=0.6\columnwidth]{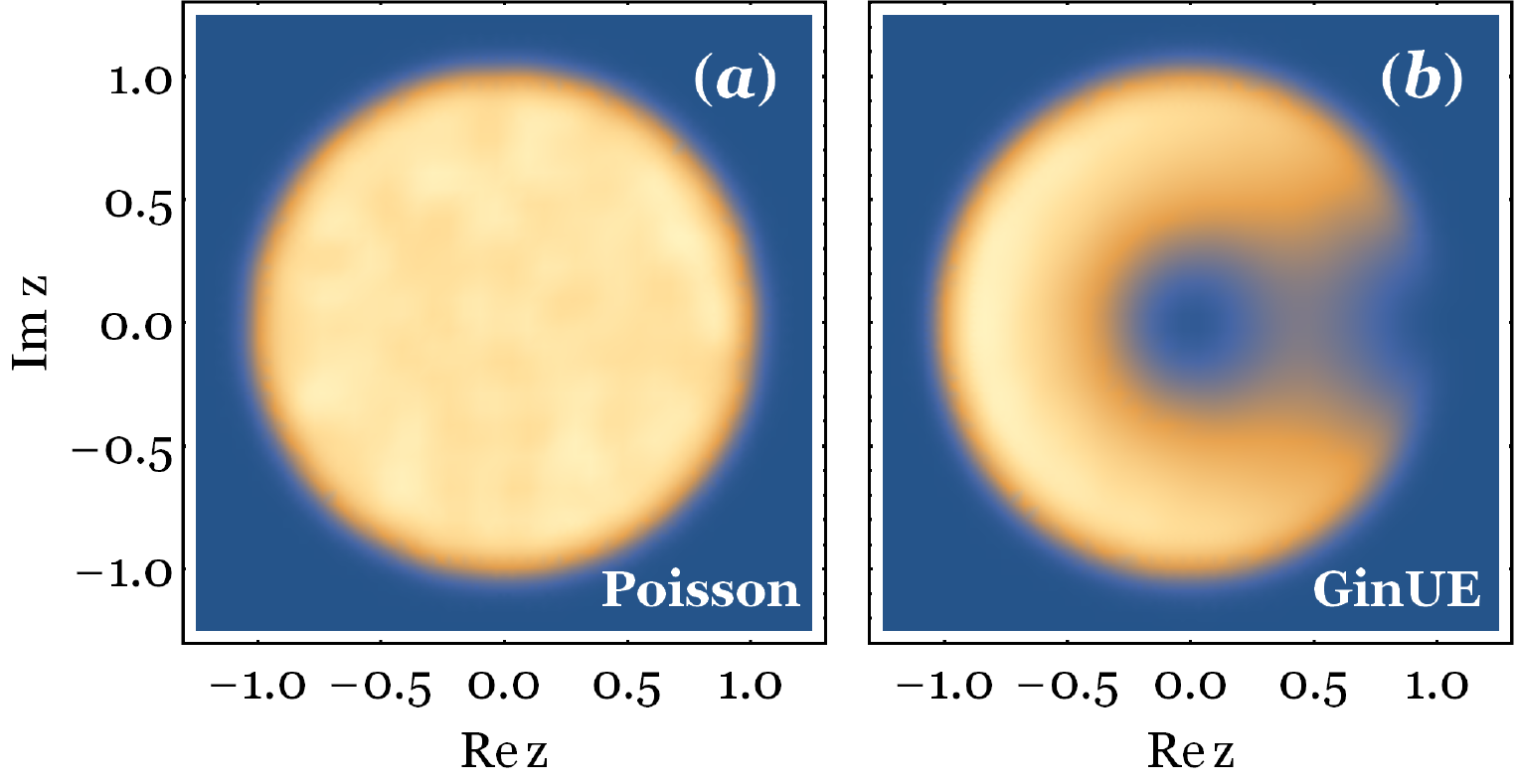}
    \caption{Complex spacing ratio density in the complex plane for $(a)$ a Poisson process with $10^5$ independent levels and $(b)$ the eigenvalues of 100 $10^4\times 10^4$ random matrices drawn from the GinUE.}
    \label{fig:Poisson_vs_GinUE}
\end{figure}

Let the set $\{\lambda_k\}_{k=1}^N$ form the spectrum of some hermitian or nonhermitian operator or a synthetic Poisson spectrum. The levels $\lambda_k$ may be real or complex. For each $\lambda_k$, we find its NN (with respect to the norm-distance), $\lambda_k^\mathrm{NN}$, and its NNN, $\lambda_k^\mathrm{NNN}$, and define the (in general complex) ratio $z_k=(\lambda_k^\mathrm{NN}-\lambda_k)/(\lambda_k^\mathrm{NNN}-\lambda_k)$. We then seek the probability distribution function $\varrho^{(N)}(z)$ of finding any spacing ratio with value $z$. If the spectrum is real, $z\equiv r$ satisfies $-1<r<1$ and may not coincide with the ratio of consecutive spacings. If the spectrum is complex, $z\equiv re^{i\theta}\equiv x+i y$, with $r<1$, and the distribution is not necessarily isotropic. We also consider the radial and angular marginal distributions, $\varrho(r)=\int \d \theta\,r\varrho(r,\theta)$ and $\varrho(\theta)=\int \d r\,r\varrho(r,\theta)$, respectively.

We start by considering two paradigmatic cases: Poisson processes and the Ginibre Ensembles. By natural extensions of the Berry-Tabor and Bohigas-Giannoni-Schmit conjectures, integrable systems have Poisson ratio statistics, while systems with a chaotic semiclassical limit follow Ginibre statistics. Due to the independence of levels in the Poisson spectrum, the presence of a reference level does not influence its two nearest neighbors and hence all ratios $z$ have the same probability and the distribution is flat. In contrast, for random matrices, we expect the usual repulsion, with two immediate consequences. First, the ratio density should vanish at the origin; second, the repulsion should spread all the neighbors of the reference level evenly around it, leading to a suppression of the ratio density for small angles. 

Figure~\ref{fig:Poisson_vs_GinUE} shows the ratio density $\varrho(z)$ in the complex plane for Poisson processes, $(a)$, and GinUE matrices, $(b)$, and confirms the expectations above. The Poisson ratio is indeed flat inside the unit circle, $\varrho_\mathrm{Poi}(z)=(1/\pi)\Theta(1-r^2)$, with $\Theta$ the Heaviside step-function. It immediately follows that the radial and angular marginal distributions are, respectively, $\varrho_\mathrm{Poi}(\theta)=1/(2\pi)$ and $\varrho_\mathrm{Poi}(r)=2r$. The independence of Poisson levels allows for the exact computation of all quantities of interest in $d$-dimensional space (not just real, $d=1$, and complex, $d=2$, spectra). For details, see Section~\ref{section:ratio_Poisson}. GinUE random matrices, on the contrary, have cubic level repulsion, $\varrho_\mathrm{GinUE}(r)\propto r^3$ as $r\to0$ (one power of $r$ comes from the area element on the plane), and the distribution shows some anisotropy, measured, for instance, by $\av{\cos\theta}=\int\d \theta\,\cos\theta\varrho_\mathrm{GinUE}(\theta)\simeq -0.24$ (for Poisson processes, obviously, $\av{\cos\theta}=0$).

\begin{figure}[tbp]
    \centering
    \includegraphics[width=\textwidth]{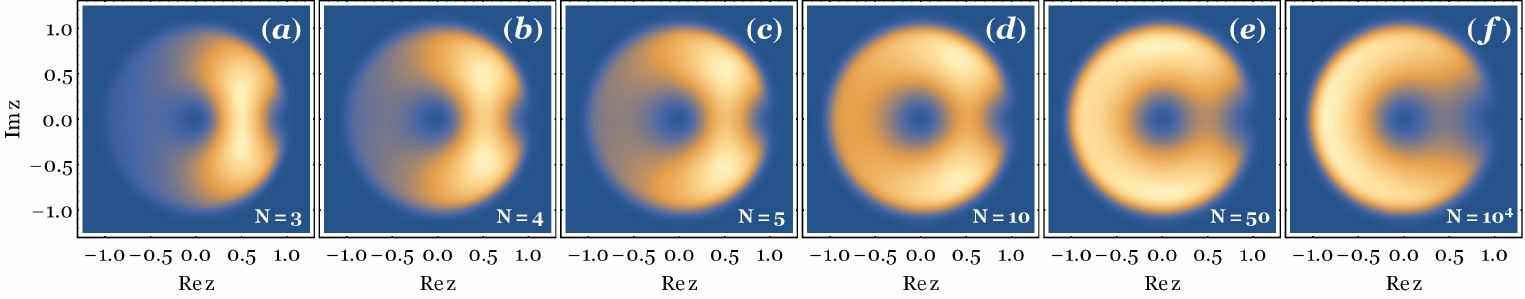}
    \caption{Complex spacing ratio density in the complex plane for $10^6$ eigenvalues of $N\times N$ random matrices drawn from the GinUE, for various $N$. Besides ubiquitous cubic level repulsion, there is a enhancement (suppression) of small angles at small (large) $N$.}
    \label{fig:GinUE_finite_N}
\end{figure}

Here, contrary to the case of consecutive spacing ratios, small-size matrices, say $N=3$ or $N=4$, \emph{do not} correctly capture large-$N$ asymptotics. Indeed, for small $N$, there is an enhancement of small angles instead of the suppression at large $N$, see Fig.~\ref{fig:GinUE_finite_N}. The same issue can be seen for real spectra: for large $N$, there is a high probability of finding negative ratios, while for small $N$, the probability of positive $r$ is higher. 

The \textit{peak inversion} of Fig.~\ref{fig:GinUE_finite_N} can be understood as a boundary effect. For definiteness, consider matrices drawn from a hermitian ensemble. For $N=3$, the sign of the ratios is completely fixed: the two levels at the edges must, by definition, have both neighbors to the same side and hence $r>0$; the central level has one neighbor to each side and hence $r<0$; it follows that the area below the peak for negative $r$ is exactly $1/3$ and below positive $r$ is $2/3$ (the analytical expressions below confirm this reasoning exactly). As $N$ increases, the edge levels, which always have positive ratios, loose importance relative to the growing number of bulk levels, which tend to have negative ratios, and peak inversion follows. The argument for nonhermitian matrices is analogous: bulk levels favor large angles, boundary levels lead to small angles; at small $N$, boundary levels dominate, but they cannot compete in number with bulk levels at large $N$. Although it may seem, at first, that this strong $N$-dependence precludes one from obtaining small-size Wigner-like surmises, we will show that boundary effects can be overcome by considering the circular ensemble from a given universality class, and a new nonhermitian generalization, the toric ensemble, see Sections~\ref{section:hermitian_RMT} and \ref{section:ratio_nonhermitian}, respectively.

\section{Chaos and integrability in Liouvillian spectra}
\label{section:ratio_Louvillian}
The random Liouvillians considered in preceding chapters have, naturally, chaotic spectra, since they were constructed out of random matrices. To test complex spacing ratios, we apply them not only to the random Liouvillian but also to a family of Liouvillians describing boundary-driven spin chains, whose spectrum is integrable or chaotic depending on the values of the parameters of the model. We next briefly describe this family of Liouvillians.

\subsection{Boundary-driven spin chains}
Let us consider a chain of $N$ spins-$1/2$ described by Pauli operators $\sigma_\ell^\alpha$, $\alpha\in\{\mathrm{x},\mathrm{y},\mathrm{z}\}$, $\ell\in\{1,\dots,N\}$. We take the Hamiltonian $H$ from a family of next-to-nearest-neighbor Heisenberg XXZ Hamiltonians,
\begin{equation}
    H=J\sum_{\ell=1}^{N-1}\left(\sigma_\ell^\mathrm{x}\sigma_{\ell+1}^\mathrm{x}+\sigma_\ell^\mathrm{y}\sigma_{\ell+1}^\mathrm{y}+\Delta\sigma_\ell^\mathrm{z}\sigma_{\ell+1}^\mathrm{z}\right)+J'\sum_{\ell=1}^{N-2}\left(\sigma_\ell^\mathrm{x}\sigma_{\ell+2}^\mathrm{x}+\sigma_\ell^\mathrm{y}\sigma_{\ell+2}^\mathrm{y}+\Delta'\sigma_\ell^\mathrm{z}\sigma_{\ell+2}^\mathrm{z}\right),
\end{equation}
and we consider two types of incoherent jump processes (in total $r=N+4$):
\begin{enumerate}[(i)]
    \item dephasing of all (bulk) spins,
    \begin{equation}
        W_\ell=\sqrt{\gamma}\sigma_\ell^\mathrm{z},\quad\ell\in\{1,\dots,N\}\,;
    \end{equation}
    \item amplitude damping (spin polarization) processes at the boundaries,
    \begin{equation}
        W_{N+1}=\sqrt{\gamma_\mathrm{L}^+}\sigma_1^+\,,\quad 
        W_{N+2}=\sqrt{\gamma_\mathrm{L}^-}\sigma_1^-\,,\quad 
        W_{N+3}=\sqrt{\gamma_\mathrm{R}^+}\sigma_N^+\,,\quad
        W_{N+4}=\sqrt{\gamma_\mathrm{R}^-}\sigma_N^-\,.
    \end{equation}
\end{enumerate}
The model is, thus, characterized by the nine parameters $J,J',\Delta,\Delta',\gamma,\gamma_\mathrm{L,R}^\pm$, which allow us to tune its integrability or chaoticity.

We again formulate the spectral problem for the Liouvillian superoperator $\mathcal{L}$ in terms of its tensor-product representation, i.e.\ as a $4^N\times 4^N$ matrix acting on a $4^N$-dimensional Liouville space $\mathcal{K}$ of density operators $\rho$,
\begin{equation}\label{eq:spin_chain_Liouvillian}
\begin{split}
    \mathcal{L}&=-i\left\{\left(H-\frac{i}{2}\sum_{\mu=1}^rW_\mu^\dagger W_\mu\right)\otimes\mathbbm{1}-\mathbbm{1}\otimes\left(H+\frac{i}{2}\sum_{\mu=1}^rW_\mu^\dagger W_\mu\right)^{\sf{T}}\right\}+\sum_{\mu=1}^rW_\mu\otimes W_\mu^*\\
    &\equiv-i\left(K\otimes \mathbbm{1}-\mathbbm{1}\otimes K^*\right)+\sum_\mu W_\mu\otimes W_\mu^*\,,
\end{split}
\end{equation}
where we have again introduced the $2^N\times2^N$ nonhermitian Hamiltonian $K$.

The Hilbert space states are $\ket{s_1,\dots,s_N}$, the quantum numbers being the spin of each site, $s_\ell$. The total spin (magnetization) operator $S=\sum_{\ell=1}^N\sigma_\ell^\mathrm{z}$ acts on the states as $S\ket{s_1,\dots,s_N}=(\sum_{\ell=1}^Ns_\ell)\ket{s_1,\dots,s_N}\equiv s\ket{s_1,\dots,s_N}$, i.e.\ $\ket{s_1,\dots,s_N}$ are eigenstates of $S$ with eigenvalue $s$. The Liouville space states (density matrices) are $\kket{s_1,\dots,s_N;s'_1,\dots,s'_N}=\ket{s_1,\dots,s_N}\otimes\bra{s'_1,\dots,s'_N}^{\sf{T}}$. The total spin superoperator $\mathcal{S}=\comm{S}{\cdot\,}=S\otimes \mathbbm{1}-\mathbbm{1}\otimes S^{\sf{T}}$ acts on the states as $\mathcal{S}\kket{s_1,\dots,s_N;s'_1,\dots,s'_N}=(s-s')\kket{s_1,\dots,s_N;s'_1,\dots,s'_N}\equiv N-M\kket{s_1,\dots,s_N;s'_1,\dots,s'_N}$, i.e.\ $\kket{s_1,\dots,s_N;s'_1,\dots,s'_N}$ are eigenstates of $\mathcal{S}$ with eigenvalue $N-M$ (the magnetization). 

In the Hilbert space, the total spin magnetization is conserved by all bulk processes, both the unitary ones, $\comm{H}{S}=0$, and by the dephasing noise, $\comm{W_\ell}{S}=0$. Additionally, the Liouvillian satisfies a so-called Liouvillian weak symmetry~\cite{buca2012} in Liouville space, $\comm{\mathcal{L}\left(\rho\right)}{S}=\mathcal{L}\left(\comm{\rho}{S}\right)$, which can also be stated as the spin superoperator being conserved by Liouvillian evolution $\comm{\mathcal{L}}{\mathcal{S}}=0$.

The Liouvillian weak symmetry implies that the Liouville space $\mathcal{K}$ is split into sectors $\mathcal{K}_M$ of conserved quantum number $M$, each $\mathcal{K}$ spanned by $\binom{2N}{M}$ states $\kket{s_1,\dots,s_N;s'_1,\dots,s'_N}$ with $s-s'=\sum_\ell(s_\ell-s'_\ell)=N-M$. The tensor-product representation of the Liouvillian block-diagonalizes into $2N+1$ sectors $\mathcal{L}_M$, $\mathcal{L}=\bigoplus_{M=0}^{2N} \mathcal{L}_M$, each block an $\binom{2N}{M}\times\binom{2N}{M}$ matrix. The symmetric sector $M=N$ contains all states with vanishing magnetization (as many up spins as down spins), including the steady-state. Each complex conjugate pair of eigenvalues of the Liouvillian is divided across two sectors of symmetric magnetization, i.e.\ if sector $M$ contains the eigenvalue $\Lambda$, then sector $\abs{N-M}$ contains the eigenvalue $\Lambda^*$. The different sectors $M$ must be analyzed separately because spectra corresponding to different conserved quantum numbers form independent level sequences which superimpose without interacting~\cite{guhr1998}. By proceeding this way, we guaranty that all states belong to the same symmetry class.

The numerical results were obtained by diagonalizing the tensor-product representation, Eq.~(\ref{eq:spin_chain_Liouvillian}), for different chain lengths $N$ and in specific sectors $M$. The largest system we diagonalized was $N=10$ spins in the sector with magnetization $M=7$, which corresponds to a $77520\times 77520$ matrix. Note that, here, no ensemble averaging is to be performed since the systems are not random.

\begin{figure}[tbp]
    \centering
    \includegraphics[width=0.99\textwidth]{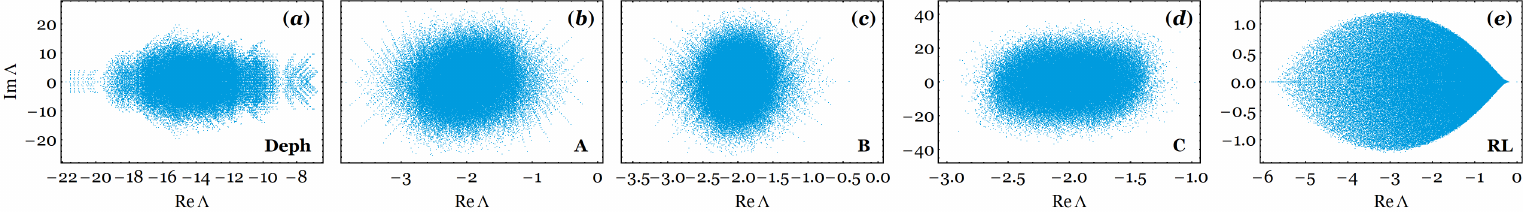}
    \caption{Liouvillian spectra for the five different models. The boundary driven spin-chains are diagonalized in the sector with $N=10$ and $M=7$. $(a)$ Deph--boundary driven XX chain with bulk dephasing; $(b)$ A--XXX chain with pure-source/pure-sink driving; $(c)$ B--XXX chain with arbitrary polarizing boundary driving; $(d)$ C--XXZ chain with nearest neighbor and next-to-nearest neighbor interactions; $(e)$ RL--random Liouvillian at strong dissipation.}
    \label{fig:Liouvillian_spectra}
\end{figure}

The following four cases of parameters were studied:
\begin{itemize}
    \item \textbf{(Deph)} Boundary driven XX chain with bulk dephasing. Numerical parameters chosen as $J=1$, $J'=\Delta=\Delta'=0$, $\gamma=1$, $\gamma_\mathrm{L}^+=0.5$, $\gamma_\mathrm{L}^-=1.2$, $\gamma_\mathrm{R}^+=1$, $\gamma_\mathrm{R}^-=0.8$. This model can be mapped onto the Fermi-Hubbard model with imaginary interaction $U=i\gamma$~\cite{medvedyeva2016} and hence is Bethe ansatz integrable.
    \item \textbf{(A)} Isotropic Heisenberg (XXX) chain  with pure-source/pure-sink driving and no dephasing. Numerical parameters chosen as $J=\Delta=1$, $J'=\Delta'=0$, $\gamma=\gamma_\mathrm{L}^-=\gamma_\mathrm{R}^+=0$, $\gamma_\mathrm{L}^+=0.6$, $\gamma_\mathrm{R}^-=1.4$. The steady-state of this model is known to be integrable~\cite{prosen2011}, but the bulk of the spectrum is expected to be nonintegrable.
    \item \textbf{(B)} XXX chain  with arbitrary boundary-driving and no dephasing. Numerical parameters chosen as $J=\Delta=1$, $J'=\Delta'=0$, $\gamma=0$, $\gamma_\mathrm{L}^+=0.5$, $\gamma_\mathrm{L}^-=0.3$ $\gamma_\mathrm{R}^+=0.3$, $\gamma_\mathrm{R}^-=0.9$. The bulk Hamiltonian of this model is integrable, but, by adding the boundary-driving, not even the steady-state is expected to be exactly-solvable.
    \item \textbf{(C)} XXZ chain with next-to-nearest-neighbor interactions, arbitrary boundary-driving, and no dephasing. Numerical parameters chosen as $J=J'=1$, $\Delta=0.5$, $\Delta'=1.5$, $\gamma=0$, $\gamma_\mathrm{L}^+=0.5$, $\gamma_\mathrm{L}^-=0.3$ $\gamma_\mathrm{R}^+=0.3$, $\gamma_\mathrm{R}^-=0.9$. For this model, not even the bulk Hamiltonian is integrable.
\end{itemize}
Additionally, we considered a fifth model for comparison:
\begin{itemize}
    \item \textbf{(RL)} Random Liouvillian, Eq.~(\ref{eq:Liouvillian_tensor}), at strong dissipation. Numerical parameters chosen as $N=80$, $\beta=2$, $r=2$, $g=100$.
\end{itemize}
The spectra of the four boundary-driven spin-chains, for the case of $N=10$ spins in the sector of magnetization $M=10$, and of the random Liouvillian, are depicted in Fig.~\ref{fig:Liouvillian_spectra}. Note that the spectra of the driven spin-chains, Figs.~\ref{fig:Liouvillian_spectra} $(a)$--$(d)$, are not symmetric around the real axis, since the complex conjugate pairs are \emph{not} in the same sector. The exception is the dephasing model, Fig.~\ref{fig:Liouvillian_spectra}~$(a)$, for which all eigenstates with nonzero magnetization are doubly degenerate.

\subsection{Complex spacing ratios for Liouvillian spectra}
\begin{figure}[htbp]
    \centering
    \includegraphics[width=0.99\textwidth]{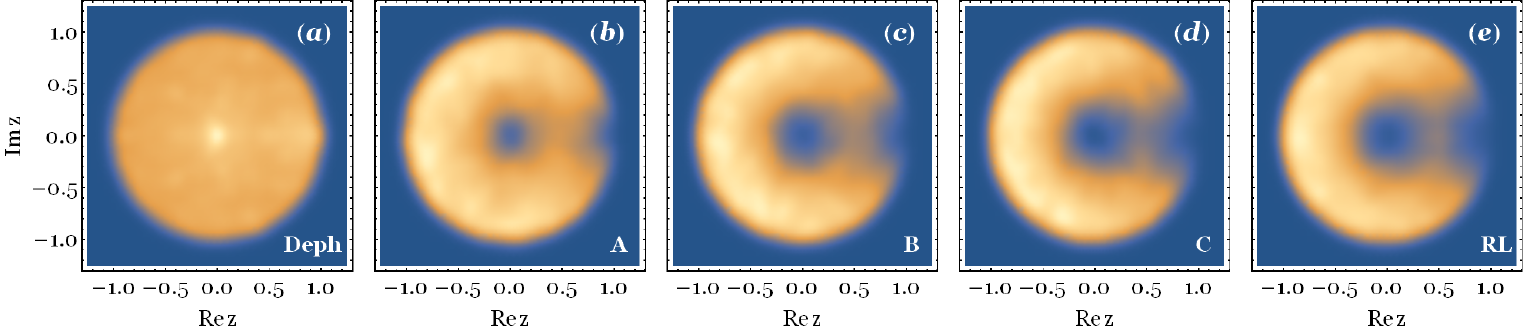}
    \caption{Complex spacing ratio density for different Liouvillian spectra. $(a)$ Deph--boundary driven XX chain with bulk dephasing; $(b)$ A--XXX chain with pure-source/pure-sink driving; $(c)$ B--XXX chain with arbitrary polarizing boundary driving; $(d)$ C--XXZ chain with nearest neighbor and next-to-nearest neighbor interactions; $(e)$ RL--random Liouvillian at strong dissipation.}
    \label{fig:Liouvillian_ratio}
\end{figure}
Applying the numerical procedure described at the start of Section~\ref{section:ratio_definitions_overview}, we computed the distribution of the complex spacing ratios for the five spectra of Fig.~\ref{fig:Liouvillian_spectra}, see Fig.~\ref{fig:Liouvillian_ratio}. It shows a striking difference between known integrable and chaotic models and indicates that complex spacing ratios indeed offer a clean signature of quantum chaos. The dephasing-XX model, Fig.~\ref{fig:Liouvillian_ratio}~$(a)$, displays Poisson statistics arising from its known integrability. Models B, C and RL, Figs.~\ref{fig:Liouvillian_ratio}~$(c)$, $(d)$, $(e)$, respectively, clearly have RMT statistics (in the former two because of their chaotic nature, in the latter by construction). Model A, Fig.~\ref{fig:Liouvillian_ratio}~$(b)$, on the other hand, shows an intermediate behavior between Poisson and RMT statistics, both in terms of radial level repulsion and of anisotropy of the angular distribution. This could arise either from actual intermediate statistics of the spectrum or from finite-size effects. The latter are investigated in the next section, where we also analyze the peak at the origin for the dephasing-XX model, which is absent for Poisson processes. On the contrary, model C already displays the universal large-$N$ behavior, with no noticeable finite-size effects, as will be confirmed in Section~\ref{subsection:ratio_finite_size} below.

We next try to capture the main features of the distribution of complex spacing ratios through a reduced set of numbers, which we call \emph{single-number signatures}. A popular single-number signature, used for the ratio of undirected spacings, is the degree of level repulsion $\alpha$, i.e.\ the exponent describing the power-law behavior of the radial marginal distribution, $\varrho(r)\propto r^\alpha$, as $r\to0$ or, equivalently, $\alpha=\lim_{r\to0}\log\varrho(r)/\log r$. For hermitian random matrices it is given by the Dyson index, $\alpha=\beta$, for nonhermitian random matrices from the universality class of either GinOE, GinUE, or GinSE it is $\alpha=3$, while for real Poisson processes it is $\alpha=0$, and for complex Poisson processes $\alpha=1$. Although the degrees of repulsion $\alpha$ just stated can be easily checked against the numerical spectra and the above predictions confirmed, an actual computation of $\alpha$ for a given spectrum introduces a large relative error. An alternative measure of the radial distribution is its moments, for instance, the mean $\av{r}$. For Poisson processes, we can compute exactly $\av{r}=2/3$ (see Section~\ref{section:ratio_Poisson}), while for GinUE matrices we numerically find $\av{r}\approx0.74$. To measure the anisotropy of the angular marginal distribution, we consider $\av{\cos\theta}$, which is zero for a flat distribution and positive (negative) when small angles are enhanced (suppressed), in particular, $\av{\cos\theta}\approx-0.24$ for large-$N$ GinUE matrices. 

We give the values of $\av{\cos\theta}$ and $\av{r}$ for the five Liouvillians in Table~\ref{tab:Liouvillian_signatures}. From the radial measure $\av{r}$ it is difficult to discern the integrability or chaoticity of the different models. Indeed, the values for all four models A, B, C, RL are within $3\%$ of each other, despite the different degrees of chaos. On the other hand, as anticipated in Section~\ref{section:ratio_definitions_overview}, the angular distribution offers a more sensitive signature. From the value of $\av{\cos\theta}$, the dephasing-XX model clearly supports Poisson statistics and models C and RL are very close to RMT statistics. Model B, which seemed also very close to RMT statistics from Fig.~\ref{fig:Liouvillian_ratio} and from the value of $\av{r}$ here shows a more significant deviation. Finally, model A has a value of $\av{\cos\theta}$ almost exactly halfway between Poisson and RMT statistics, attesting its intermediate behavior, at least for the sector dimensions considered.

\begin{table}[tbp]
    \centering
    \caption{Single-number signatures of integrability/chaos for different Liouvillians, models Deph, A, B, C, RL. They are compared with exact analytical results for Poisson processes (see Section~\ref{section:ratio_Poisson}) and numerical results for random GinUE matrices.}
    \label{tab:Liouvillian_signatures}
    \vspace{+0.7ex}
    \resizebox{\textwidth}{!}{
    \begin{tabular}{ccllllll}
        \hline
        & Poisson & \multicolumn{1}{c}{Deph} & \multicolumn{1}{c}{A} & \multicolumn{1}{c}{B} & \multicolumn{1}{c}{C} & \multicolumn{1}{c}{RL} & \multicolumn{1}{c}{Ginibre}  \\
        \hline
        $-\langle\cos\theta\rangle$ &
        $0$ & $-0.0305(26)$ & $0.1293(24)$ & $0.1890(23)$ & $0.2349(7)$ & $0.2287(20)$ & $0.24051(61)$ \\
        $\langle r \rangle$ &
        $2/3$ & \ \ \,$0.6537(9)$ & $0.7122(7)$ & $0.7292(7)$ & $0.7368(7)$ & $0.7373(6)$ & $0.73810(18)$ \\
        \hline
    \end{tabular}
    }
\end{table}

\subsection{Finite-size analysis}
\label{subsection:ratio_finite_size}
We now provide a finite-size analysis of the dephasing-XX model, which conforms to Poisson statistics, model A, with intermediate statistics, and model C, with RMT statistics.

\subsubsection{\textit{Dephasing-XX model}}
\begin{figure}[htbp]
    \centering
    \includegraphics[width=\textwidth]{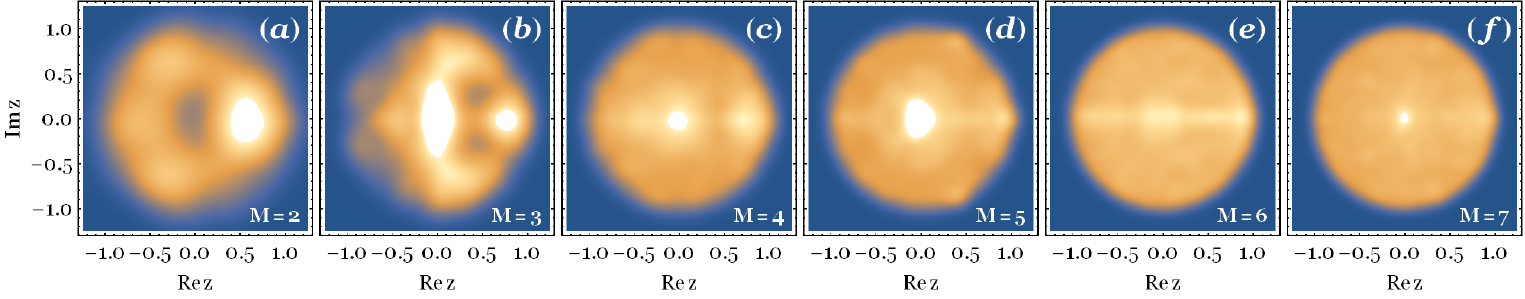}
    \caption{Complex spacing ratio density for spectra of Liouvillians from the dephasing-XX model, for $N=10$ spins and various spin sectors $M$. The sector dimensions are $(a)$ $k_{N\!M}=190$, $(b)$ $k_{N\!M}=1140$, $(c)$ $k_{N\!M}=4845$, $(d)$ $k_{N\!M}=15504$, $(e)$ $k_{N\!M}=38760$, $(f)$ $k_{N\!M}=77520$.}
    \label{fig:LDeph_ratio_M_N10}
\end{figure}

\begin{figure}[htbp]
    \centering
    \includegraphics[width=\textwidth]{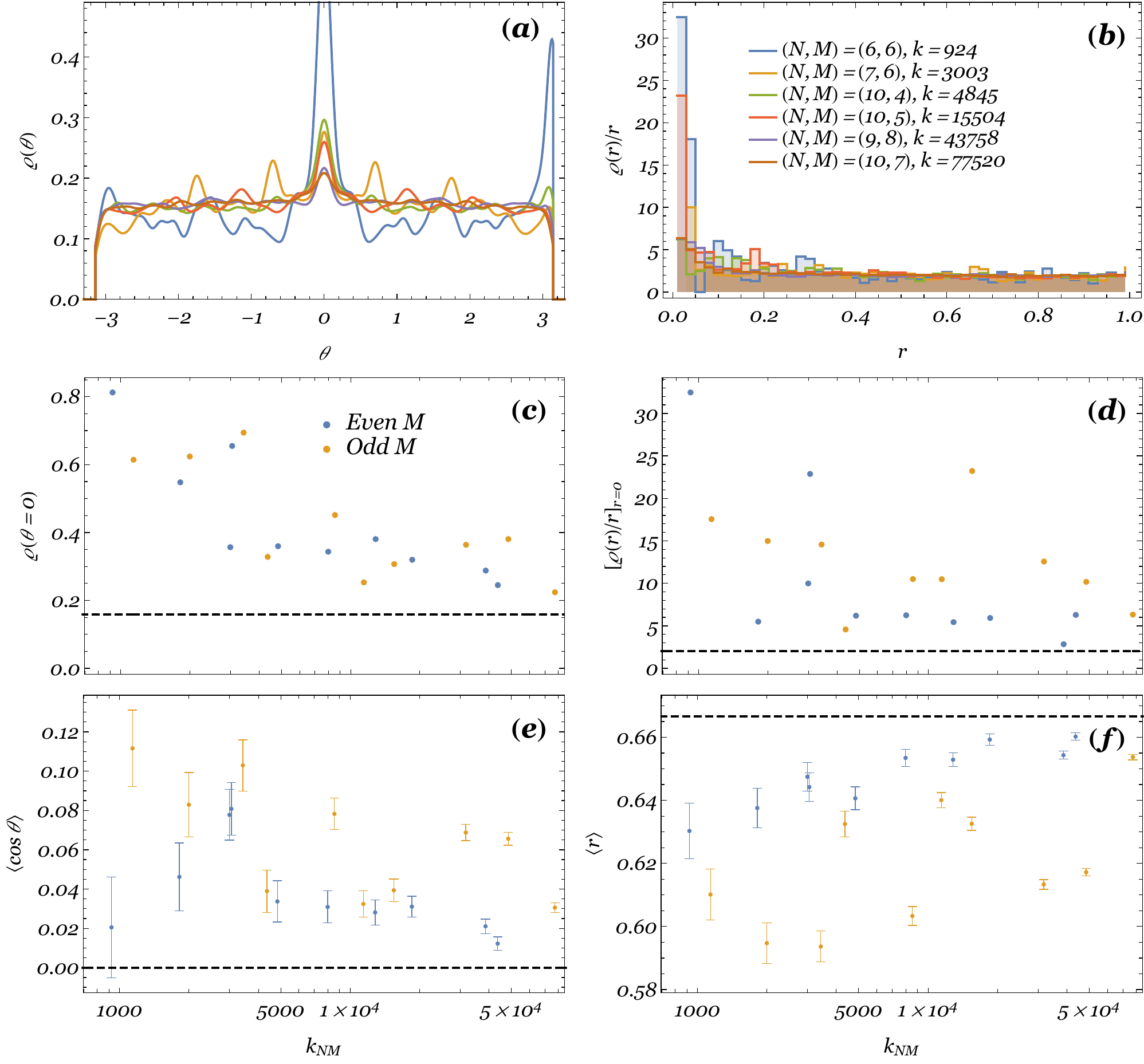}
    \caption{Finite-size effects on the complex spacing ratios of a spin-chain Liouvillian of the dephasing-XX model, for different chain lengths $N$ and spin sectors $M$, the sector dimension being $k_{N\!M}=\binom{2N}{M}$. $(a)$: angular marginal distribution; $(b)$: radial marginal distribution normalized by $r$ to flatten it out; $(c)$: height of the peak at $\theta=0$; $(d)$: height of the peak at $r=0$; $(e)$: average value of $\cos\theta$; $(f)$: average value of $r$. In $(c)$--$(f)$, blue points correspond to even $M$, orange points to odd $M$, dashed black line to the limiting value for Poisson processes.}
    \label{fig:LDeph_finite_size}
\end{figure}

To see the dependence of the spectral statistics on sector dimension, we plot, in Fig.~\ref{fig:LDeph_ratio_M_N10}, the complex spacing ratio density for the Liouvillian of the dephasing-XX model with increasing $M$ (and hence increasing $k_{N\!M}=\binom{2N}{M}$) at fixed $N=10$. Once again, we see that universality only emerges at large dimension, with the small sector distributions not close to neither Poisson nor RMT statistics. As the sector dimension increases, the distribution tends to Poisson statistics (recall Fig.~\ref{fig:Poisson_vs_GinUE}~$(a)$) although some non-Poisson features still survive, namely, an accumulation of ratios near the origin ($r=0$) and along the real axis ($\theta=0$). In the limit of infinitely large spectra, the complex spacing ratio statistics should conform exactly to those of Poisson processes.

A more quantitative analysis can be provided by looking at single-number signatures as a function of sector dimension. In Figs.~\ref{fig:LDeph_finite_size}~$(a)$ and $(b)$ we plot the angular and radial marginal distributions, respectively (the latter is divided by $r$ such that it becomes flat and the small-$r$ effects are amplified). We see that, indeed, there is a sharp peak around $\theta=0$ and at $r=0$. However, the heights of both peaks decrease when the dimension of the sector in which the Liouvillian is diagonalized increases. This can be seen in Figs.~\ref{fig:LDeph_finite_size}~$(c)$ and $(d)$. Figures~\ref{fig:LDeph_finite_size}~$(e)$ and $(f)$ depict the single-number signatures $\av{\cos\theta} $ and $\av{r}$, respectively, which clearly tend to the expected value for Poisson processes (dashed line) as $k_{N\!M}$ increases. In the latter two plots, there is also a visible difference between sectors with even or odd $M$, with sectors of even $M$ tending faster to the large-dimension universal limit. This aspect is also visible in Fig.~\ref{fig:LDeph_finite_size}~$(d)$.

\subsubsection{\textit{Model A}}
\begin{figure}[htbp]
    \centering
    \includegraphics[width=\textwidth]{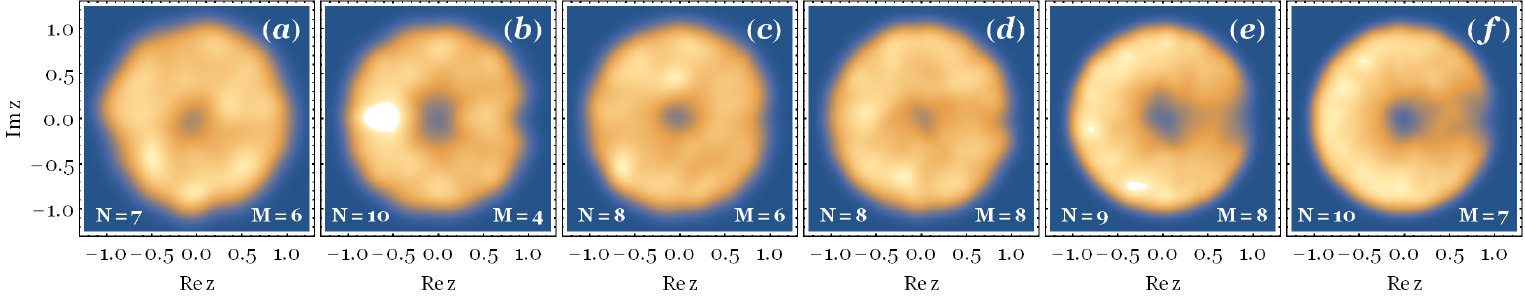}
    \caption{Complex spacing ratio density for spectra of Liouvillians from model A, for various chain lengths $N$ and spin sectors $M$. The sector dimensions are $(a)$ $k_{N\!M}=3003$, $(b)$ $k_{N\!M}=4845$, $(c)$ $k_{N\!M}=8008$, $(d)$ $k_{N\!M}=12870$, $(e)$ $k_{N\!M}=43758$, $(f)$ $k_{N\!M}=77520$.}
    \label{fig:LA_ratio_NM}
\end{figure}

Figure~\ref{fig:LA_ratio_NM} shows the complex spacing ratio density for different sector dimensions. One immediately sees that there is a smaller degree of level repulsion than for fully chaotic systems and which does not increase substantially when $k_{N\!M}$ grows by nearly two orders of magnitude. Some anisotropy is developing as $k_{N\!M}$ increases but even for $k_{N\!M}=77520$, the density along positive real semi-axis is still far from zero. 

\begin{figure}[tbp]
    \centering
    \includegraphics[width=0.8\textwidth]{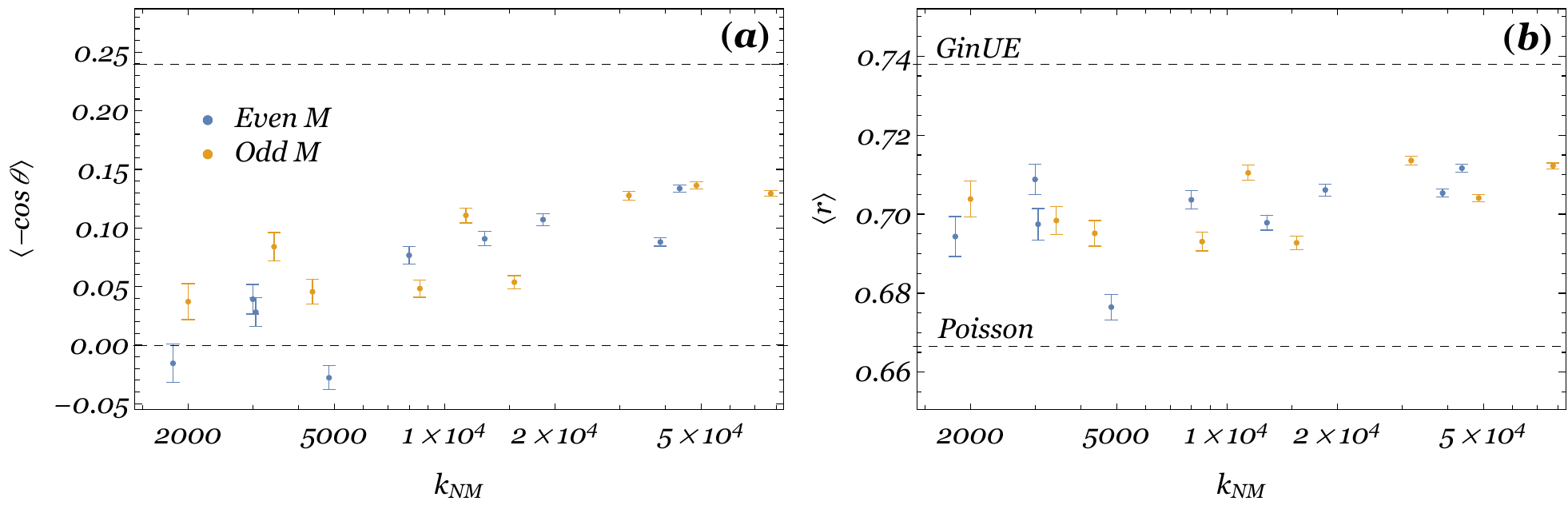}
    \caption{Finite-size effects on the complex spacing ratios of a spin-chain Liouvillian of model A. We consider different chain lengths $N$ and spin sectors $M$, the sector dimension being $k=\binom{2N}{M}$. $(a)$: average value of $\cos\theta$; $(b)$: average value of $r$. The upper (lower) dashed line corresponds to the GinUE- (Poisson-) limit.}
    \label{fig:LA_finite_size}
\end{figure}

Both observations above are confirmed from Fig.~\ref{fig:LA_finite_size}, where we plot the two single-number signatures $\av{\cos\theta}$ and $\av{r}$. While the anisotropy indeed grows (slowly) with $k_{N\!M}$, the average of the radial marginal is approximately flat. No difference between even $M$ and odd $M$ is visible in this case. Contrary to the dephasing-XX model above, for which the $N=10$, $M=7$ sector is already very close to the limiting GinUE statistics, the convergence of model A towards either Poisson or GinUE statistics is much slower. 

From these results it is, therefore, inconclusive whether the model is tending very slowly to RMT statistics (as favored by Fig.~\ref{fig:LA_finite_size}~$(a)$) or if it follows some type of intermediate statistics. Considerably larger sector dimensions are, unfortunately, out of reach of current computational capabilities.

\subsubsection{\textit{Model C}}
\begin{figure}[htbp]
    \centering
    \includegraphics[width=\textwidth]{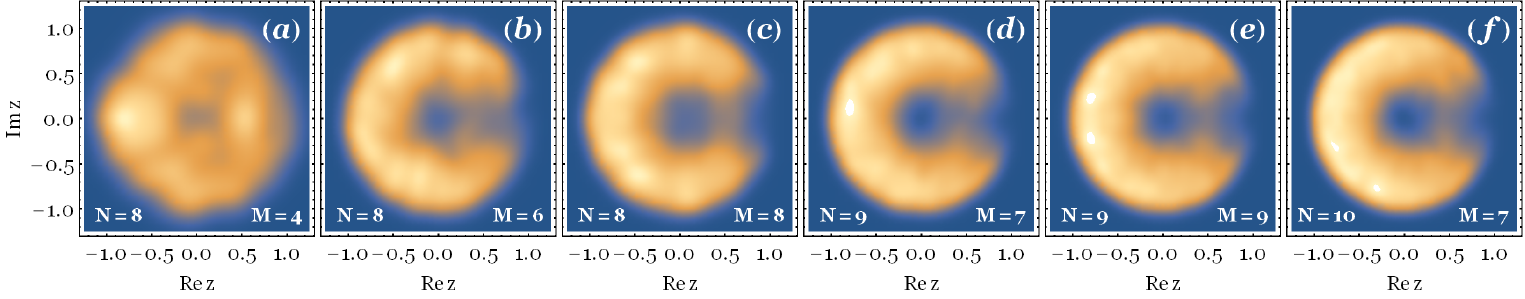}
    \caption{Complex spacing ratio density for spectra of Liouvillians from model C, for various chain lengths $N$ and spin sectors $M$. The sector dimensions are $(a)$ $k_{N\!M}=1820$, $(b)$ $k_{N\!M}=8008$, $(c)$ $k_{N\!M}=12870$, $(d)$ $k_{N\!M}=31824$, $(e)$ $k_{N\!M}=48620$, $(f)$ $k_{N\!M}=77520$.}
    \label{fig:LC_ratio_NM}
\end{figure}

\begin{figure}[htbp]
    \centering
    \includegraphics[width=0.8\textwidth]{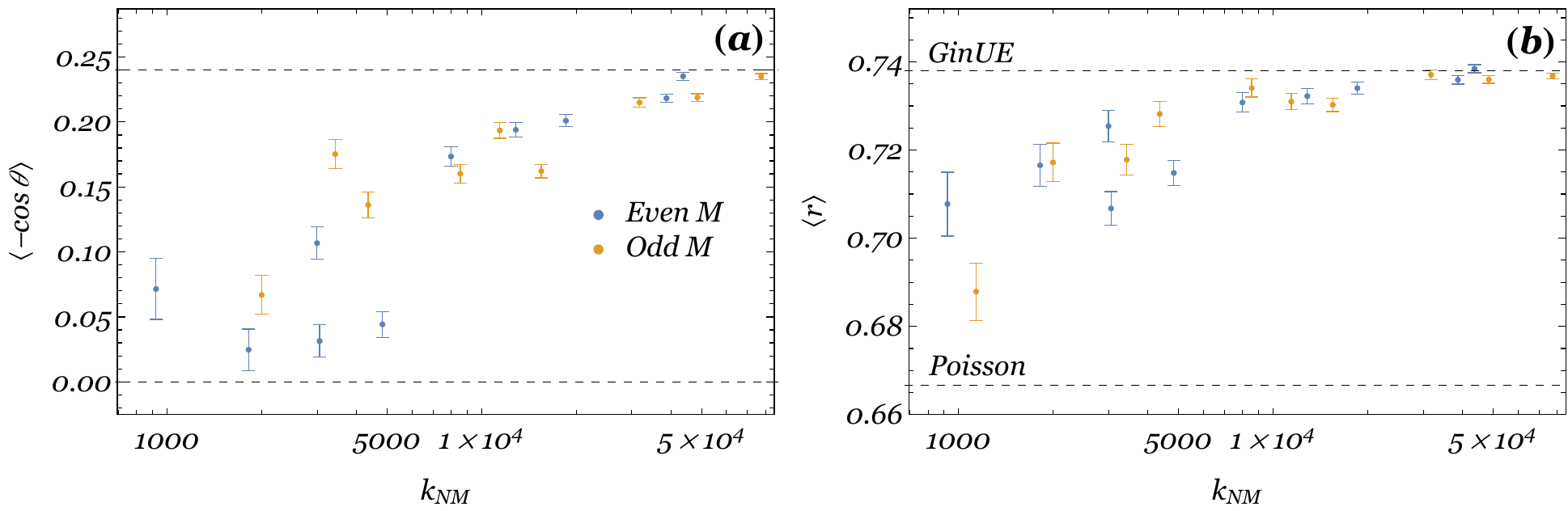}
    \caption{Finite-size effects on the complex spacing ratios of a spin-chain Liouvillian of model C, as a function of sector dimension $k_{N\!M}$ $(a)$ average value of $\cos\theta$; $(b)$: average value of $r$. The upper (lower) dashed line corresponds to the GinUE- (Poisson-) limit.}
    \label{fig:LC_finite_size}
\end{figure}

Finally, we consider a chaotic Liouvillian, model C. The complex spacing ratio density for different sector dimensions is shown in Fig.~\ref{fig:LC_ratio_NM}. Here, the universal limit of RMT statistics is quickly attained. Figure~\ref{fig:LC_finite_size} depicts the two single-number signatures $\av{\cos\theta}$ and $\av{r}$ and confirms the fast convergence. For the largest sectors diagonalized, the results are already compatible, within their errors, with the infinite-size limit.

\section{Theoretical analysis I.\ Poisson processes}
\label{section:ratio_Poisson}

As discussed before, the independence of the levels of a Poisson process implies that all complex spacings are equiprobable. It follows that the ratio distribution is isotropic and, more specifically, flat. Taking isotropy as a starting point, i.e.\ assuming that the distribution only depends on the absolute value of the ratio, $r$, we now show that it is, indeed, flat. Furthermore, the independence of the levels simplifies the problem enough so that we are able to compute not only the spacing ratio we have described in Section~\ref{section:ratio_definitions_overview}, but also the more general ratio of the distance to the $m^\mathrm{th}$-nearest-neighbor ($m$NN) by the distance to the $k^\mathrm{th}$-nearest-neighbor ($k$NN), which reduces to the previous ratio when $m=1$, $k=2$. We do all calculations in $d$-dimensional space. Real, complex, and quaternionic spectra correspond to $d=1,2,4$, respectively, but our results also apply to other cases, say, uncorrelated random vectors in three-dimensional space.

\subsection{Joint spacing distributions}
\label{subsection:ratio_joint_spacings}
By translational invariance, we can consider the level for which the ratio is being computed (the reference level) at the origin. To compute the probability $\hat{P}(s)\d s$ of finding its NN at a distance $s$, we introduce the conditional probability $g(s)\d s$ of finding a level in $[s,s+\d s]$ given our reference level at the origin, and the probability $H(s)=\int_s^\infty\d s'\hat{P}(s')=1-\int_0^s\d s'\hat{P}(s')$ of having no level in $[0,s]$ (the hole probability). By independence of Poisson levels, the probability $g(s)\d s$ is actually independent of the presence of the reference level. For the NN to be at $s$ we must verify that \textit{(i)} there is a level at $s$ and \textit{(ii)} there are no levels in $[0,s]$, whence we conclude that
\begin{equation}\label{eq:P=gH}
    \hat{P}(s)\d s=g(s)\d s\,H(s)\,,\quad \text{i.e.} \quad  \hat{P}(s)=g(s)\,H(s)\,.
\end{equation}
Noting that the hole probability is equal to $1-F(s)$, where $F(s)$ is the cumulative distribution of $\hat{P}(s)$, we can equally well express $\hat{P}(s)$ solely in terms of $H(s)$, 
\begin{equation}\label{eq:P=f(H)}
    \hat{P}(s)=-\frac{\d H}{\d s}\,.
\end{equation}
Alternatively, by inserting Eq.~(\ref{eq:P=f(H)}) into Eq.~(\ref{eq:P=gH}), we obtain
\begin{equation}\label{eq:relation_g_and_H}
    g(s)=-\frac{1}{H}\frac{\d H(s)}{\d s}=-\frac{\d\log H}{\d s}\Rightarrow H(s)\propto e^{-\int_0^s\d s'g(s')}\,,
\end{equation}
whence we can express $\hat{P}(s)$ solely in terms of $g(s)$ as
\begin{equation}\label{eq:P=f(g)}
    \hat{P}(s)\propto g(s)\,e^{-\int_0^s\d s'g(s')}\,.
\end{equation}

Now, Eq.~(\ref{eq:P=gH}) is easily generalized to give the joint distribution of the NN- and NNN-spacing, i.e.\ of the probability density $\hat{P}(s_1,s_2)$ of having the NN at a distance $s_1$ and the NNN at a distance $s_2$, which we need to compute the distribution of their ratio. It is given by considering one level each at $s_1$ and $s_2$ and all remaining levels beyond $s_2$, i.e.
\begin{equation}\label{eq:NN_NNN_joint_spacing}
    \hat{P}(s_1,s_2)=g(s_1)\,g(s_2)\heav{s_2-s_1}H(s_2).
\end{equation}
Analogously, the joint distribution of the first-$k$NN spacings is 
\begin{equation}\label{eq:kNN_joint_spacings}
    \hat{P}(s_1,\dots,s_k)=\prod_{j=1}^kg(s_j)\heav{s_{j+1}-s_j}H(s_k)\,.
\end{equation}

It is worthwhile to note that, using Eqs.~(\ref{eq:P=f(H)}), (\ref{eq:relation_g_and_H}), (\ref{eq:P=f(g)}), we can express the whole hierarchy of joint probabilities solely in terms of the single-variable functions $\hat{P}(s)$, $g(s)$ or $H(s)$, whichever is easier to compute in a given situation. This is, of course, a particularity of Poisson processes, and does not carry over to random matrix ensembles.

The distribution of the (absolute value) of the ratio $r=s_1/s_2$ is given in terms of the joint distribution $\hat{P}(s_1,s_2)$, and, hence, it is also completely determined by the single spacing distribution $\hat{P}(s)$:
\begin{equation}\label{eq:P(r)_definition}
    \begin{split}
        \varrho(r)&=\int \d s_1\d s_2\, \hat{P}(s_1,s_2)\dirac{r-\frac{s_1}{s_2}}\\
        &=\int \d s\, s\, \hat{P}(rs,s)\\
        &=\Theta(1-r)\int_0^\infty \d s\, \frac{s\,\hat{P}(s)\,\hat{P}(rs)}{\int_{rs}^\infty\d s'\, \hat{P}(s')}\,.
    \end{split}
\end{equation}
In the last line, we have expressed the ratio distribution solely in terms of the single spacing probability. It only remains to compute one of $\hat{P}(s)$, $H(s)$, $g(s)$, which, as alluded above, we will do in $d$-dimensions in the next section. On the real line, we will recover the standard result, $\hat{P}(s)=\exp{-s}$ (on the unfolded scale).

Likewise, the $m$NN by $k$NN ratio, $r_{mk}\equiv s_m/s_k$, is defined in terms of the joint spacing distribution $\hat{P}\left(s_1,\dots,s_k\right)$ and is fully determined by the single spacing distribution:
\begin{equation}\label{eq:ratio_r_mk_def}
\begin{split}
    \varrho_{mk}\left(r_{mk}\right)
    &=\int\d s_1\cdots\d s_m\cdots \d s_k\,\hat{P}\left(s_1,\dots,s_m,\dots,s_k\right)\dirac{r-\frac{s_m}{s_k}}\\
    &=\int\d s_1\cdots\d s_{m-1}\d s_{m+1}\cdots \d s_k\,\hat{P}\left(s_1,\dots,s_{m-1},rs_k,s_{m+1},\dots,s_k\right)\\
    &=\heav{1-r}\int_0^\infty\d s_1\cdots\d s_{m-1}\d s_{m+1}\cdots \d s_k\frac{s_k\hat{P}\left(s_1\right)\cdots\hat{P}\left(s_{m-1}\right)}{\int_{s_1}^\infty\d s_1'\hat{P}\left(s'_1\right)\cdots\int_{s_{m-1}}^\infty\d s_{m-1}'\hat{P}\left(s'_{m-1}\right)}\\
    &\times\frac{\hat{P}\left(rs_k\right)\hat{P}\left(s_{m+1}\right)\cdots\hat{P}\left(s_k\right)}{\int_{rs_k}^\infty\d s'\hat{P}\left(s'\right)\int_{s_{m+1}}^\infty\d s_{m+1}'\hat{P}\left(s'_{m+1}\right)\cdots\int_{s_{k-1}}^\infty\d s_{k-1}'\hat{P}\left(s'_{k-1}\right)}\\
    &\times\heav{s_{k}-s_{k-1}}\cdots\heav{s_{m+1}-rs_k}\heav{rs_k-s_{m-1}}\cdots\heav{s_2-s_1}.
\end{split}
\end{equation}

\subsection{Poisson processes in $d$-dimensions}
\begin{figure}[htbp]
    \centering
    \includegraphics[width=0.45\textwidth]{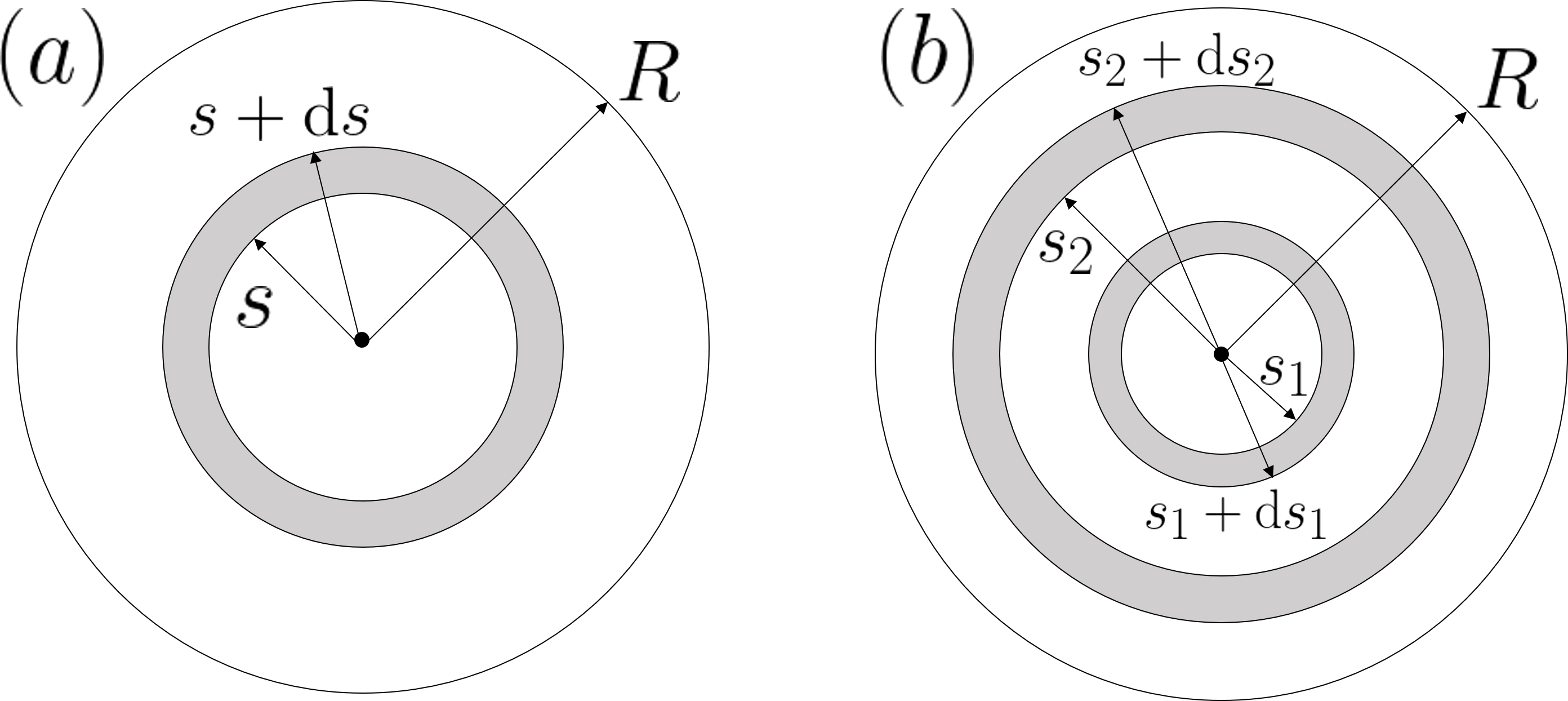}
    \caption{Schematic setup for the computation of spacing distributions for Poisson processes, here in $d=2$ (but straightforwardly generalized for $d\neq2$). $(a)$: setup for computation of the single spacing density $\hat{P}(s)$, with one level at the origin (reference level), no levels in a ball of radius $s$, one level (NN) in the gray shell of radii $s$ and $s+\d s$ and all remaining levels in the outer shell, between $s$ and $R\to\infty$; $(b)$: setup for computation of $\hat{P}(s_1,s_2)$.}
    \label{fig:poisson_sketch}
\end{figure}

We consider the Poisson process to be composed by $N$ iid random variables, supported in a $d$-dimensional ball of radius $R$. At a later point, we shall take the limits $N,R\to\infty$ with constant mean density $NR^{-d}=1$. The probabilities $g(s)\d s$ and $H(s)$ are then given by ratios of $d$-dimensional volumes $V_d(L)=\pi^{d/2}/\Gamma(d/2+1)L^d$, where $L$ is a length, see the schematic representation of Fig.~\ref{fig:poisson_sketch}.

To determine $g(s)$, consider the setup of Fig.~\ref{fig:poisson_sketch}~$(a)$. We note that any one of $N-1$ levels can be the NN if it falls inside the interval $[s,s+\d s]$, whence it follows that
\begin{equation}\label{eq:Poi_g_finiteN}
    g(s)\d s=(N-1)\frac{V_d(s+\d s)-V_d(s)}{V_d(R)}=\frac{N-1}{R^d}d\,s^{d-1}\d s+\mathcal{O}(\d s^2)\,.
\end{equation}
Taking the limits $N,R\to\infty$, we obtain immediately that $g(s)\propto s^{d-1}$.

Regarding $H(s)$, since all other $(N-2)$ levels are independent and must lie beyond a distance $s$, we have
\begin{equation}\label{eq:Poi_H_finiteN}
    H(s)=\left(1-\frac{V_d(s)}{V_d(R)}\right)^{N-2}=\left(1-\frac{s^d}{R^d}\right)^{N-2}\,.
\end{equation}

To be able to properly take limits, we need to unfold the spectrum to unit mean, i.e.\ we change variables to $\mathfrak{s}=s/\langle s\rangle$. Note that in the computation of $g(s)$ the unfolding would only give an overall constant, so we did not need it to proceed. Using Eqs.~(\ref{eq:P=gH}), (\ref{eq:Poi_g_finiteN}), (\ref{eq:Poi_H_finiteN}), we have
\begin{equation}
    \hat{P}(s)=d\,\frac{N-1}{R^d}s^{d-1}\left(1-\frac{s^d}{R^d}\right)^{N-2}\,,
\end{equation}
which is correctly normalized as it should. We then have 
\begin{equation}\label{eq:<s>}
    \langle s \rangle=\int_0^\infty\d s\, s\, \hat{P}(s)=\Gamma(1+1/d)\frac{\Gamma(N)}{\Gamma(N+1/d)}R\xrightarrow[N\to\,\infty]{}\Gamma(1+1/d)N^{-1/d}R\,,
\end{equation}
where we have used the asymptotic behavior of the gamma function, $\lim_{N\to\infty}N^\alpha\Gamma(N)/\Gamma(N+\alpha)=1$ for any $\alpha\in\mathbb{C}$.

In terms of the unfolded variable $\s$, the hole probability reads
\begin{equation}\label{eq:Poi_unfolded_H}
  H(\s)=\left(1-\frac{\Gamma(1+1/d)^d\,\s^d}{N}\right)^{N-2}\xrightarrow[N\to\,\infty]{}e^{-\Gamma(1+1/d)^d\,\s^d}\,,
\end{equation}
and the (unfolded) spacing distribution is given by
\begin{equation}\label{eq:Poi_unfolded_P}
    \hat{P}(\s)=d\,\Gamma(1+1/d)^d\s^{d-1}e^{-\Gamma(1+1/d)^d\,\s^d}\,.
\end{equation}
Note that, for $d=1$, we recover the standard exponential distribution, $\hat{P}(\s)=e^{-\s}$. In $d$-dimensions, the spacing follows, instead, a Brody distribution~\cite{brody1973}.

Although the NN- and NNN-joint spacing distribution $\hat{P}\left(\s_1,\s_2\right)$ can be written solely in terms of the single spacing distribution $\hat{P}(\s)$ by inserting Eq.~(\ref{eq:Poi_unfolded_P}) into Eq.~(\ref{eq:NN_NNN_joint_spacing}),
\begin{equation}\label{eq:joint_NN_NNN_ddim}
    \hat{P}\left(\s_1,\s_2\right)=d^2\Gamma(1+1/d)^{2d}\s_1^{d-1}\s_2^{d-1}e^{-\Gamma(1+1/d)^d\s_2^d}\heav{\s_2-\s_1},
\end{equation}
we can also compute it explicitly by the same geometric arguments as above, to confirm that the reasoning employed in Section~\ref{subsection:ratio_joint_spacings} leading to Eq.~(\ref{eq:NN_NNN_joint_spacing}) is indeed correct. To that end, consider the setup of Fig.~\ref{fig:poisson_sketch}~$(b)$. With one of the $N$ levels at the origin as the reference level, there are $N-1$ levels which can be in $[s_1,s_1+\d s_1]$, $N-2$ levels in $[s_2,s_2+\d s_2]$ and $N-3$ levels outside of $s_2$. It follows that
\begin{equation}
    \hat{P}\left(s_1,s_2\right)=(N-1)\frac{V_d\left(s_1+\d s_1\right)-V_d\left(s_1\right)}{V_d\left(R\right)}(N-2)\frac{V_d\left(s_2+\d s_2\right)-V_d\left(s_2\right)}{V_d\left(R\right)}\left(1-\frac{V_d\left(s_2\right)}{V_d\left(R\right)}\right)^{N-3}.
\end{equation}
Expanding $V_d(s_{1,2}+\d s_{1,2})$ to first order in $\d s_{1,2}$, changing to the unfolded scale and taking limits, we arrive at Eq.~(\ref{eq:joint_NN_NNN_ddim}).

Finally, the joint distribution of the $k$NN spacings, Eq.~(\ref{eq:kNN_joint_spacings}), reads, in $d$ dimensions:
\begin{equation}\label{eq:joint_k_ddim}
     \hat{P}(\s_1,\dots,\s_k)=d^k\Gamma(1+1/d)^{k d}\prod_{j=1}^{k-1}\s_j^{d-1}e^{-\Gamma(1+1/d)^d\s_k^d}\,\prod_{j=1}^k\Theta(\s_{j+1}-\s_j)\,.
\end{equation}

\subsection{Ratio distribution}
We now turn to the ratio distributions. Henceforth, we always assume we are at the unfolded scale and denote the spacings by $s$ instead of $\s$.
Inserting Eq.~(\ref{eq:Poi_unfolded_P}) into the last equality of Eq.~(\ref{eq:P(r)_definition}), we obtain
\begin{equation}\label{eq:NN_NNN_ratio_r_ddim}
    \begin{split}
        \varrho(r)&=\heav{1-r}d\, \Gamma(1+1/d)\,r^{d-1}\int_0^\infty \d s\,\frac{s^{2s-1}e^{-\Gamma(1+1/d)^ds^d(1+r^d)}}{\int_{rs}^\infty s'^{d-1}e^{-\Gamma(1+1/d)^ds'^d}}\\
        &=\heav{1-r}d\, \Gamma(1+1/d)^d\,r^{d-1}\int_0^\infty \d s\, s^{2s-1}e^{-\Gamma(1+1/d)^ds^d}\\
        &=d\, r^{d-1}\heav{1-r}.
    \end{split}
\end{equation}
The constraint enforced by the $\Theta$-function implies that the distribution is supported in the $d$-dimensional unit ball, which we parameterize by the radial distance $r$ and the $(d-1)$-dimensional solid angle $\Omega_{d-1}$. By recalling that $\int_0^1\d r\, r^{d-1}\int\d\Omega_{d-1}\varrho\left(r,\Omega_{d-1}\right)=\int_0^1\d r\varrho(r)$ and that the distribution is isotropic and hence $\varrho(r,\Omega_{d-1})$ is independent of $\Omega_{d-1}$, by using Eq.~(\ref{eq:NN_NNN_ratio_r_ddim}), and by noting that $\int\d\Omega_{d-1}=S_{d-1}$ gives the area of the unit sphere in $d$ dimensions and that $S_{d-1}/V_d(1)=d$, we conclude that
\begin{equation}\label{eq:poisson_ddim_ratio_final}
    \varrho(r,\Omega_{d-1})=\heav{1-r}\frac{d}{\int\d\Omega_{d-1}}=\heav{1-r}\frac{d}{S_{d-1}}=\heav{1-r}\frac{1}{V_d(1)},
\end{equation}
i.e.\ distribution is indeed flat since it is given by the inverse of the volume of its support. Specializing to the complex plane, we have shown that the distribution is flat on the unit disk,
\begin{equation}\label{eq:poisson_complex_ratio_final}
    \varrho_\mathrm{Poi}(z)=\frac{1}{\pi}\heav{1-\abs{z}^2}.
\end{equation}

We next consider the distribution of the $m$NN by $k$NN ratio, of which the above result is a special case ($m=1$, $k=2$). Inserting Eq.~(\ref{eq:Poi_unfolded_P}) into the last equality of Eq.~(\ref{eq:ratio_r_mk_def}), we obtain
\begin{equation}\label{eq:ratio_r_mk_final}
\begin{split}
    \varrho_{mk}(r_{mk})&=d^k\Gamma(1+1/d)^{kd}(r_{mk})^{d-1}\int\d s_1\cdots \d s_{m-1}\d s_{m+1}\cdots \d s_k\, s_1^{d-1}\cdots s_{m-1}^{d-1}s_{m+1}^{d-1}\cdots s_{k-1}^{d-1}s_k^{2d-1}\\
    &\times e^{-\Gamma(1+1/d)^ds_k^d}\heav{s_{k}-s_{k-1}}\cdots\heav{s_{m+1}-rs_k}\heav{rs_k-s_{m-1}}\cdots\heav{s_2-s_1}\\
    &=d^k\Gamma(1+1/d)^{kd}(r_{mk})^{d-1}\int_0^\infty\d s_k\, s_k^{2d-1} e^{-\Gamma(1+1/d)^ds_k^d}\\
    &\times\left(\int_{rs_k}^{s_k}\d s_{k-1}s_{k-1}^{d-1}\cdots\int_{rs_k}^{s_{m+2}}\d s_{m+1}s_{m+1}^{d-1}\right)\left(\int_0^{rs_k}\d s_{m-1}s_{m-1}^{d-1}\cdots\int_0^{s_2}\d s_1\,s_1^{d-1}\right)\\
    &=d^2\Gamma(1+1/d)^{kd}\frac{(r_{mk})^{md-1}(1-(r_{mk})^d)^{k-m-1}}{(m-1)!(k-m-1)!}\int_0^\infty \d s_k\,s_k^{kd-1}e^{-\Gamma(1+1/d)^ds_k^d}\\
    &=\binom{k-1}{m}d\,m\,(r_{mk})^{dm-1}(1-(r_{mk})^d)^{k-m-1}\,.
\end{split}
\end{equation}

\begin{figure}[tbp]
	\centering
	\includegraphics[width=0.95\textwidth]{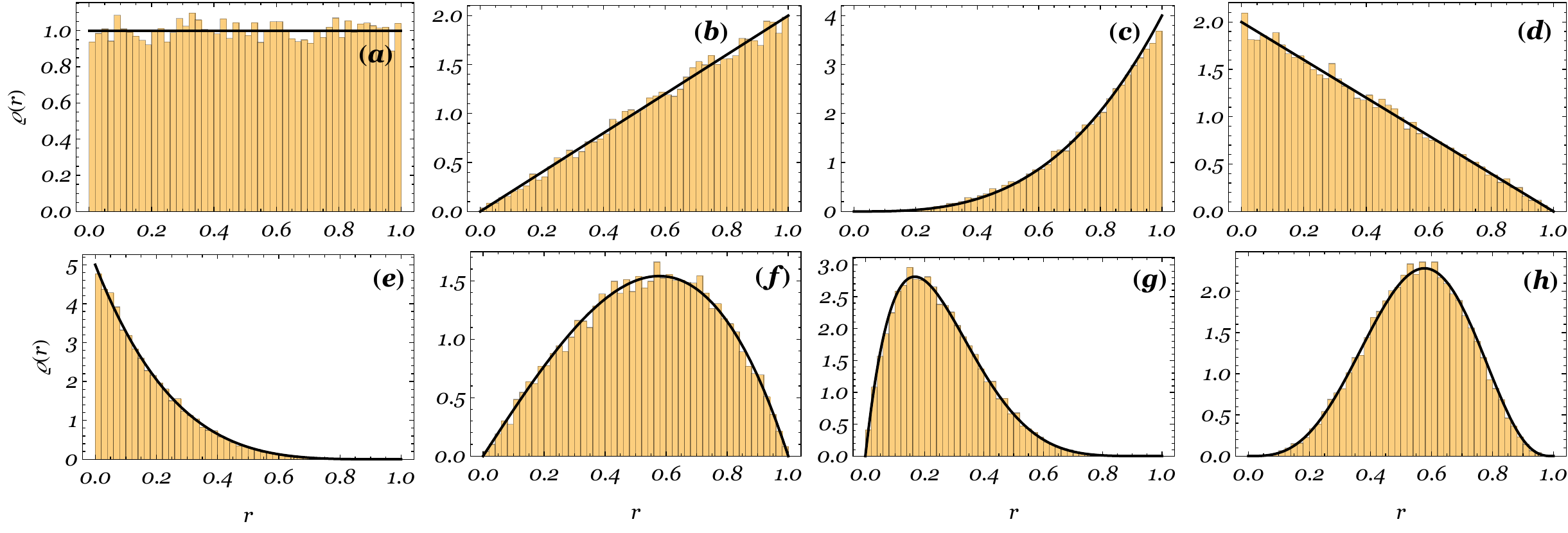}
	\caption{Comparison of analytic prediction for the $m$NN by $k$NN spacing ratio, Eq.~(\ref{eq:ratio_r_mk_final}), black line, with numerical results. Each histogram is obtained by computing the ratios for $20\,000$ iid levels. The numerical parameters are: $(a)$ $d=1$, $m=1$, $k=2$; $(b)$ $d=2$, $m=1$, $k=2$; $(c)$ $d=4$, $m=1$, $k=2$; $(d)$ $d=1$, $m=1$, $k=3$; $(e)$ $d=1$, $m=1$, $k=6$; $(f)$ $d=2$, $m=1$, $k=3$; $(g)$ $d=1$, $m=2$, $k=8$; $(h)$ $d=2$, $m=2$, $k=6$. $(a)$, $(b)$, $(c)$ correspond to the NN by NNN spacing ratio, Eq.~(\ref{eq:NN_NNN_ratio_r_ddim}), for real, complex and quaternionic spectra, respectively.}
	\label{fig:analytics_Poisson}
\end{figure}

Equation~(\ref{eq:ratio_r_mk_final}) constitutes the most general distribution for Poisson spacing ratios in $d$-dimensions. Figure~\ref{fig:analytics_Poisson} shows a comparison of Eq.~(\ref{eq:ratio_r_mk_final}) with numerical spacing ratios of $20\,000$ iid levels, for eight different combinations of $d,m,k$, showing perfect agreement in all cases.

Finally, note that, a flat distribution in the $d$-dimensional unit ball is only possible if $\varrho(r_{mk})\propto (r_{mk})^{d-1}$ for all $r_{mk}$, as in Eq.~(\ref{eq:NN_NNN_ratio_r_ddim}). This implies $m=1$ and $k-m=1\Rightarrow k=2$. We thus see that a flat distribution is a peculiarity of the NN by NNN ratio and is not achieved by any other combination of $m,k$. 

\section{Theoretical analysis II.\ Hermitian RMT}
\label{section:ratio_hermitian}

We now consider the NN by NNN spacing ratio for random matrices, addressing first hermitian ensembles. Analytical computations are more intricate here, because of the correlations of the levels. 

\subsection{Arbitrary hermitian ensembles}
Let again the spectrum of an (arbitrary for now) hermitian-RMT matrix be composed of levels $\{\lambda_k\}_{k=1}^N$, which are \emph{not} taken to be ordered. Since the joint eigenvalue distribution function $P^{(N)}(\{\lambda_k\})$ is invariant under permutations of levels, we reorder the set such that our reference level is $\lambda_1$, its NN $\lambda_2$ and its NNN $\lambda_3$.\footnote{Contrary to the previous section, we may not, in general, set the reference level $\lambda_1=0$, since $P^{(N)}(\{\lambda_k\})$ may not be invariant under translations.} These choices (together with the immediate implication that all other $N-2$ levels must be further away from $\lambda_1$ than $\lambda_3$ is) are enforced via the constraint
\begin{equation}
    \heav{(\lambda_3-\lambda_1)^2-(\lambda_2-\lambda_1)^2}\prod_{j>3}^{N}\heav{(\lambda_j-\lambda_1)^2-(\lambda_3-\lambda_1)^2}.
\end{equation}
The NN by NNN ratio is $r=(\lambda_2-\lambda_1)/(\lambda_3-\lambda_1)$. We can then immediately write down the expression for its distribution,
\begin{equation}
\begin{split}
    \varrho^{(N)}(r)=\int&\d\lambda_1\cdots\d\lambda_NP^{(N)}(\lambda_1,\dots,\lambda_N)\dirac{r-\frac{\lambda_2-\lambda_1}{\lambda_3-\lambda_1}}\\ \times&\heav{(\lambda_3-\lambda_1)^2-(\lambda_2-\lambda_1)^2}\prod_{j>3}^{N}\heav{(\lambda_j-\lambda_1)^2-(\lambda_3-\lambda_1)^2}.
\end{split}
\end{equation}

We next change variables to $u\equiv \lambda_1$, $v\equiv\lambda_3-\lambda_1$, $s_n=\lambda_{n+3}-\lambda_1$ ($n=1,\dots,N-3$), perform the integration in $\lambda_2$ using the $\delta$-function, and obtain
\begin{equation}\label{eq:hermitian_ratio_general}
    \varrho^{(N)}(r)=\heav{1-r^2}\int \d u\d v\prod_{j=1}^{N-3}\d s_j\heav{s_j^2-v^2}\, \abs{v}\,P^{(N)}\left(u,u+rv,u+v,u+s_1,\dots,u+s_{N-3}\right).
\end{equation}

\subsection{Gaussian ensembles}
\label{subsection:ratio_hermitian_gaussian}
Equation (\ref{eq:hermitian_ratio_general}) is valid for an arbitrary hermitian ensemble. We now specialize for the case of the Gaussian ensembles, GO/U/SE, labeled by the Dyson index $\beta$. The joint eigenvalue distribution reads
\begin{equation}
    P^{(N)}_\mathrm{GE}(x_1,\dots,x_N)\propto\exp{-\frac{1}{2}\sum_{j=1}^Nx_j^2}\prod_{j>k}^{N}\abs{x_j-x_k}^\beta\,.
\end{equation}
In terms of the variables of Eq.~(\ref{eq:hermitian_ratio_general}), we have
\begin{equation}
\begin{split}
    P^{(N)}_\mathrm{GE}&(u,u+rv,u+v,u+s_1,\dots,u+s_{N-3})\propto\\
    &\abs{r}^\beta\abs{1-r}^\beta\abs{v}^{3\beta}\prod_{j=1}^{N-3}\abs{s_k}^\beta\abs{s_k-v}^\beta\abs{s_k-rv}^\beta\prod_{j<k}^{N-3}\abs{s_j-s_k}^\beta\\
    &\times\,\exp{-\frac{1}{2}\left[Nu^2+2u\left((1+r)v+\sum_{j=1}^{N-3}s_j\right)+(1+r^2)v^2+\sum_{j=1}^{N-3}s_j^2\right]}\,.
\end{split}
\end{equation}

The integration in $u$ is Gaussian and can be readily performed, yielding
\begin{equation}
    \int\d u \exp{-\frac{1}{2}\left[Nu^2+2u\left((1+r)r+\sum_{j=1}^{N-3}s_j\right)\right]}\propto\exp{\frac{1}{2N}\left[(1+r)v+\sum_{j=1}^{N-3}s_j\right]^2}\,.
\end{equation}
We finally obtain the distribution of the ratio as an $(N-2)$-fold integral:
\begin{equation}\label{eq:hermitian_ratio_gaussian}
\begin{split}
    \varrho^{(N)}_\mathrm{GE}(r)\propto&\heav{1-r^2}\abs{r}^\beta\abs{1-r}^\beta\int \d v\,\abs{v}^{3\beta+1}\exp{-\frac{1}{2}v^2\left(1+r^2-\frac{(1+r)^2}{N}\right)}\\
    &\times\int\prod_{j=1}^{N-3}\d s_j \heav{s_j^2-v^2} \abs{s_j}^\beta \abs{s_j-v}^\beta \abs{s_j-rv}^\beta \exp{-\frac{1}{2}\left(s_j^2- \frac{1+r}{N}vs_j\right)}\\
    &\times\exp{-\frac{1}{N}\sum_{k,\ell=1}^{N-3}s_ks_\ell}\prod_{k<\ell}^{N-3}\abs{s_k-s_\ell}^\beta\,.
\end{split}
\end{equation}

For small $N$ ($N=3,4$) the integrals in Eq.~(\ref{eq:hermitian_ratio_gaussian}) can be computed exactly. Unfortunately, contrary to the consecutive spacings ratio, for NN by NNN ratios, the small-size expressions do not accurately describe the large-$N$ asymptotics.

For $N=3$, no $s_j$-integrals exist in Eq.~(\ref{eq:hermitian_ratio_gaussian}). Furthermore, the $r$-dependence can be factored out of the $v$-integral and no integrals have to be performed at all:
\begin{equation}\label{eq:gaussian_ratio_N3}
    \varrho^{(3)}_\mathrm{GE}(r)\propto\heav{1-r^2}\abs{r}^\beta\abs{1-r}^\beta\int \d v\,v^{3\beta+1}e^{-\frac{1}{3}v^2\left(1-r+r^2\right)}=\mathcal{N}\frac{\abs{r}^\beta\abs{1-r}^\beta}{(1-r+r^2)^{1+3\beta/2}}\heav{1-r^2},
\end{equation}
where the $\beta$-dependant normalization is $\mathcal N=9/4$ for $\beta=1$, $\mathcal N=27\sqrt{3}/(2\pi)$ for $\beta=2$ and $\mathcal N=243\sqrt{3}/(2\pi)$ for $\beta=4$. The distribution of Eq.~(\ref{eq:gaussian_ratio_N3}) for $\beta=2$ is plotted in Fig.~\ref{fig:analytics_GUE}~$(a)$ below, in comparison with exact diagonalization results.

For $N=4$, we must perform an additional integral in $s$ (here for $\beta=2$),
\begin{equation}\label{eq:GUE_N=4_start}
\begin{split}
    \varrho^{(4)}_\mathrm{GUE}(r)\propto\heav{1-r^2}r^2(1-r)^2\int_{-\infty}^{+\infty}\d v\, \abs{v}^7\exp{-\frac{3}{8}v^2\left(1+r^2-\frac{2}{3}r\right)}\\
    \times\int_{-\infty}^{+\infty}\d s\, s^2(s-v)^2(s-rv)^2\exp{-\frac{3}{8}s^2}\exp{\frac{1}{4}(1+r)vs}\heav{s^2-v^2}.
\end{split}
\end{equation}
If we denote 
\begin{equation}
    f(s,v,r)=\int\d s\, s^2(s-v)^2(s-rv)^2\exp{-\frac{3}{8}s^2}\exp{\frac{1}{4}(1+r)vs}\,,
\end{equation}
then Eq.~(\ref{eq:GUE_N=4_start}) reads
\begin{equation}
\begin{split}
    \varrho^{(4)}_\mathrm{GUE}&(r)\propto\heav{1-r^2}r^2(1-r^2)\\
     \times\bigg[
    &\int_0^{+\infty}\d v\abs{v}^7\exp{-\frac{3}{8}v^2\left(1+r^2-\frac{2}{3}r\right)}\left(f(s,v,r)\Big|^{s=-v}_{s=-\infty}+f(s,v,r)\Big|^{s=\infty}_{s=v}\right)\\
    +&\int_{-\infty}^0\d v\abs{v}^7\exp{-\frac{3}{8}v^2\left(1+r^2-\frac{2}{3}r\right)}\left(f(s,v,r)\Big|^{s=v}_{s=-\infty}+f(s,v,r)\Big|^{s=\infty}_{s=-v}\right)\bigg]\,,
\end{split}
\end{equation}
and is evaluated (with the correct normalization) to 
\begin{equation}\label{eq:GUE_N4}
\begin{split}
    \varrho^{(4)}_\mathrm{GUE}(r)=&\,\frac{1}{4\pi}
    \frac{r^2(1-r)^2\heav{1-r^2}}{(1-r+r^2)^7(8+3r^2)^{13/2}(4-4r+3r^3)^{9/2}}\\
    &\times\left(\sqrt{8+3r^2}B_1^{(4)}(r)+\sqrt{4-4r+3r^2}B_2^{(4)}(r)+\sqrt{3}\sqrt{8+3r^2}\sqrt{4-4r+3r^2}B_3^{(4)}(r)\right),
\end{split}
\end{equation}
with the polynomials $B^{(4)}_k$ given in Appendix~\ref{appendix:polynomials}, Eqs.~(\ref{eq:B4}).

When $N\to\infty$, we can rewrite Eq.~(\ref{eq:hermitian_ratio_gaussian}), discarding all exponentials suppressed by $1/N$:
\begin{equation}\label{eq:gaussian_ratio_N_infty}
\begin{split}
    \varrho^{(N\to\infty)}_\mathrm{GE}(r)\propto\heav{1-r^2}&\frac{\abs{r}^\beta\abs{1-r}^\beta}{(1+r^2)^{1+3\beta/2}}\int\d v\, v^{3\beta+1}e^{-v^2}\int\prod_{j=1}^N\d s_j\heav{s_j^2-\frac{v^2}{1+r^2}}\\
    &\times\abs{s_j}^\beta\abs{s_j-\frac{rv}{\sqrt{1+r^2}}}^\beta\abs{s_j-\frac{v}{\sqrt{1+r^2}}}^\beta e^{-s_j^2}\prod_{j<k}^N\abs{s_j-s_k}^\beta\,.
\end{split}
\end{equation}

Although we cannot compute this integral exactly, note that its multiplying prefactor gives the exact distribution of $r$ for $r\to0$. In fact, it also qualitatively describes the distribution for all $r$, albeit missing the exact heights of the peaks of positive and negative $r$. We thus obtain a loosely approximating distribution,
\begin{equation}\label{eq:gaussian_ratio_N_infty_approx}
    \varrho^{(N\to\infty)}_\mathrm{GE}(r)\approx\mathcal{N}\frac{r^\beta\abs{1-r}^\beta}{(1+r^2)^{1+3\beta/2}}\heav{1-r^2}.
\end{equation}
At any rate, the absence of a term $-r$ inside the denominator (which was killed by the limit $N\to\infty$) completely distinguishes this result from the case $N=3$, see Fig.~\ref{fig:analytics_GUE}~$(b)$ below (the black line is the approximation of Eq.~(\ref{eq:gaussian_ratio_N_infty_approx})).

\subsection{Circular ensembles}
\label{subsection:ratio_hermitian_circular}
We discussed in Section~\ref{section:ratio_definitions_overview} how the difference between the $N=3$ and $N\to\infty$ statistics is due to boundary effects. To eliminate them we should consider periodic boundary conditions, i.e.\ identify the ends of the spectrum. Hence, we consider the circular ensembles, whose spectrum is supported on the unit circle and whose joint eigenvalue distribution is 
\begin{equation}
    P^{(N)}_\mathrm{CE}\left(\phi_1,\dots,\phi_N\right)\propto\prod_{j<k}\abs{e^{i\phi_j}-e^{i\phi_k}}^\beta\,.
\end{equation}

Note that, although the eigenvalues are complex ($e^{i\phi_j}$), they are fully described by real angles $\phi_j\in(-\pi,\pi]$. The spacing ratio is defined in terms of the real variables, $r=(\phi_2-\phi_1)/(\phi_3-\phi_1)$, i.e.\ we are measuring the spacings on the circle, not in the embedding space, $\mathbb{C}$. By rotational invariance of the circle, we may set $\phi_1=0$. 
We can rewrite the Vandermonde interaction as $\abs{e^{i\phi_j}-e^{i\phi_k}}^\beta=\sin^\beta(\abs{\phi_j-\phi_k}/2)$. The general result of Eq.~(\ref{eq:hermitian_ratio_general}), applied to the circular ensembles, reads
\begin{equation}
\begin{split}
    \varrho^{(N)}_\mathrm{CE}(r)&\propto\heav{1-r^2}\int_{-\pi}^{\pi} \d v\abs{v}\sin^\beta\frac{\abs{v}}{2}\sin^\beta\frac{\abs{rv}}{2}\sin^\beta\frac{\abs{(1-r)v}}{2}\\
    &\times\int_{-\pi}^{\pi}\prod_j\d s_j\heav{s_j^2-v^2}\sin^\beta\frac{\abs{s_j}}{2}\sin^\beta\frac{\abs{s_j-rv}}{2}\sin^\beta\frac{\abs{s_j-v}}{2}\prod_{j<k}^{N-3}\sin^\beta\frac{\abs{s_j-s_k}}{2}\,.
\end{split}
\end{equation}

We now evaluate the preceding integral for $N=3$ and $N=4$, restricting ourselves to the complex case, $\beta=2$. We have $\sin^2((\phi_j-\phi_k)/2)=2(1-\cos(\phi_k-\phi_j))$.

For $N=3$, a single integral in $v$ is to be performed,
\begin{equation}
    \varrho^{(3)}_\mathrm{CUE}(r)\propto\heav{1-r^2}\int_{-\pi}^\pi\d v \abs{v}\left(1-\cos v\right)\left(1-\cos rv\right)\left(1-\cos(r-1)v\right),
\end{equation}
which yields, after normalization, the result
\begin{equation}\label{eq:CUE_N3}
\begin{split}
    \varrho^{(3)}_\mathrm{CUE}(r)=&\,\frac{1}{48\pi^2}\frac{\heav{1-r^2}}{(r-2)^2(r-1)^2(r-\frac{1}{2})^2r^2(r+1)^2}\\
    \times&\left(Q_1^{(3)}(r)+Q_2^{(3)}(r)\cos(\pi r)+Q_3^{(3)}(r)\cos(2\pi r)+Q_4^{(3)}(r)\sin(\pi r)+Q_5^{(3)}(r)\sin(2\pi r)\right),
\end{split}
\end{equation}
where the polynomials $Q^{(3)}_k(r)$ are given in Appendix~\ref{appendix:polynomials}, Eqs.~(\ref{eq:Q3}).
The distribution of Eq.~(\ref{eq:CUE_N3}) is plotted in red in Fig.~\ref{fig:analytics_GUE}~$(b)$.

For $N=4$, we have an additional integral in $s$ to perform,
\begin{equation}\label{eq:CUE_N=4_start}
\begin{split}
    \varrho^{(4)}_\mathrm{CUE}(r)\propto\heav{1-r^2}\int_{-\pi}^\pi\d v \abs{v}\left(1-\cos v\right)\left(1-\cos rv\right)\left(1-\cos(r-1)v\right)\\
    \times
    \int_{-\pi}^\pi\d s\left(1-\cos s\right)\left(1-\cos(s-rv)\right)\left(1-\cos(s-v)\right)\heav{s^2-v^2}.
\end{split}
\end{equation}
If we denote 
\begin{equation}
    f(s,v,r)=\int\d s\left(1-\cos s\right)\left(1-\cos(s-rv)\right)\left(1-\cos(s-v)\right),
\end{equation}
then Eq.~(\ref{eq:CUE_N=4_start}) reads
\begin{equation}
\begin{split}
     \varrho^{(4)}_\mathrm{CUE}&(r)\propto\heav{1-r^2}\\
     \times\bigg[
     &\int_0^\pi\d v\abs{v}\left(1-\cos v\right)\left(1-\cos rv\right)\left(1-\cos(r-1)v\right)\left(f(s,v,r)\Big|^{s=-v}_{s=-\pi}+f(s,v,r)\Big|^{s=\pi}_{s=v}\right)\\
     +&\int_{-\pi}^0\d v\abs{v}\left(1-\cos v\right)\left(1-\cos rv\right)\left(1-\cos(r-1)v\right)\left(f(s,v,r)\Big|^{s=v}_{s=-\pi}+f(s,v,r)\Big|^{s=\pi}_{s=-v}\right)\bigg]\,,
\end{split}
\end{equation}
and is evaluated (with the correct normalization) to
\begin{equation}\label{eq:CUE_N4}
\begin{split}
    \varrho^{(4)}_\mathrm{CUE}(r)=&\,\frac{1}{2^{19} 3^{17} \pi ^3}
    \frac{1}{(r-6)^2 (r-5)^2 (r-4)^2 (r-3)^3 \left(r-\frac{5}{2}\right)^2 (r-2)^3 \left(r-\frac{3}{2}\right)^3 \left(r-\frac{4}{3}\right)^2}\\
    \times&\frac{1}{(r-1)^3\left(r-\frac{2}{3}\right)^3 \left(r-\frac{1}{2}\right)^3 \left(r-\frac{1}{3}\right)^3 r^3 \left(r+\frac{1}{3}\right)^2 \left(r+\frac{1}{2}\right)^3 \left(r+\frac{2}{3}\right)^2 (r+1)^3}\\
    \times&\frac{1}{\left(r+\frac{4}{3}\right)^2 \left(r+\frac{3}{2}\right)^2 (r+2)^3\left(r+\frac{5}{2}\right)^2 (r+3)^2 (r+4)^2 (r+5)^2 (r+6)^2}\\
    \times&\Big(Q_1^{(4)}(r)+Q_2^{(4)}(r)\cos(\pi r)+Q_3^{(4)}(r)\cos(2\pi r)+Q_4^{(4)}(r)\cos(3\pi r)\\
    &+Q_5^{(4)}(r)\sin(\pi r)+Q_6^{(4)}(r)\sin(2\pi r)+Q_7^{(4)}(r)\sin(3\pi r)\Big)\,,
\end{split}
\end{equation}
where $Q^{(4)}_k$ are polynomials which we do not detail here.
The distribution of Eq.~(\ref{eq:CUE_N4}) is plotted in blue in Fig.~\ref{fig:analytics_GUE}~$(b)$.

\subsection{Comparison with numerical results}

\begin{figure}[htbp]
	\centering
	\includegraphics[width=\textwidth]{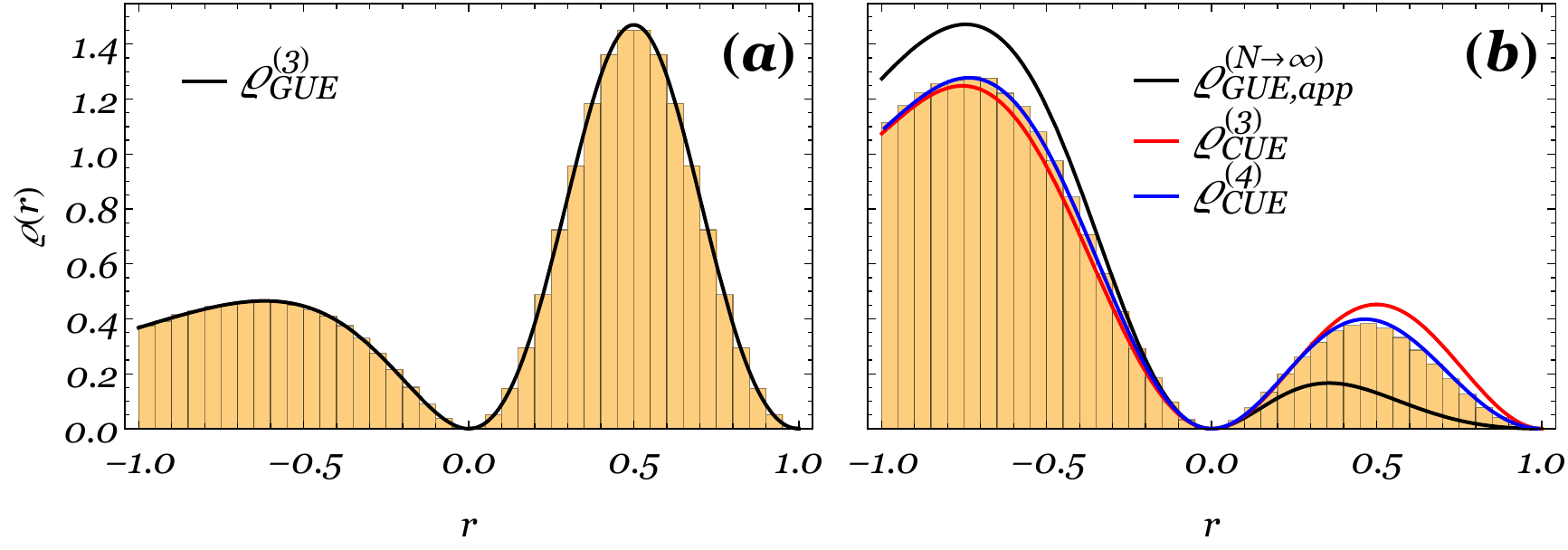}
	\caption{Comparison of exact diagonalization (ED) of GUE-drawn random matrices with analytical results. $(a)$: ED for $N=3$, black line given by exact expression, Eq.~(\ref{eq:gaussian_ratio_N3}), with $\beta=2$; $(b)$: ED for $N=10^4$, black line given by approximate GUE result for $N\to\infty$, Eq.~(\ref{eq:gaussian_ratio_N_infty}), red line given by exact result for $N=3$ CUE, Eq.~(\ref{eq:CUE_N3}), blue line given by exact result for $N=4$ CUE, Eq.~(\ref{eq:CUE_N4}).}
	\label{fig:analytics_GUE}
\end{figure}

In Sections~\ref{subsection:ratio_hermitian_gaussian} and \ref{subsection:ratio_hermitian_circular}, we made several analytical predictions for the ratio distribution, which are compared to numerical results from exact diagonalization of GUE-drawn matrices in Fig.~\ref{fig:analytics_GUE}. For small size-matrices ($N=3$ represented in Fig.~\ref{fig:analytics_GUE}~$(a)$, $N=4$ not shown) the agreement is perfect, which was to be expected since the computation is exact. These results are not, however, particularly useful, since universality is only displayed for large $N$. In that limit, see Fig.~\ref{fig:analytics_GUE}~$(b)$, we can only give approximate distributions. The best approximation obtained from the GUE joint eigenvalue distribution (given by the prefactor of the exact distribution, Eq.~(\ref{eq:gaussian_ratio_N_infty_approx})), although capturing the essential features, is not accurate enough for quantitative purposes. On the other hand, the small-size surmises computed for the circular ensembles describe, quantitatively, very well the universal large-$N$ asymptotics. Indeed, the distribution for the $N=4$ CUE, Eq.~(\ref{eq:CUE_N4}), is already indistinguishable from the $N\to\infty$ limit.

\section{Theoretical analysis III.\ Nonhermitian RMT}
\label{section:ratio_nonhermitian}

\subsection{Arbitrary nonhermitian ensembles}
We now turn to complex spectra. We consider an $N\times N$ arbitrary matrix from a nonhermitian ensemble whose complex eigenvalues are $\{\lambda_k\}_{k=1}^{N}$, $\lambda_k=x_k+iy_k$, and their joint distribution function is $P^{(N)}(\lambda_1,\dots,\lambda_N)= P^{(N)}(x_1,\dots,x_N;y_1,\dots,y_N)$.
We again consider the first level $\lambda_1$ to be the reference level, its NN to be $\lambda_2$ and its NNN to be $\lambda_3$. The complex NN by NNN spacing ratio is 
\begin{equation}
\begin{split}
    z\equiv&\, r e^{i\theta}\equiv x+iy=\frac{\lambda_2-\lambda_1}{\lambda_3-\lambda_1}\\
    =&\,\frac{(x_2-x_1)(x_3-x_1)+(y_2-y_1)(y_3-y_1)}{(x_3-x_1)^2+(y_3-y_1)^2}+
    i\,\frac{(x_3-x_1)(y_2-y_1)-(x_2-x_1)(y_3-y_1)}{(x_3-x_1)^2+(y_3-y_1)^2}\,.
\end{split}
\end{equation}
We introduce new variables $u\equiv x_1$, $v\equiv y_1$, $p\equiv x_2-x_1$, $q\equiv y_2-y_1$, $s\equiv x_3-x_1$, $t\equiv y_3-y_1$, $a_n\equiv x_{n+3}-x_1$, $b_n\equiv y_{n+3}-y_1$, $n=1,\dots,{N-3}$. In terms of these new variables, the $\delta$-function constraints (fixing the real and imaginary parts of $z$) are
\begin{equation}\label{eq:delta_constraints}
\begin{split}
    &\dirac{x-\frac{ps+qt}{s^2+t^2}}\dirac{y-\frac{sq-pt}{s^2+t^2}}\\
    =(s^2+t^2)&\,\delta\Big(p-\left(sx-ty\right)\Big)\,\delta\Big(q-\left(tx+sy\right)\Big)\,,
\end{split}
\end{equation}
and the $\Theta$-function constraints (requiring all $\lambda_n$ with $n>3$ to be further away from $\lambda_1$ than $\lambda_3$) are
\begin{equation}\label{eq:theta_constraints}
    \heav{(s^2+t^2)-(p^2+q^2)}\prod_{j=4}^{N}\\\heav{(a_j^2+b_j^2)-(s^2+t^2)}.
\end{equation}
The distribution function of $z$ is again obtained by integrating the joint eigenvalue distribution multiplied by the constraints of Eqs.~(\ref{eq:delta_constraints}) and (\ref{eq:theta_constraints}). Integrating over $p$ and $q$ using the $\delta$-functions, we arrive at the distribution for $z$ for an arbitrary nonhermitian ensemble:
\begin{equation}\label{eq:nonhermitian_ratio_general}
\begin{split}
    &\varrho^{(N)}(x,y)=\heav{1-(x^2+y^2)}\int\d u\d v\d s\d t\prod_{j=1}^{N-3}\d a_j \d b_j\heav{(a_j^2+b_j^2)-(s^2+t^2)}\,(s^2+t^2)\\
    &\times P^{(N)}(u,u+sx-ty,u+s,u+a_1,\dots,u+a_{N-3};v,v+tx+sy,v+t,v+b_1,\dots,v+b_{N-3})\,.
\end{split}
\end{equation}

\subsection{Ginibre Unitary Ensemble}
\label{subsection:ratio_nonhermitian_ginibre}
We now restrict ourselves to the GinUE (complex Gaussian iid entries, $\beta=2$), whose joint eigenvalue distribution reads:
\begin{equation}
    P^{(N)}_\mathrm{GinUE}(x_1,\dots,x_N;y_1,\dots,y_N)=\prod_{j<k}\left[(x_j-x_k)^2+(y_j-y_k)^2\right]\exp{-\sum_{j=1}^N\left(x_j^2+y_j^2\right)}\,.
\end{equation}

Replacing the $x_j$, $y_j$ by the variables of Eq.~(\ref{eq:nonhermitian_ratio_general}) and performing the Gaussian integration over the two variables $u$, $v$, we arrive at the ratio distribution for the Ginibre ensemble,
\begin{equation}\label{eq:complex_ratio_N}
\begin{split}
    &\varrho^{(N)}_\mathrm{GinUE}(x,y)\propto\heav{1-(x^2+y^2)}(x^2+y^2)(1+x^2+y^2-2x)\\
    &\times\int \d s\d t\,(s^2+t^2)^4\exp{-(s^2+t^2)\left[(1+x^2+y^2)\left(1-\frac{1}{N}\right)-\frac{2}{N}x\right]}\\
    &\times\int\prod_{j=1}^{N-3}\d a_j\d b_j\heav{a_j^2+b_j^2-(s^2+t^2)}\left(a_j^2+b_j^2\right)\left((a_j-s)^2+(b_j-t)^2\right)\\
    &\times\left[(a_j-sx+ty)^2+(b_j-tx-sy)^2\right]\prod_{j<k}^{N-3}\left[(a_j-a_k)^2+(b_j-b_k)^2\right]\\
    &\times\exp{-\sum_{j=1}^{N-3}\left[a_j^2\left(1-\frac{1}{N}\right)-\frac{1}{N}\sum_{k\neq j}a_ja_k-\frac{2}{N}a_j\left\{s(1+x)-t y\right\}\right]}\\
    &\times\exp{-\sum_{j=1}^{N-3}\left[b_j^2\left(1-\frac{1}{N}\right)-\frac{1}{N}\sum_{k\neq j}b_jb_k-\frac{2}{N}b_j\left\{sy+t(1+x\right\}\right]}\,.
\end{split}
\end{equation}

As before, the distribution for $N=3$ follows from Eq.~(\ref{eq:complex_ratio_N}) without the need to perform any integrals explicitly. Indeed, in polar coordinates $x=r\cos\theta$, $y=r\sin\theta$, $s=\xi\cos\varphi$, $t=\xi\sin\varphi$, we have
\begin{equation}
    \varrho^{(3)}_\mathrm{GinUE}(r,\theta)\propto \heav{1-r}r^2(1+r^2-2r\cos\theta)\int \d \xi\,\xi^9 \exp{-\frac{2}{3}\xi^2(1+r^2-r\cos\theta)}\,.
\end{equation}
$z$ can be scaled out of the remaining integral, and we get, after normalization,
\begin{equation}\label{eq:GinUE_N3}
    \varrho^{(3)}_\mathrm{GinUE}(r,\theta)=\frac{81}{8\pi}\frac{r^2(1+r^2-2r\cos\theta)}{(1+r^2-r\cos\theta)^5}\heav{1-r}.
\end{equation}
Equation~(\ref{eq:GinUE_N3}) describes exactly the exact diagonalization results for $N=3$, see Figs.~\ref{fig:analytics_GinUE}~$(a)$--$(d)$. The distribution for other small-$N$ could also be computed explicitly.

As for the hermitian case, the leading order behavior (i.e.\ the first term in a power expansion in $r$) of the distribution for $N\to\infty$ can be obtained without carrying out any integral. Although it does not give a good quantitative match, it captures the high (low) density at large (small) angles.
By discarding all exponential suppressed by $1/N$ from Eq.~(\ref{eq:complex_ratio_N}), factoring out terms containing $z$ from the $s$ and $t$ integrals, we obtain the prefactor,
\begin{equation}\label{eq:GinUE_Ninfty_polar}
    \varrho^{(N\to\infty)}_\mathrm{GinUE}(r,\theta)\approx\frac{12}{\pi}\frac{r^2(1+r^2-2r\cos\theta)}{(1+r^2)^5}\heav{1-r}.
\end{equation}

\subsection{Toric Unitary Ensemble}
\label{subsection:ratio_nonhermitian_toric}

We now want to eliminate boundary effects from a complex spectrum, by considering a 2-dimensional analog of the circular ensembles. Recall that the circular ensembles had eigenvalues on the unit circle $\mathbb{S}^1$. A possible generalization would be to consider eigenvalues on the sphere $\mathbb{S}^2\subset\mathbb{R}^3$, which would be provided by the Spherical Unitary Ensemble (SUE)~\cite{forrester1992,brouwer1995,krishnapur2009,forrester2010,forrester2016}, of matrices $A^{-1}B$ with both $A$, $B$ GinUE matrices. However, while belonging to the same universality class as the GinUE, also for the SUE the convergence to the large-$N$ limit is quite slow. Instead, we consider eigenvalues on the 2-dimensional (Clifford) torus $\mathbb{T}^2=\mathbb{S}^1\times\mathbb{S}^1\subset \mathbb{S}^3\subset\mathbb{R}^4$. 

We parameterize the torus by two angles $\vartheta\in(-\pi,\pi]$, $\varphi\in(-\pi,\pi]$, with a generic point $P\in\mathbb{T}^2$ given by $P=(1/\sqrt{2})\left(\cos\vartheta,\sin\vartheta,\cos\varphi,\sin\varphi\right)$. In analogy with the CUE, we want to construct a flat joint eigenvalue distribution on $\mathbb{T}^2$, which we call the distribution of the toric unitary ensemble (TUE). Since the torus has no curvature, the distribution is simply given by the Vandermonde interaction, $P^{(N)}_\mathrm{TUE}(\vartheta_1,\dots,\vartheta_N;\varphi,\dots,\varphi_N)\propto \abs{\Delta_{\mathbb{T}^2}}^2$. $\Delta_{\mathbb{T}^2}$ is given by the distance between points in the embedding space parameterized by the eigenvalues. That is, if $P_j\in\mathbb{T}^2$ is parameterized by the angles $(\vartheta_j,\varphi_j)$ then the Vandermonde interaction is $\abs{\Delta_{\mathbb{T}^2}}=\prod_{j<k}\norm{P_j-P_k}_{\mathbb{R}^4}$. One can check that, with this reasoning, the usual Vandermonde terms for the Gaussian, Ginibre, circular and spherical ensembles coincide with, respectively, $\abs{\Delta_{\mathbb{R}}}=\prod_{j<k}\norm{P_j-P_k}_{\mathbb{R}}$, $\abs{\Delta_{\mathbb{R}^2}}=\prod_{j<k}\norm{P_j-P_k}_{\mathbb{R}^2}$, $\abs{\Delta_{\mathbb{S}^1}}=\prod_{j<k}\norm{P_j-P_k}_{\mathbb{R}^2}$, $\abs{\Delta_{\mathbb{S}^2}}=\prod_{j<k}\norm{P_j-P_k}_{\mathbb{R}^3}$ (with $P_j$ in the respective embedding spaces). Using our parameterizantion of the torus and considering only $\beta=2$, the Vandermonde interaction reads $\abs{\Delta_{\mathbb{T}^2}}^2=\prod_{j<k}\left[2-\cos(\vartheta_j-\vartheta_k)-\cos(\varphi_j-\varphi_k)\right]$.

We can then write down the joint eigenvalue distribution for the toric unitary ensemble (TUE),
\begin{equation}\label{eq:TUE_joint}
    P^{(N)}_\mathrm{TUE}(\vartheta_1,\dots,\vartheta_N;\varphi_1,\dots,\varphi_N)\propto \prod_{j<k}\left[2-\cos(\vartheta_j-\vartheta_k)-\cos(\varphi_j-\varphi_k)\right].
\end{equation}

Two interesting questions are whether there is a nonhermitian matrix ensemble whose eigenvalues are distributed according to Eq.~(\ref{eq:TUE_joint}) and whether there are physical problems for which the TUE arises naturally.

Having introduced the relevant joint eigenvalue distribution, the remaining procedure is straightforward. By rotational invariance in both factors $\mathbb{S}^1$, we can set $\vartheta_1=0$ and $\varphi_1=0$. The complex spacing ratio is, accordingly, $z=(\vartheta_2+i\varphi_2)/(\vartheta_3+i\varphi_3)$. If we then insert Eq.~(\ref{eq:TUE_joint}) into the general ratio distribution, Eq.~(\ref{eq:nonhermitian_ratio_general}), we obtain
\begin{equation}\label{eq:TUE_finite_N}
\begin{split}
    &\varrho^{(N)}_\mathrm{TUE}(x,y)\propto\int_{-\pi}^\pi\d s\d t \prod_{j=1}^{N-3}\d a_j \d b_j \heav{(a_j^2+b_j^2)-(s^2+t^2)}(s^2+t^2)^2\left[2-\cos s-\cos t\right]\\
    &\times\left[2-\cos(sx-ty)-\cos(t x+s y)\right]\left[2-\cos(s(x-1)-ty)-\cos(t (x-1)+s y)\right]\\
    &\times\prod_{j=1}^{N-3}\left[2-\cos a_j-\cos b_j\right]
    \left[2-\cos(s-a_j)-\cos(t-b_j)\right]\\
    &\times\left[2-\cos(sx-ty-a_j)-\cos(t x+s y-b_j)\right]\prod_{j<k}\left[2-\cos(a_j-a_k)-\cos(b_j-b_k)\right].
\end{split}
\end{equation}

For $N=3$, the double integral to be performed is
\begin{equation}\label{eq:TUE_N3}
\begin{split}
    \varrho^{(3)}_\mathrm{TUE}(x,y)\propto\int_{-\pi}^\pi&\d s\d t (s^2+t^2)^2\left[2-\cos s-\cos t\right]\left[2-\cos(sx-ty)-\cos(t x+s y)\right]\\
    &\times\left[2-\cos(s(x-1)-ty)-\cos(t (x-1)+s y)\right].
\end{split}
\end{equation}
The integral of Eq.~(\ref{eq:TUE_N3}), and its generalizations for $N=4,5,\dots$, can be numerically integrated (the analytic expression is far too involved to be useful) and describe very well the large-$N$ asymptotics of the GinUE universality class, see Figs.~\ref{fig:analytics_GinUE} $(e)$--$(h)$.

\subsection{Comparison with numerical results}

\begin{figure}[htbp]
	\centering
	\includegraphics[width=\textwidth]{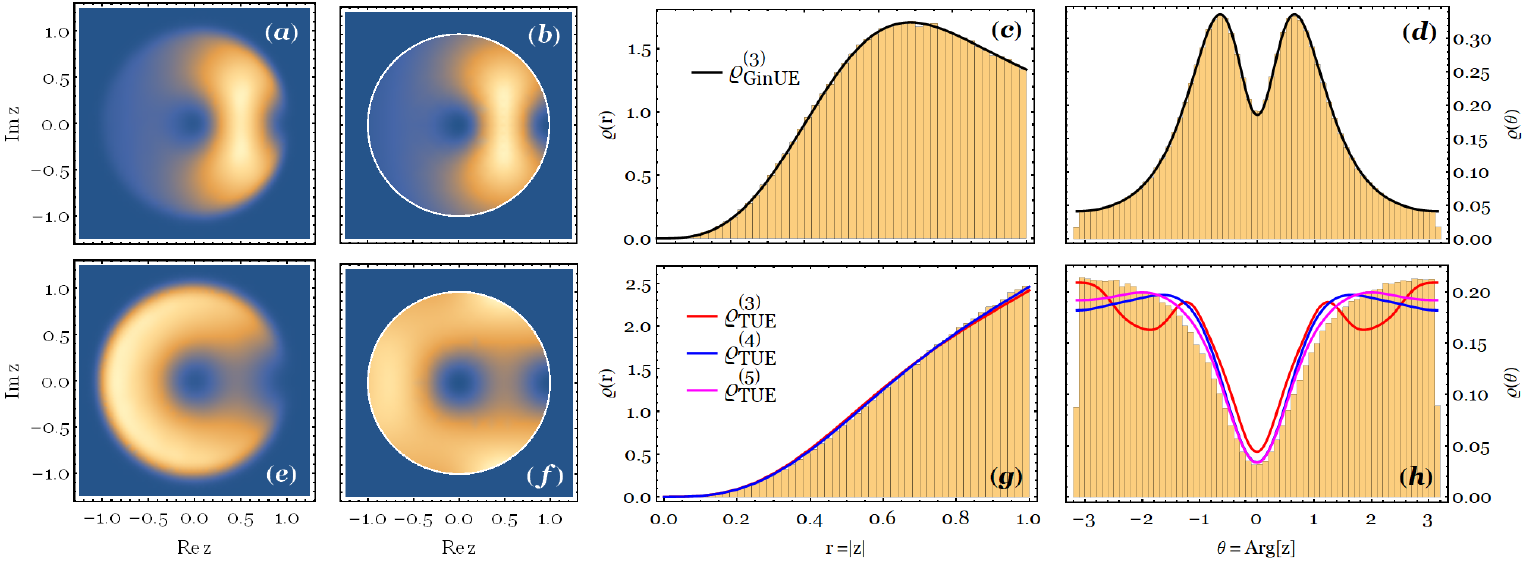}
	\caption{Comparison of exact diagonalization (ED) of GinUE-drawn matrices with analytical results, for $N=3$ in $(a)$--$(d)$ and $N\to\infty$ in $(e)$--$(h)$. $(a)$: ED for $N=3$; $(b)$: exact distribution for $N=3$ GinUE, Eq.~(\ref{eq:GinUE_N3}); $(c)$: histogram of absolute value of ratios from ED; $(d)$: histogram of argument of ratios from ED; black line computed from Eq.~(\ref{eq:GinUE_N3}); $(e)$ ED for GinUE matrices with $N=10^4$; $(f)$: surmise from $N=3$ TUE, Eq.~(\ref{eq:TUE_N3}); $(g)$: histogram of absolute value of ratios from ED for $N=10^4$ GinUE-matrices; $(h)$: histogram of argument of ratios from ED; red, blue and magenta curves computed from Eq.~(\ref{eq:TUE_finite_N}) for $N=3$, $4$, $5$, respectively.}
	\label{fig:analytics_GinUE}
\end{figure}

The various results obtained in Sections~\ref{subsection:ratio_nonhermitian_ginibre} and \ref{subsection:ratio_nonhermitian_toric} are summarized and compared to numerical results from exact diagonalization of GinUE-drawn matrices in Fig.~\ref{fig:analytics_GinUE}. Once again, the exactness of the computation for $N=3$ leads to perfect agreement of both approaches, Figs.~\ref{fig:analytics_GinUE}~$(a)$--$(d)$. The universal large-$N$ asymptotics, Figs.~\ref{fig:analytics_GinUE}~$(e)$--$(h)$, are well described by the surmises obtained from the TUE. The convergence of the radial marginal distribution is similar to the real case: both $N=3$ and $N=4$ are really good approximations, the latter being already almost indistinguishable from exact diagonalization. The angular marginal distribution has a much slower convergence, especially near the edges (the backward direction). Although the qualitative description is acceptable for $N=3$, quantitatively even $N=5$ does not yet describe it perfectly, although the agreement does improve as $N$ increases.

\cleardoublepage


\chapter{Conclusions}
\label{chapter:conclusions}

\section{Summary of achievements}
\label{section:achievements}
In this thesis, we addressed the critical problem of establishing the spectral and steady-state properties of stochastic Liouvilians of the Lindblad type. By analyzing an ensemble of stochastic Liouvillians where unitary dynamics coexists with $r$ independent dissipation channels, we found that the dispersion of the decay rates, the spectral gap, and the properties of the steady-state are divided into different regimes, namely a regime of weak-dissipation, a crossover, and a strong dissipative regime. However, the boundaries of each regime do not necessarily coincide for the different observables. Each of the scaling regimes, as well as their boundaries, can be characterized by a set of exponents which rule the dependence of the observable on the effective dissipation strength, $g_{\text{eff}}$, and on the system size $N$. We found that these exponents are independent of the universality index $\beta$ of random matrices and determined their scaling exponents. 

The main identified spectral and steady-state properties are as follows. The support of the spectrum passes from an ellipsoid to the lemon-like shape reported in Ref.~\cite{denisov2018} for the case of $r=N^{2}-1$, as dissipation increases. For fixed $g_{\text{eff}}$ and $r>1$, the spectral gap increases with $N$ and its distribution becomes peaked in the $N\to\infty$ limit. The mean gap has a single scaling regime, which we were able to compute explicitly using holomorphic Green's function methods, and is strongly influence by finite-size effects. The steady-state is close to a fully mixed state and its spectral statistics exhibit a crossover from Poissonian to GUE (GOE) for $\beta=2$ $(1)$ as a function of $g_{\text{eff}}N^{1/2}$. 

The case of a single dissipation channel, $r=1$, is qualitatively different. For $r=1$, $\Delta$ vanishes with increasing $N$ in the strong dissipative regime. Also, with increasing system size and $r>1$, the steady-state purity approaches that of the maximally mixed state, $\mathcal{P}_{\text{FM}}=1/N$, in the week dissipative regime, while for strong dissipation it attains a value larger than, though proportional to, $\mathcal{P}_{\text{FM}}$. 

The results summarized above, which are closely related to spectral densities and edge statistics of complex spectra, were complemented by a study of the spectral correlations in the bulk of complex spectra. This was achieved by introducing complex spacing ratios. We found that angular correlations between levels in dissipative systems provide a clean signature of quantum chaos: Poisson processes, which describe integrable systems, have a flat, and hence isotropic, ratio distribution in the complex plane, while for RMT ensembles from the Ginibre universality class there is a suppression of small angles in the large-$N$ limit. We also reencountered the familiar cubic level repulsion in the latter case. 

Our results show that complex spacing ratios allow one to clearly distinguish (known or conjectured) integrable systems from chaotic ones. In particular, we looked at the case of boundary-driven spin-chain Liouvillians, including a finite-size numerical analysis of the problem.

Exact, closed-form, expressions for the ratio distribution were computed for Poisson processes in $d$ dimensions. For RMT ensembles, both hermitian and nonhermitian, exact expressions could only be given in terms of multiple integrals over the joint eigenvalue distributions, which were evaluated for small-size matrices. However, due to strong finite-size effects, the latter expressions were of no practical use. By noting that the strong finite-size dependence was actually a boundary effect in the Gaussian ensembles, we overcame it by considering the circular ensemble of the same universality class, which has eigenvalues on a boundaryless support. For the nonhermitian case, a two-dimensional generalization of the CUE, the Toric Unitary Ensemble was introduced with the same purpose and its joint eigenvalue distribution determined. Using the CUE and TUE, Wigner-like surmises for the complex spacing ratios of $3\times3$ and $4\times4$ matrices were written down and were found to describe very well the large-$N$ universal behavior of the GUE and GinUE universality classes, respectively. The angular marginal distribution was found to have, however, a slower convergence towards that limit.

\section{Open questions and future work}
\label{section:future_work}
A natural question our results raise is---which regime characterizes a particular physical system? As in the case of chaotic Hamiltonian dynamics, this has to be determined on a case-by-case basis, which we will delegate to future works. The predictive power of the present analysis relies on the fact that, once this regime is determined, the system's universal properties are obtained solely by symmetry arguments.

Furthermore, several technical challenges remain to be addressed, the primary one being, arguably, obtaining the spectral density of a random Liouvillian inside the bulk of the spectrum. A direct, brute-force, diagrammatic expansion of the nonholomorphic Green's function could work, although the combinatorial complexity of the expansion may render this approach unfeasible. If one restricts oneself to the weak dissipation limit, the diagrammatic expansion can be performed exclusively in the degenerate subspace, simplifying somewhat the computation. Also of interest would be to compute the spectral density of the steady-state, which, for the case of a single dissipation channel, can be written down exactly. A theoretical explanation of the results of Sections~\ref{section:steady_state} and \ref{section:r=1} could be pursued in this manner. 

Regarding complex spacing ratios, some generalizations could be addressed, for instance, higher-order spacings or intermediate statistics.
\cleardoublepage


\phantomsection
\addcontentsline{toc}{chapter}{\bibname}

\bibliographystyle{apsrev4-1}
{\small\bibliography{Thesis_Bib_DB}}
\cleardoublepage

\appendix





\chapter{Polynomials for Chapter~\ref{chapter:spacings}}
\label{appendix:polynomials}

\begingroup
\allowdisplaybreaks

In this appendix, we provide explicit expressions for the polynomials $B_k^{(4)}$ and $Q_k^{(3)}$ appearing in Chapter~\ref{chapter:spacings}. The polynomials $Q_k^{(4)}$ can also be computed analytically, but are far too long to present here.

The polynomials $B^{(4)}_k$, appearing in Eq.~(\ref{eq:GUE_N4}), are given by:
\begin{subequations}\label{eq:B4}
\begin{align}
    \begin{split}
    B_1^{(4)}(r)
    &=12951414 r^{27}-116562726 r^{26}+716326335 r^{25}-3277037250 r^{24}+12270189015 r^{23}\\
    &-38963484108 r^{22}+107498737062 r^{21}-262084488084 r^{20}+570149074368 r^{19}\\
    &-1117174783248 r^{18}+1980961864272 r^{17}-3199495203936 r^{16}+4722358240224 r^{15}\\
    &-6412609449216 r^{14}+8048567177536 r^{13}-9427117015424 r^{12}+10375815396352 r^{11}\\
    &-10842021943296 r^{10}+10772209840128 r^9-10161095221248 r^8+8951997923328 r^7\\
    &-7224758108160 r^6+5174767386624 r^5-3185572315136 r^4+1605155946496 r^3\\
    &-621654573056 r^2+163829514240 r-21843935232\,,
    \end{split}\\
    \begin{split}
    B_2^{(4)}(r)
    &=-12951414 r^{27}+107928450 r^{26}-655886403 r^{25}+2951065872 r^{24}-10985595759 r^{23}\\
    &+32119902594 r^{22}-70513221474 r^{21}+95491032120 r^{20}+14317577808 r^{19}\\
    &-526540502544 r^{18}+1795990853552 r^{17}-4065272606784 r^{16}+7116794858016 r^{15}\\
    &-10182070683712 r^{14}+12140824025664 r^{13}-12296549315328 r^{12}+10764201081344 r^{11}\\
    &-8626103116800 r^{10}+6961675011072 r^9-6347896606720 r^8+6303744909312 r^7\\
    &-6100007436288 r^6+5151738871808 r^5-3634899517440 r^4+2036060258304 r^3\\
    &-866165456896 r^2+248843599872 r-37235982336\,,
    \end{split}\\
    \begin{split}
    B_3^{(4)}(r)
    &=14880348 r^{26}-124002900 r^{25}+755709102 r^{24}-3386420784 r^{23}+12623875758 r^{22}\\
    &-39842512308 r^{21}+110119167900 r^{20}-269048029248 r^{19}+589943145024 r^{18}\\
    &-1165945663872 r^{17}+2094745567104 r^{16}-3426572906496 r^{15}+5131663704576 r^{14}\\
    &-7036436210688 r^{13}+8870013361152 r^{12}-10269680615424 r^{11}+10956130172928 r^{10}\\
    &-10742680977408 r^9+9693073833984 r^8-7989878587392 r^7+5986043559936 r^6\\
    &-3999516327936 r^5+2336231522304 r^4-1140917796864 r^3+440301256704 r^2\\
    &-118380036096 r+16911433728\,.
    \end{split}
\end{align}
\end{subequations}

The polynomials $Q^{(3)}_k$, appearing in Eq.~(\ref{eq:CUE_N3}), are given by:
\begin{subequations}\label{eq:Q3}
\begin{align}
    \begin{split}
    Q_1^{(3)}(r)
    &=8 \left(8+3 \pi ^2\right) r^{10}-40 \left(8+3 \pi ^2\right) r^9+50 \left(8+3 \pi ^2\right) r^8+40 \left(8+3 \pi ^2\right) r^7\\
    &-2 \left(653+174 \pi ^2\right) r^6+2 \left(727+60 \pi ^2\right) r^5+25 \left(6 \pi ^2-29\right) r^4-40 \left(3 \pi ^2-1\right) r^3\\
    &+\left(253+24 \pi ^2\right) r^2-180 r+36\,,
    \end{split}\\
    \begin{split}
    Q_2^{(3)}(r)&=256 r^7-896 r^6+960 r^5-160 r^4-64 r^3-192 r^2+160 r-32\,,
    \end{split}\\
    \begin{split}
    Q_3^{(3)}(r)&=-16 r^8+64 r^7-38 r^6-110 r^5+117 r^4+24 r^3-61 r^2+20 r-4\,,
    \end{split}\\
    \begin{split}
    Q_4^{(3)}(r)&=128 \pi  r^8-512 \pi  r^7+480 \pi  r^6+352 \pi  r^5-688 \pi  r^4+192 \pi  r^3+80 \pi  r^2-32 \pi  r\,,
    \end{split}\\
    \begin{split}
    Q_5^{(3)}(r)&=-32 \pi  r^9+144 \pi  r^8-132 \pi  r^7-210 \pi  r^6+350 \pi  r^5-14 \pi  r^4-178 \pi  r^3+80 \pi  r^2-8 \pi  r\,.
    \end{split}
\end{align}
\end{subequations}

\endgroup 
\cleardoublepage

\end{document}